\documentclass[usenatbib]{mnras}
\usepackage{setspace}
\usepackage{epsf}
\usepackage{graphicx}
\usepackage{color}
\usepackage{txfonts}
\usepackage{tabularx}
\usepackage{here}
\usepackage{float}
\usepackage{threeparttable}
\usepackage{tabu}

\usepackage[T1]{fontenc}
\usepackage{aecompl}

\newcommand{\oi}{[O\,{\sc i}]}
\newcommand{\oii}{[O\,{\sc ii}]}
\newcommand{\oiii}{[O\,{\sc iii}]}

\newcommand{\NI}{[N\,{\sc i}]}
\newcommand{\nii}{[N\,{\sc ii}]}

\newcommand{\sii}{[S\,{\sc ii}]}
\newcommand{\siii}{[S\,{\sc iii}]}
\newcommand{\siv}{[S\,{\sc iv}]}

\newcommand{\hei}{He\,{\sc i}}

\newcommand{\neii}{[Ne\,{\sc ii}]}
\newcommand{\neiii}{[Ne\,{\sc iii}]}

\newcommand{\arii}{[Ar\,{\sc ii}]}
\newcommand{\ariii}{[Ar\,{\sc iii}]}
\newcommand{\ariv}{[Ar\,{\sc iv}]}

\newcommand{\clii}{[Cl\,{\sc ii}]}
\newcommand{\cliii}{[Cl\,{\sc iii}]}

\newcommand{\feiii}{[Fe\,{\sc iii}]}

\newcommand{\ha}{H$\alpha$}
\newcommand{\hb}{H$\beta$}
\newcommand{\hi}{H\,{\sc i}}

\usepackage{color} 
\newcommand{\kms}{km s$^{-1}$}

\newcommand{\te}{$T_{\rm e}$}
\newcommand{\Ne}{$n_{\rm e}$}

\newcolumntype{C}{>{\centering\arraybackslash}X}
\newcolumntype{L}{>{\raggedright\arraybackslash}X}
\newcolumntype{R}{>{\raggedleft\arraybackslash}X}

\title[XSHOOTER spectroscopy of \mbox{Lin49}]
{
XSHOOTER spectroscopy of the enigmatic planetary nebula \mbox{Lin49} in
the Small Magellanic Cloud\thanks{
Based on observations made with ESO Telescopes at the Paranal
Observatory under program ID 091.C-0934(B) (PI: Lex Kaper) during Amsterdam GTO time.}
}

\author[Otsuka et al.]
{
\begin{minipage}{1.0\linewidth}
Masaaki Otsuka$^{1}$\thanks{E-mail:otsuka@asiaa.sinica.edu.tw},
F.~Kemper$^{1}$, 
M.~L.~Leal-Ferreira$^{2}$\thanks{CNPq/Brazil Fellow},
I.~Aleman$^{2}$,
J.~Bernard-Salas$^{3}$,
J.~Cami$^{4,5}$,
B.~Ochsendorf$^{2,6}$,
E.~Peeters$^{4,5}$,
P.~Scicluna$^{1}$
\end{minipage}
\\
\\
\begin{minipage}{1.0\linewidth}
\begin{itemize}
\setlength{\itemsep}{0cm}
\item[] \noindent$^{1}$Academia Sinica, Institute of Astronomy and
	Astrophysics,
	11F Astronomy-Mathematics Building, NTU/AS campus, No. 1, Sec. 4, Roosevelt Rd., Taipei 10617, Taiwan, Republic of China
\item[] \noindent$^{2}$Leiden Observatory, University of Leiden, PO Box 9513, 2300 RA,
         Leiden, The Netherlands
\item[] \noindent$^{3}$Department of Physical Sciences, The Open University, Milton
 Keynes, MK7 6AA, UK
\item[] \noindent$^{4}$Department of Physics and Astronomy, The University of Western Ontario, London, ON N6A 3K7, Canada
\item[] \noindent$^{5}$SETI Institute, 189 Bernardo Ave, Suite 100,
	Mountain View, CA 94043, USA
\item[] \noindent$^{6}$Space Telescope Science Institute, 3700 San Martin Drive,
Baltimore, MD 21218, USA
\end{itemize}
\end{minipage}
}

\begin{document}

\date{}

\pagerange{\pageref{firstpage}--\pageref{lastpage}} \pubyear{2016}

\maketitle

\label{firstpage}

\begin{abstract}
We performed a detailed spectroscopic analysis of the fullerene 
C$_{60}$-containing planetary nebula (PN) \mbox{Lin49} in the Small
Magellanic Cloud using XSHOOTER at the ESO VLT and the \emph{Spitzer}/IRS
 instruments. We derived nebular abundances for nine elements. We used {\sc
 TLUSTY} to derive photospheric parameters for the central star. 
\mbox{Lin49} is C-rich and metal-deficient PN
 ($Z$$\sim$0.0006). The nebular abundances are in good agreement with
 Asymptotic Giant Branch nucleosynthesis models for stars with initial
 mass 1.25\,M$_{\odot}$ and metallicity $Z$ = 0.001. Using the {\sc
 TLUSTY} synthetic spectrum of the central star to define the
 heating and ionising source, we constructed
 the photoionisation model with {\sc CLOUDY} 
that matches the observed spectral energy distribution (SED) and the
 line fluxes in the UV to far-IR wavelength ranges simultaneously.
We could not fit the $\sim$1-5\,$\mu$m SED using a model with
0.005-0.1\,$\mu$m-sized graphite grains and a constant hydrogen density
 shell owing to the prominent near-IR excess, while at other wavelengths 
the model fits the observed values reasonably well. We argue that the near-IR 
excess might indicate either (1) the presence of very small particles in the form
of small carbon clusters, small graphite sheets, or 
fullerene precursors, or (2) the presence of a high-density structure 
surrounding the central star. We found that SMC C$_{60}$ PNe 
show a near-IR excess component to lesser or greater degree. 
This suggests that these C$_{60}$ PNe might maintain a structure 
nearby their central star.
\end{abstract}

\section{Introduction}

\begin{figure}
\centering
\includegraphics[width=0.8\columnwidth]{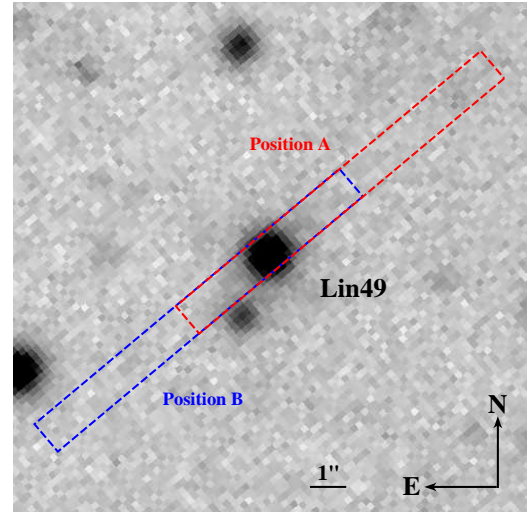}
\caption{Image of \mbox{Lin49} in the $z^{'}$-band and the slit positions used in the
 XSHOOTER observations. We observed \mbox{Lin49} on the slit positions A and B. The
 averaged FWHM amongst nine nearby stars is $\sim$0.69{\arcsec}.
}
\label{image}
\end{figure}

The discovery of C$_{60}$ in the C-rich 
planetary nebula (PN) Tc1 \citep{Cami:2010aa} confirmed the presence outside the 
Solar System of the enigmatic molecule buckminsterfullerene C$_{60}$, 
first discovered by \citet{Kroto_85_C60}. Since then, C$_{60}$ has been
identified towards ten other PNe in the Milky Way
\citep{Cami:2010aa,Garcia-Hernandez:2010aa,Garcia-Hernandez:2011aa,
Garcia-Hernandez:2012aa,Otsuka:2013aa,Otsuka:2014aa}, 
bringing the total to 11 detections out of a sample of 338, both C-rich
and O-rich, PNe
observed with the Infrared Spectrograph \citep[IRS;][]{Houck:2004aa} on
the \emph{Spitzer} Space Telescope. Assuming that the evolved star content of 
the Milky Way is 1/3 C-rich and 2/3 O-rich
\citep{Ishihara:2011aa}, it can be inferred that fullerenes occur in
about 10\,\% of the Galactic C-rich PNe, although this number may
be lower if a larger fraction of Galactic PNe are C-rich. 
For C-rich PNe, \citet{Garcia-Hernandez:2015aa} reports a detection rate
of $\sim$5\,\%, $\sim$20\,\%, and $\sim$44\,\% in the Milky Way, the Large
Magellanic Cloud (LMC), and Small Magellanic Cloud (SMC), respectively. 
This indicates that the processing of fullerenes may depend on the 
metallicity, with fullerenes being more often detected in low-metallicity 
environments. In most cases, even the two strongest C$_{60}$ 
resonances at 17.4\,\micron\ and 18.9\,\micron\ are rather weak with respect to
the local continuum emission around these wavelengths, with the notable
exception of the PN \mbox{Lin49} (Fig.~\ref{image}) in the SMC, 
which appears to have C$_{60}$ 17.4\,\micron\ and 18.9\,\micron\ features of 
very similar strength and appearance to what is seen towards
Tc1. The similarities in their infrared spectra and the similar C$_{60}$
band strengths motivated us to know more about physical properties of
\mbox{Lin49}.

However, little is known about \mbox{Lin49}. 
Prior to its \emph{Spitzer}/IRS observation, 
\mbox{Lin49} only occurs in some catalogues as an SMC
PN
\citep{Lindsay_61_new,Dopita_85_kinematics,Meyssonnier_93_New,Morgan_95_UKST}
until recently. The source was selected for spectroscopic follow-up
with \emph{Spitzer} based on its mid-infrared IRAC photometric colours, 
which suggested a pre-main sequence nature (G.~Sloan, private
communication). The \emph{Spitzer}/IRS spectrum 
revealed that \mbox{Lin49} is a C-rich dust PN, showing strong C$_{60}$ 
resonances at 17.4\,$\mu$m and 18.9\,$\mu$m and similar dust features such as the
broad 11\,$\mu$m and 30\,$\mu$m bands seen in the other
C$_{60}$-containing LMC and SMC PNe 
\citep{Sloan:2014aa,Ruffle_15_Spitzer}, but the physical properties of the
central star and dusty nebula remain unknown. 
Therefore, we wanted to further characterise \mbox{Lin49} 
using the XSHOOTER UV-near-IR spectrograph \citep{Vernet:2011aa} on the ESO Very 
Large Telescope (VLT) UT2 (Kueyen), in combination with
the \emph{Spitzer}/IRS spectrum. In the case of \mbox{Lin49}, the well 
determined distance to the SMC allows us to accurately 
determine the luminosity of the central star, the size of the nebula, 
and the total gas and dust masses in the nebula, and then clarify the 
current evolutionary stage of the central star and estimate the initial mass.

In this study, we present a spectroscopic analysis of \mbox{Lin49} in
order to study the physical conditions and chemical properties 
of this interesting PN. This is part of an ongoing study to understand 
in more depth the physical and chemical properties of
fullerene-containing PNe. Although we expect that these studies give 
us information on why fullerenes formed and exists in these PNe, 
the aim of this specific paper is not to investigate the formation and 
processing of fullerene molecules.

The remainder of this paper is organised as follows. In Section~\ref{S:S2}, we
describe our XSHOOTER observation and the data reduction of the XSHOOTER
spectrum and the archived \emph{Spitzer}/IRS spectrum. The results of 
plasma-diagnostic and ionic and elemental abundance derivations using
nebular lines, derivations of photospheric properties, and spectral
energy distribution (SED) fitting are described in Section~\ref{S:S3}. 
In Section~\ref{S:S4}, we discuss the prominent near-IR excess 
found in \mbox{Lin49}, and we give interpretations of this feature. We 
discuss the SEDs of SMC C$_{60}$ PNe and non-C$_{60}$ C-rich PNe 
in the SMC by comparing with the SED of \mbox{Lin49}. 
We compare physical properties of the C$_{60}$-containing PNe 
and counterparts in the SMC. Finally, we summarise the works in Section~\ref{S:S5}.

\section{Observations and data reductions \label{S:S2}}
\subsection{ESO/VLT XSHOOTER spectroscopy \label{S:XSHOOTER}}

\begin{figure}
\centering
\includegraphics[height=\columnwidth,angle=-90]{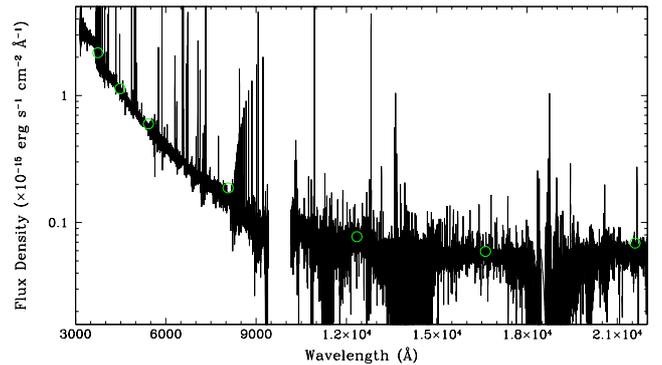}
\caption{The XSHOOTER spectrum of Lin49. The flux density was
 scaled to the $V$-band magnitude $m_{V}$ = 17.225 from the Magellanic
 Clouds Photometric Survey \citep[MCPS,][]{Zaritsky:2002aa} in the UVB
 and VIS spectra and the Two Micron All Sky Survey
 \citep[2MASS,][]{Skrutskie:2006aa} 
$J$-band magnitude $m_{J}$ = 16.58 $\pm$ 0.08 \citep{Sloan:2014aa} 
in the NIR spectrum. The green circles are these photometry results. 
Interstellar extinction was corrected for both the XSHOOTER spectrum and the photometry.}
\label{xsh}
\end{figure}

We obtained a UV to near-infrared spectrum using the medium resolution
spectrograph XSHOOTER, attached to the Cassegrain
focus of the 8.2-m VLT UT2 at the ESO Paranal
observatory, in Chile, on 2013 July 17 (UT). The XSHOOTER instrument consists of 
three spectroscopic arms: UVB, VIS, and NIR; and it covers the wavelength range from 
2936\,{\AA} to 24800\,{\AA}. The weather conditions during the exposure were stable, and the 
seeing recorded in the DIMM seeing monitor was $\sim$0.65-1.04{\arcsec}. 
For the UVB and VIS arms, we inserted the atmospheric dispersion correctors
(ADCs) in front of the slits in order to minimise the differential
atmospheric dispersion throughout the broad wavelength range. We used a slit size of
1.0{\arcsec}$\times$11{\arcsec} in the UVB arm and
0.9{\arcsec}$\times$11{\arcsec} in the other arms. We selected the 1$\times$1 
binning mode in each detector. The difference of the slit width in each
arm\footnote{The slit width is 1.0$\arcsec$ in the UVB and 0.9$\arcsec$ in the VIS and NIR arms,
respectively. The {\hb}\,4861\,{\AA} line is detected in the UVB arm.} 
and the difference of plate scale along the spatial direction on
each detector in each echelle order\footnote{These were measured directly from
the observed spectra; 0.16$''$-0.17$''$, 0.15$''$-0.17$''$,
and 0.24$''$-0.26$''$ in the UVB, VIS, and NIR arms, respectively.} have been taken into account 
in the normalisation of the emission line fluxes $F$($\lambda$) with respect to the {\hb} 
flux $F$({\hb}). We observed \mbox{Lin49} and the flux standard star GD153 \citep{Bohlin:1995aa} 
in the two different locations on the slit with a position angle of 219$^{\circ}$, i.e., 
using an AB sequence in series of 120~sec exposures and an ABBA sequence in exposures of 600~sec 
(the separation between A and B positions is 5{\arcsec}). In
Fig.~\ref{image}, we show the slit positions on the
$z^{'}$-band ($\lambda_{\rm c}$ = 8897\,{\AA}) image taken 
by the acquisition and guiding camera.

We reduced the data using the echelle 
spectra reduction package {\sc ECHELLE} and the two-dimensional 
spectra reduction package {\sc TWODSPEC} in {\sc IRAF}\footnote{IRAF 
is distributed by the National Optical Astronomy Observatories, 
operated by the Association of Universities for Research in
Astronomy (AURA), Inc., under a cooperative agreement with the National
Science Foundation.}. We subtracted the sky background and the bias 
current directly from the object frames. In the sequence, we subtracted the scattered light using the {\sc
IRAF} task {\sc APSCATTER}. We used the intensity normalised instrumental 
flat frame to correct the sensitivity of each pixel in the residual
frames and grating blaze function in each echelle order. 
We extracted the spectra between 3161\,{\AA} and 5904\,{\AA} in the UVB
arm, 5578\,{\AA} and 10255\,{\AA} in the VIS arm, and 9919\,{\AA} and 24791\,{\AA} in the NIR
arm. For the wavelength calibration of the UVB and VIS spectra, we used the Th-Ar 
comparison lines, and for the calibration of the NIR spectra, we used the OH lines 
recorded in the object frames in addition to Hg/Ar/Ne/Xe 
comparison lines.  The resulting resolving power 
($\lambda$/$\Delta\lambda$) is 8663-9650 in the UVB arm, 8409-8473 in
the VIS arm, and 4289-5417 in the NIR arm, measured from 
the full width at half maxima (FWHMs) of over 400 comparison lines in
each arm. After we corrected the count-rates for airmass 
and median combined the frames of \mbox{Lin49} and GD153, 
we performed flux calibration and telluric corrections. 
The resulting XSHOOTER spectrum is displayed in Fig.~\ref{xsh}.

The resulting signal-to-noise (S/N) ratios measured in the continuum of the
resultant spectrum are $\gtrsim$~10. Fringes appear in the
UVB spectrum with amplitudes $\sim$4-6\,$\%$ of the local
continuum intensity. These fringes pose a problem in determining
the baseline of the continuum and subsequent equivalent width
measurements and line-profile fittings in the stellar absorption analysis. Therefore, in order to minimise the 
fringing effect, we derived a smoothed spectrum using 9-pixel medians. As a result, 
the fringe amplitude decreased to $\sim$2\,$\%$ and the spectral resolution decreased 
to $\sim$1/3 of the original value. We used this smoothed spectrum in the stellar 
absorption analysis.

\subsection{\emph{Spitzer}/IRS spectroscopy \label{S:Spitzer}}

\begin{figure}
\centering
\includegraphics[width=\columnwidth,clip]{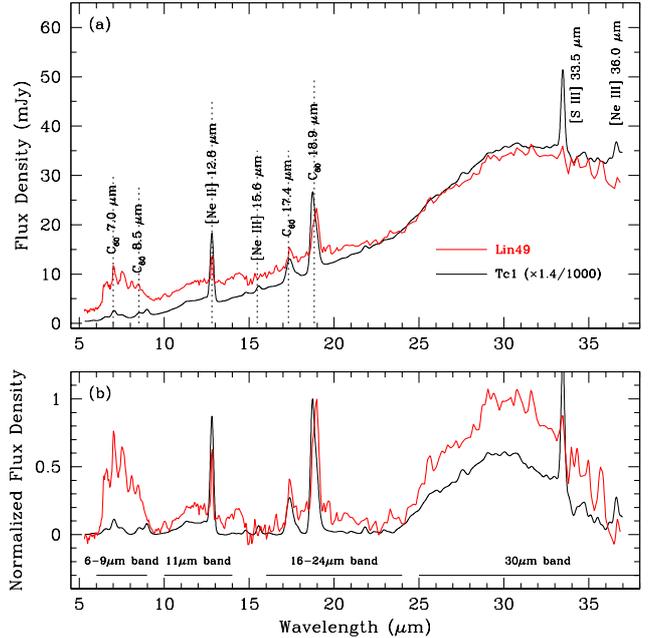}
\caption{
({\it panel~a}) \emph{Spitzer}/IRS spectra of \mbox{Lin49} and Tc1. The
  spectral resolution of the Tc 1 spectrum was reduced to match 
  that of the \mbox{Lin49} spectrum. The positions of
 prominent
 atomic gas emission lines as well as C$_{60}$ bands are indicated. ({\it panel~b}) Comparison between the
  intensity normalised spectra of \mbox{Lin49} and Tc1. 
We subtracted the local continuum by spline fitting
 in order to highlight the emission from dust grains and molecules, and
 then we normalised the resulting spectra to 
the peak flux density of the C$_{60}$ 18.9\,$\mu$m band.
\label{spt_spec}
}
\end{figure}

We analysed the archival mid-infrared \emph{Spitzer}/IRS spectra taken with the SL 
(5.2-14.5\,$\mu$m) and the LL modules (13.9-39.9\,$\mu$m). 
The data were originally taken by G.~Sloan (Program ID: 50240, AOR Key: 27537664) 
on 2008 August 4 and presented in \cite{Sloan:2014aa}.
We processed them using the data reduction packages 
{\sc SMART} v.8.2.9 \citep{Higdon:2004aa} and IRSCLEAN v.2.1.1,
provided by the \emph{Spitzer} Science Center. Since the flux density of the 
\emph{Spitzer}/MIPS \citep[Multiband Imaging Photometer for
Spitzer;][]{2004ApJS..154...25R} spectrum at the band 24\,$\mu$m
($\lambda_{\rm centre}$ = 23.84\,$\mu$m) is
9.77(--14) $\pm$ 3.90(--15)
erg~s$^{-1}$~cm$^{-2}$~$\mu$m$^{-1}$\footnote{Here
and henceforth we use the notation 9.77(--14) to mean $9.77 \times 10^{-14}$} \citep{Sloan:2014aa}, 
and 
this value is consistent with the corresponding band flux density
in the \emph{Spitzer}/IRS spectrum, we do not 
perform flux density correction.

In Fig.~\ref{spt_spec}, we present the resulting spectrum (red line)
along with the spectrum of Tc1 (black line). The spectral resolution of the Tc1 data
taken by the short-high and long-high resolution modules was 
reduced to match that of \mbox{Lin49}'s. We did not remove 
atomic gas lines from the Tc1 spectrum, so the C$_{60}$
18.9\,$\mu$m and [S\,{\sc iii}]\,18.7\,$\mu$m line complex in Tc1 is
shifted towards the blue relative to the same complex in \mbox{Lin49}.

\mbox{Lin49} and Tc1 show a broad 6--9\,$\mu$m band, and broad 11\,$\mu$m and 
30\,$\mu$m bands. The 17.4\,$\mu$m and 18.9\,$\mu$m C$_{60}$ resonances are very strong with
respect to the local continuum. The band profiles and strengths of these C$_{60}$ features in
both PNe are very similar. The 6-9\,$\mu$m profiles in \mbox{Lin49} and Tc1 are similar to the 6-9\,$\mu$m 
thermal emission from hydrogenated amorphous carbon (HAC) as displayed
in \citet{Scott:1997aa}. HAC is a generic name for 
a mixture of aliphatic and aromatic carbon, consisting of polycyclic aromatic
hydrocarbon (PAH) clusters embedded within a matrix of aliphatically bonded material.

The differences between \mbox{Lin49} and Tc1 
are the degree of excitation of the nebula (the [Ne\,{\sc
iii}]\,15.55/36.01\,$\mu$m lines are too weak to be clearly seen in
\mbox{Lin49}, suggesting that the
excitation degree of the \mbox{Lin49}'s nebula is significantly lower
than that of Tc1; indeed, we could not detect the [Ne\,{\sc iii}]
nebular lines in the XSHOOTER spectrum) and the broad
16--24\,$\mu$m band. As far as we know, the broad 16--24\,$\mu$m feature
has been seen in C-rich PNe and it is not limited to 
fullerene-containing C-rich PNe.  Although the nature of this feature has been discussed by 
\citet{2009ApJ...699.1541B}, \citet{Garcia-Hernandez:2012aa},
\citet{Otsuka:2013aa}, and \citet{Otsuka:2015ab},
the carrier is still under debate.

\section{Results\label{S:S3}}

\subsection{Nebular line analysis}

\subsubsection{Flux measurements and interstellar extinction}

\begin{table}
\centering
\caption{The calculated $c$({\hb}). We used $c$({\hb}) for
 each spectral band. By adopting the
$f$($\lambda$) of \citet{1989ApJ...345..245C} with $R_{V}$ = 3.1 and the
 average $c$({\hb}), we derived $E$($B$-$V$) = 0.07 $\pm$ 0.01 toward
\mbox{Lin49} (including the extinction in the Milky Way) using the relation: 
$c$({\hb}) = 1.45$E$($B$-$V$).
\label{T:chb}}
\begin{tabu} to \columnwidth {@{}LCL@{}}
\hline
\multicolumn{1}{c}{Band}    &$c$({\hb})   &Using lines\\
\hline
XSHOOTER-UVB	&0.10 $\pm$ 0.04&	H$\gamma$\\	 
XSHOOTER-VIS	&0.12 $\pm$ 0.02&	H$\alpha$\\
XSHOOTER-NIR-$J$	&0.10 $\pm$ 0.01&	Pa$\gamma$, Pa$\beta$\\			
XSHOOTER-NIR-$H$	&0.10 $\pm$ 0.02&	Br10\\
XSHOOTER-NIR-$K$	&0.11 $\pm$ 0.02&	Br$\gamma$\\
Average &0.11 $\pm$ 0.01\\
\hline
\end{tabu}
\end{table}

We identified 186 atomic emission lines in the XSHOOTER and \emph{Spitzer}/IRS data of \mbox{Lin49}.
From Gaussian fits, we obtained central wavelengths and fluxes for
these emission lines. Dereddened line fluxes $I$($\lambda$) were calculated
using the following formula: 
\begin{equation}
I(\lambda) = F(\lambda)~\cdot~10^{c({\rm H}\beta)(1+f(\lambda))},
\end{equation}
\noindent 
where $F$($\lambda$) is the observed flux, $c$(H$\beta$) is the reddening coefficient 
normalised by H$\beta$, and $f$($\lambda$) is the interstellar extinction function at 
$\lambda$ computed from the reddening law. 
Several extinction functions for the Milky Way and the
Magellanic Clouds are available
\citep[e.g.,][]{Savage:1979aa,Seaton:1979aa,Howarth:1983aa,Prevot:1984aa,
Fitzpatrick:1986aa,1989ApJ...345..245C}, 
with no significant difference in the value for XSHOOTER wavelengths. In the present work, 
we adopted the $f$($\lambda$) from \citet{1989ApJ...345..245C} with $R_{V}$=3.1.

We derived $c$({\hb}) from the comparison of the observed ratios of H$\gamma$, H$\alpha$,
Pa$\gamma$, Pa$\beta$, Br10\,1.736\,$\mu$m, and Br$\gamma$ to
H$\beta$ with the corresponding 
theoretical ratios given by \citet{1995MNRAS.272...41S} for an electron temperature 
{\te} = 10$^{4}$\,K and electron density {\Ne} = 10$^{4}$\,cm$^{-3}$,
under the Case B assumption. We list the calculated
$c$({\hb}) values and their 1-$\sigma$ uncertainty in Table~\ref{T:chb}. For each spectral band 
we adopt its corresponding value of $c$({\hb}) to perform the extinction correction. 
The fluxes of the detected lines in appendix Table~\ref{xstab} are normalised to $I$({\hb}) = 100.

\subsubsection{Flux normalisation of \emph{Spitzer}/IRS and the {\hb} flux of the whole PN}

Ideally, one would use the hydrogen fluxes given by the \emph{Spitzer}/IRS
observations to normalise the {\neii}\,12.81\,$\mu$m flux
($F$({\neii}\,12.81\,$\mu$m) = (1.45 $\pm$ 0.05)$\times$10$^{-14}$~erg s$^{-1}$ cm$^{-2}$).
This is preferred because there would be no need to correct for the interstellar 
reddening and for the difference in aperture sizes. However, we were not able 
to isolate the H\,{\sc i}\,7.46/11.31/12.37\,$\mu$m lines to measure 
their fluxes, as these are weak lines in the spectrum of \mbox{Lin49} and are potentially 
contaminated by the C$_{60}$ 7.0\,$\mu$m and {\arii}\,6.99\,$\mu$m lines, and 
might be blended with the 7.7/11.3/12.3\,$\mu$m PAH features. 
Therefore, we estimate $F$({\hb}) of the whole PN to be 1.02(--13)
$\pm$ 2.15(--15)\,erg~s$^{-1}$~cm$^{-2}$ using 
the $V$-band magnitude \citep[$m_{V}$ = 17.225 $\pm$ 0.026;][]{Zaritsky:2002aa} 
and scale it to the flux density of the XSHOOTER UVB spectrum to match
this band magnitude.

The $c$({\hb}) value in the last line of Table~\ref{T:chb} is the average 
value amongst the calculated $c$({\hb}) values. 
Using the average $c$({\hb}), we derived the de-reddened {\hb} flux, $I$({\hb}), in the whole nebula to be
1.30(--13) $\pm$ 4.88(--15)\,erg~s$^{-1}$~cm$^{-2}$. Thus, we obtained
the $I$({\neii}\,12.81\,$\mu$m) = 11.169 $\pm$ 0.551, where
$I$({\hb}) = 100.

\subsubsection{Electron density and temperature}

\begin{figure}
\includegraphics[width=0.9\columnwidth,bb=48 188 548 641,clip]{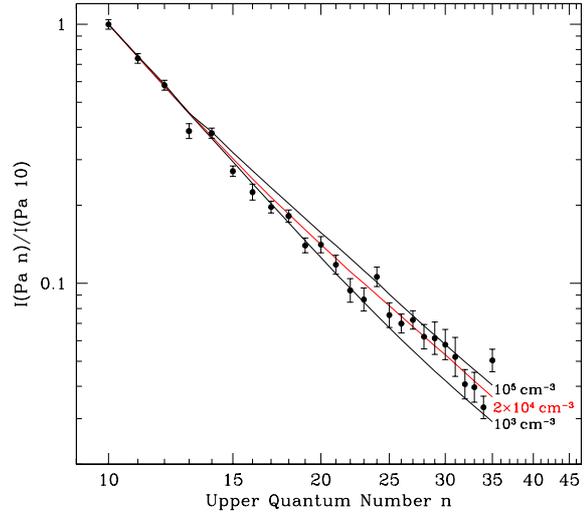}
\caption{Intensity ratio of Paschen lines to Pa10, assuming Case~B
 recombination. The theoretical intensity ratios (thick lines) 
are given for {\te} = 9260\,K determined from the Paschen Jump and 
{\Ne} = 10$^{3}$, 2$\times$10$^{4}$, and 10$^{5}$\,cm$^{-3}$.}
\label{F:paschen}
\end{figure}

\begin{figure}
\includegraphics[width=0.9\columnwidth,bb=35 194 425 507,clip]{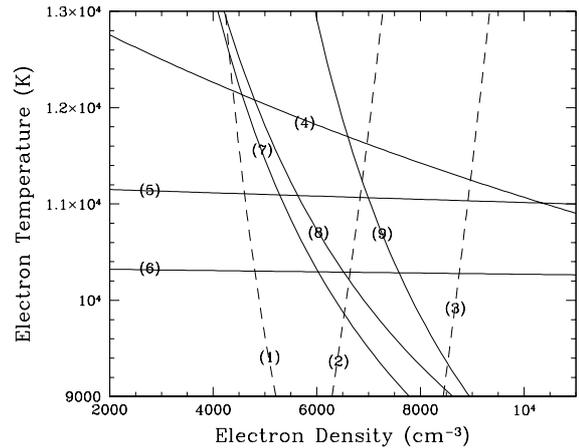}
\caption{
{\Ne}-{\te} diagram based on diagnostic CELs. The thick 
and dashed lines with the ID numbers (See Table~\ref{diagno_table}) 
are the indicators of {\te} and {\Ne}, respectively. \label{F:ne-te}
}
\end{figure}

\begin{table}
\centering
\caption{Summary of plasma diagnostics using nebular lines. 
\label{diagno_table}}
\begin{tabu} to \columnwidth {@{}cllr@{}}
\hline
ID & \multicolumn{1}{c}{{\Ne} diagnostic}&\multicolumn{1}{c}{Value}&\multicolumn{1}{c}{Result}\\
   &                          &     &\multicolumn{1}{c}{(cm$^{-3}$)}\\
 \hline

(1)&$[$N~{\sc i}$]$~(5198\,{\AA})/(5200\,{\AA}) & 2.179 $\pm$ 0.573 & 4890$_{-3460}$ \\ 
(2)&$[$O~{\sc ii}$]$~(3726\,{\AA})/(3729\,{\AA}) & 2.200 $\pm$
	 0.126 & 6830 $\pm$ 1520 \\ 
(3)&$[$S~{\sc ii}$]$~(6716\,{\AA})/(6731\,{\AA}) & 0.507 $\pm$
	 0.012 & 8910 $\pm$ 1460 \\ 
   &Paschen Decrement                                    &                 &$\sim$20000\\
\hline
ID &\multicolumn{1}{c}{{\te} diagnostic}&\multicolumn{1}{c}{Value}&\multicolumn{1}{c}{Result}\\
   &                          &     &\multicolumn{1}{c}{(K)}\\
 \hline
(4)&$[$N~{\sc ii}$]$~(6548\,{\AA}+6583\,{\AA})/(5755\,{\AA})
     & 53.89 $\pm$ 2.15 & 11\,660 $\pm$ 230 \\ 
(5)&$[$O~{\sc iii}$]$~(4959\,{\AA}+5007\,{\AA})/(4363\,{\AA})
     & 147.1 $\pm$ 12.7 & 11\,090 $\pm$ 320 \\ 
(6)&$[$S~{\sc iii}$]$~(9069\,{\AA})/(6312\,{\AA}) &
	 8.757 $\pm$ 0.419 & 10\,300 $\pm$ 220 \\ 
(7)&$[$N~{\sc i}$]$~(5198\,{\AA}/5200\,{\AA})/(1.04\,$\mu$m) &
	 3.706 $\pm$ 1.184 & 8960 $\pm$ 1650 \\ 
(8)&$[$O~{\sc ii}$]$~(3726/29\,{\AA})/(7320/30\,{\AA}) &
	 9.523 $\pm$ 0.274 & 10\,060 $\pm$ 180 \\  
(9)&$[$S~{\sc ii}$]$~(6717/31\,{\AA})/(4069/76\,{\AA} + & 1.702 $\pm$ 0.062 & 9050
	     $\pm$ 310 \\  
   & 1.029/1.034/1.037\,$\mu$m)\\
    &He~{\sc i}~(7281\,{\AA})/(5876\,{\AA}) & 0.061 $\pm$ 0.003 &
	     11\,180 $\pm$ 770 \\ 
    &He~{\sc i}~(7281\,{\AA})/(6678\,{\AA}) & 0.251 $\pm$ 0.012 &
	     11\,540 $\pm$ 620 \\ 
    &(Paschen Jump)/(Pa11)&0.102 $\pm$ 0.012&9260 $\pm$ 770\\
 \hline
\end{tabu}
\end{table}

In the following nebular line-diagnostics and subsequent ionic abundance 
calculations, the adopted transition probabilities, 
effective collision strengths, and recombination coefficients 
are the same as those listed in Tables~7 and 11 of \citet{Otsuka:2010aa}.

With recombination lines (RLs), we calculated the {\te} and {\Ne} required for the
He$^{+}$ and C$^{2+}$ abundance derivations first. Following
\citet{Zhang:2005aa}, we calculated the {\te}({\hei}) using 
the {\hei} $I$(7281\,{\AA})/$I$(5876\,{\AA}) and 
$I$(7281\,{\AA})/$I$(6678\,{\AA}) ratios 
and the emissivities of these {\hei} lines given by \citet{Benjamin:1999aa} 
for the case of {\Ne} = 10$^{4}$\,cm$^{-3}$. These three {\hei}
lines are insensitive to {\Ne} when compared to the other {\hei} lines. 
We adopted the average between the two {\te}({\hei}) results (11\,360 $\pm$ 840\,K) to 
derive the number density ratio of the He$^{+}$ to the H$^{+}$ 
$n$(He$^{+}$)/$n$(H$^{+}$). We did not detect any He\,{\sc ii} nebular
emission lines in the XSHOOTER spectrum, so $n$(He$^{2+}$)/$n$(H$^{+}$)
= 0.

The electron temperature derived from the Paschen jump {\te}(PJ) by using Equation (7) of \citet{Fang:2011aa}.

In the last RL plasma diagnostics, we estimated {\Ne} from the Paschen
decrement. The intensity ratios of the high-order hydrogen lines 
to a lower-order hydrogen line is sensitive to {\Ne}, in particular 
when {\Ne} $>$~10$^{5}$\,cm$^{-3}$. We investigated such 
higher density regions using the Paschen series Pa\,$n$ ($n$: principal
quantum number of the upper level), as presented in
Fig.~\ref{F:paschen}. We compared the observed ratios of $I$(Pa\,$n$)/$I$(Pa\,10) to the theoretical values in a range from 
10$^{3}$ to 10$^{5}$ and {\te}(PJ) = 9260\,K in the Case~B assumption, as computed by
\citet{1995MNRAS.272...41S}. In Fig.~\ref{F:paschen}, we plot the
theoretical values in the cases of {\Ne}=10$^{3}$,
2$\times$10$^{4}$, and 10$^{5}$\,cm$^{-3}$ with the observed ones. 
The 2$\times$10$^{4}$\,cm$^{-3}$ model gives the best fit to the 
observed data (indicated by the red line, reduced $\chi^{2}$ value is 0.95).

We derived {\Ne} and {\te} from collisionally excited lines (CELs) 
by solving the statistical equilibrium equation for the level
populations using a multi-level 
atomic model. The values for {\Ne} and {\te} calculated from the
diagnostic CEL ratios and the results obtained from the RL plasma
diagnostics are listed in Table~\ref{diagno_table}. 
The second, third, and last columns give the diagnostic lines, their 
line ratios, and the resulting values for {\Ne} and {\te}, respectively. 
The numbers in the first column indicate the ID of each curve in
the {\Ne}-{\te} diagram in Fig.~\ref{F:ne-te}. Using this diagram, we
determined the optimal {\Ne} and {\te} pairs.

Given that from the RL plasma diagnostics we know that in 
\mbox{Lin49} {\te} is around 10$^{4}$\,K, we assume this as a constant value to 
calculate all {\Ne}(CEL)s. Moreover, we assume a value of 6830 $\pm$ 1520\,cm$^{-3}$ for {\Ne}({\oii}) 
to derive {\te}({\nii},{\oii},{\siii},{\oiii}); and a value of 8910 $\pm$ 1460\,cm$^{-3}$ for {\Ne}({\sii})
to derive {\te}({\sii},{\NI}). Since the {\NI}\,5200\,{\AA}
line is partially affected by fringes, its flux and the {\Ne}({\NI}) are
very uncertain. Therefore, we used {\Ne}({\sii}) to calculate
{\te}({\NI}), instead of {\Ne}({\NI}). Note that the {\te}({\NI}) is
also uncertain.

Using Equation (2) from \citet{2000MNRAS.312..585L}, we calculated the recombination contamination to
the {\oii}\,7320/30\,{\AA} lines due to O$^{2+}$ assuming
{\te}=10$^{4}$\,K, and find that it is very small (0.02\,{\%} of
their observed de-reddened fluxes). In the {\te}({\oiii}) and
{\te}({\nii}) derivations, we do not subtract the recombination
contribution of the O$^{3+}$ and N$^{2+}$ from the observed {\oiii}\,4363\,{\AA} and
{\nii}\,5755\,{\AA} fluxes, because we do not
detect any O$^{3+}$ and N$^{2+}$ lines in the present spectra. As the 
O$^{2+}$/(O$^{+}$ + O$^{2+}$) ratio is small ($\sim$0.03, See next section), 
the O$^{3+}$ and N$^{2+}$ recombination contamination to
{\oiii}\,4363\,{\AA} (and perhaps
{\nii}\,5755\,{\AA}, too) is probably very small.

We derived the electron densities in the neutral to low ionisation regions using
the {\NI}, {\sii}, and {\oii} nebular line ratios, whereas the electron
density in higher ionisation regions (e.g., derived from the {\ariv}\,$I$(4711\,{\AA})/$I$(4740\,{\AA}) ratio) can not be
calculated because \mbox{Lin49} is a very low-excitation PN, indicated by the
$I$({\oiii}\,5007\,{\AA})/$I$({\hb}) = 0.16. However, we confirm that {\te}({\oiii}) and
{\te}({\siii}), and the volume emissivities of O$^{2+}$,
Ne$^{+}$, S$^{2+}$, Cl$^{2+}$, and Ar$^{2+}$ (these emissivities are calculated under
the {\te}({\oiii}) for O$^{2+}$ and {\te}({\siii}) for the other ions and a constant 
{\Ne}({\oii})) do not change significantly 
when compared to those under an {\Ne}(Paschen decrement) = 2$\times$10$^{4}$\,cm$^{-3}$
($\lesssim$~3\,$\%$). This is neither the case for the ionic abundances.

\subsubsection{Nebular abundance derivations using ICFs\label{S:neb-abund}}

\begin{table*}
\caption{Elemental abundances based on the ICFs, 
solar abundances, relative abundances to the solar values, and 
the predicted elemental abundances in the AGB nucleosynthesis models
 by \citet{Fishlock:2014aa}
 for initially 1.0\,M$_{\odot}$, 1.25\,M$_{\odot}$, and 1.5\,M$_{\odot}$ stars with $Z$ =
 0.001. The C(RL) is the C abundance derived from the C$^{2+}$ abundance using 
the recombination C\,{\sc ii}\,4267\,{\AA} line and the
 C(CEL) is an {\it expected} value when we adopted the average CEL C/O ratio
 amongst six SMC C$_{60}$ PNe. See the text in details.}
\centering
 \begin{tabu} to \textwidth {@{}lcccccc@{}}
\hline
X&\multicolumn{1}{c}{$\epsilon$(X)}&
\multicolumn{1}{c}{$\epsilon$(X$_{\odot}$)}&\multicolumn{1}{c}{[X/H]}
&\multicolumn{1}{c}{$\epsilon$(X$_{\rm model}$) for 1.0\,M$_{\odot}$}
&\multicolumn{1}{c}{$\epsilon$(X$_{\rm model}$) for 1.25\,M$_{\odot}$}
&\multicolumn{1}{c}{$\epsilon$(X$_{\rm model}$) for 1.5\,M$_{\odot}$}\\
(1) & (2) &(3) &(4)&(5)&(6)&(7)\\
\hline 
He     & 10.80 to 11.01 &10.93 $\pm$ 0.01 &--0.13 to +0.08       &10.99&11.01&11.01 \\ 
C(RL)  & ~~8.67 $\pm$ 0.09   &~~8.39 $\pm$ 0.04  &~+0.28 $\pm$ 0.10  &~~8.06 &~~8.56&~~8.89\\ 
C(CEL)       & ~~8.46 $\pm$ 0.24   &~~8.39 $\pm$ 0.04  &~+0.07 $\pm$ 0.25  &~~8.06 &~~8.56&~~8.89\\ 
N & ~~6.93 $\pm$ 0.02   &~~7.86 $\pm$ 0.12  &~--0.93 $\pm$ 0.12 &~~7.15 &~~7.26&~~7.18\\ 
O & ~~8.11 $\pm$ 0.01  &~~8.73 $\pm$ 0.07  &~--0.62 $\pm$ 0.07 &~~7.58  &~~7.68&~~7.79\\ 
Ne & ~~7.18 $\pm$ 0.05  &~~8.05 $\pm$ 0.10  &~--0.89 $\pm$ 0.11 &~~6.89  &~~7.37&~~7.72\\ 
S & ~~6.02 $\pm$ 0.01   &~~7.16 $\pm$ 0.02  &~--1.15 $\pm$ 0.02 &~~5.99 &~~6.00&~~6.00\\ 
Cl & ~~4.03 $\pm$ 0.05  &~~5.25 $\pm$ 0.06  &~--1.22 $\pm$ 0.08 &~~4.07  &~~4.08&~~4.10\\ 
Ar & ~~5.48 $\pm$ 0.11  &~~6.50 $\pm$ 0.10  &~--1.02 $\pm$ 0.15 &~~5.27  &~~5.28&~~5.28\\ 
Fe & ~~4.55 $\pm$ 0.04  &~~7.46 $\pm$ 0.08  &~--2.91 $\pm$ 0.09 &~~6.37  &~~6.38&~~6.38\\ 
\hline
\end{tabu}
\label{element-icf}
\end{table*}

We list the {\te} and {\Ne} pair adopted in each ionic abundance
calculation in appendix Table~\ref{T:tene}. The choices of {\te} and {\Ne} were driven by the
ionisation potentials of the target ions. We adopt a constant {\Ne} = 10$^{4}$\,cm$^{-3}$ 
to calculate He$^{+}$/H$^{+}$ using recombination coefficients of
\citet{Benjamin:1999aa} and C$^{2+}$/H$^{+}$ using those of
\citet{2000A&AS..142...85D} (the RL ionic abundances are not sensitive to {\Ne} with
$<$~10$^{8}$\,cm$^{-3}$). The He$^{+}$ and C$^{2+}$ abundances were derived under the Case B 
assumption for the lines with levels that have 
the same spin as the ground state, and under the Case A assumption 
for lines of other multiplicities.

The results are summarised in appendix Table~\ref{T:ionica}, where the
fifth and tenth columns show the number density
ratio of the ion X$^{\rm m+}$ relative to H$^{+}$ derived from the emission
line with wavelength listed in the third and eighth columns. The adopted
values calculated using a weighted average are listed in the last line for
each ion (in boldface). In the two consecutive lines below the results 
for each ion, the ionisation 
correction factor (ICF) and the elemental abundance are given.

The ICFs have been empirically determined based on the fraction 
of observed ion number densities with similar ionisation potentials 
to the target element, and have also been theoretically determined based 
on the fractions of the ions 
calculated by photoionisation (P-I) models. For \mbox{Lin49}, we tested 
the ICFs calculated by the P-I model of the C$_{60}$ 
PN M1-11 performed by \citet{Otsuka:2013aa}, as well as the empirically
determined ICFs. M1-11 is a Galactic C$_{60}$ PN with a central 
star with similar $T_{\rm eff}$ to our target (31\,830 K, 
while the central star of \mbox{Lin49} has $T_{\rm eff}$ = 30\,500 K --
See next section). The model of \citet{Otsuka:2013aa} includes
amorphous carbon and silicon carbide (SiC) grains and PAH molecules and
aims to fit the observed UV to far-IR SED and match observed gas 
emission line fluxes. 
The interaction between gas and dust affects the thermal structure
of the nebula. As a result, the ionisation structure will be affected. 
\mbox{Lin49} and M1-11 have similar $T_{\rm eff}$ of the central
star as the heating/ionisation source and similar C-rich dust
features. Therefore, we assume that the ICFs calculated in
the P-I model of M1-11 are reasonable values for \mbox{Lin49}. 
By adopting these ICFs, we also have the opportunity to test their
robustness in the P-I modelling as discussed later.

The resulting elemental abundances $\epsilon$(X) are 
listed in Table~\ref{element-icf}. These results are given in the form
of $\log_{10}$(X/H)+12. The fourth column is the relative abundance to the
solar abundance, taken from \citet{Lodders:2010aa}. Except for Cl, 
there is no significant difference in the solar photospheric abundances 
between \citet{Lodders:2010aa} and \citet{Asplund:2009aa}. Although the 
results for Cl abundances in these two papers are in agreement within the 
uncertainties (5.26 $\pm$ 0.06 and 5.50 $\pm$ 0.30, respectively), those are 
still large uncertainties when compared to other elements and one should 
be careful when discussing the [Cl/H] results. This is also the case for
the solar O abundance \citep[the measurement uncertainties are very small 
but the solar O abundance seems to remain under
debate. See, e.g., ][]{Asplund:2009aa}. 

Below we give a detail explanation for the C abundance.
The calculation methods of the He, N, Ne, Cl, Ar, and Fe abundances
are explained in Appendix~\ref{AS:abundances}. 
The O and S abundance calculations are explained in the course of the He
calculation.

\subsubsection*{C abundance from RLs}

Several prior studies on SMC PN abundances have reported the detection of 
RL carbon lines \citep[e.g.,][]{Tsamis:2003aa,Tsamis:2004aa,
Leisy:2006aa,Shaw:2010aa}. As far as we know, the RL C$^{2+}$ and
C abundance derivations in \mbox{Lin49} are only the second derivation for a  
SMC PN.

In \mbox{Lin49}, we need to take care when determining the C$^{2+}$
abundance. The C$^{2+}$/H$^{+}$ determined from C\,{\sc
ii}\,3918.98/20.69\,{\AA} (3$p$$^{2}$P-4$s$$^{2}$S),
is much higher than those obtained from other detected lines. This is 
due to intensity enhancement by resonant 
absorption of C\,{\sc ii}\,635.25/636.99\,{\AA}
(2$p$$^{2}$P$^{\rm o}$-4$s$$^{2}$S) and then fluorescence by decay of the
4$s$$^{2}$S level. C\,{\sc ii}\,7231.32/36.42\,{\AA}
(3$p$$^{2}$P-3$d$$^{2}$D) may also be enhanced by such a resonance and 
fluorescence of C\,{\sc ii}\,687\,{\AA} (2$p$$^{2}$P$^{\rm o}$-3$d$$^{2}$D) 
and the 3$d$$^{2}$D decay. The 2$p$$^{2}$P$^{\rm o}$ level of the C\,{\sc
ii}\,6578.05\,{\AA} (2$p$$^{2}$P$^{\rm o}$-2$s$$^{2}$S) 
could be affected by the C\,{\sc ii}\,3918.98/20.69\,{\AA} and
7231.32/36.42\,{\AA}. Thus, C$^{2+}$ abundances except
for the value derived from C\,{\sc ii}\,4267\,{\AA} (3$d$$^{2}$D-4$f$$^{2}$F)
would be overestimated. Following a 
detailed report on fluorescence and recombination lines in 
the PN IC418 by \citet{Escalante:2012aa}, we supposed the
C$^{2+}$/H$^{+}$ obtained from C\,{\sc ii}\,4267\,{\AA} 
to be the most reliable, 
as this line has no paths directly connected to the 2$p$$^{2}$P$^{\rm o}$
level. We should note that C\,{\sc ii} line fluorescence enhancement 
is not common in low-excitation PNe. For example,
\citet{Otsuka:2013aa,Otsuka:2015aa} do not observe such 
enhancements in M1-11 or in the C-rich PN K648 ($T_{\rm eff}$ = 36\,360\,K).

As discussed below, we test three different ICFs to derive the C 
abundance. The equation proposed by \citet{Kingsburgh:1994aa}
%
\begin{eqnarray}
 \rm C      &=& \rm ICF(C)~{\cdot}~\frac{C^{2+}}{H^{+}},\nonumber\\
 \rm ICF(C) &=& \rm \frac{O}{O^{2+}} \label{EQ:C1}
\end{eqnarray}
%
\noindent gives a value of ICF(C) = 32.2 $\pm$ 2.2 and $\epsilon$(C) =
9.5 $\pm$ 0.09. \citet{Delgado-Inglada:2014ab} calculated the 
C/O ratio in PNe from a P-I grid modelling, 
obtaining the following equation to derive the C abundance: 
%
\begin{eqnarray}
 \rm C         &=& \rm ICF(C)~{\cdot}~\frac{C^{2+}}{H^{+}}, \nonumber\\
 \rm ICF(C)    &=& \rm
  \frac{O}{O^{2+}}~\cdot~\left(0.05+2.21{\omega}-2.77{\omega}^2+1.74{\omega}^3\right),\nonumber\\
 \rm {\omega}  &=& \rm \frac{O^{2+}}{O^{+}+O^{2+}} \label{EQ:C2}.
\end{eqnarray}
%
Note that Equation (\ref{EQ:C2}) is valid in the range 0.05 $<$ $\omega$ $<$
0.97 and, therefore, is not valid for \mbox{Lin49}, for which $\omega$ is 0.031 $\pm$ 0.002. 
Nevertheless, we applied Equation (\ref{EQ:C2}) to our data, and obtained ICF(C) = 37.8 $\pm$ 3.4 and 
$\epsilon$(C) = 9.65 $\pm$ 0.09. The uncertainty in the C/O ratio (i.e., the C abundance) is 
higher near the lower limit of the valid $\omega$ interval. \citet{Delgado-Inglada:2014aa} 
estimated a confidence interval from --1 to +0.26 dex in the
low-excitation PN NGC40,
which has an $\omega$ (0.03) very similar to \mbox{Lin49}. In
low-ionisation PNe such as NGC40 and \mbox{Lin49}, the same applies
for the ICF(C) given by \citet{Kingsburgh:1994aa}. Thus, Equations
(\ref{EQ:C1}) and (\ref{EQ:C2}) are not ideal to determine $\epsilon$(C)
in \mbox{Lin49}, and the result would lie in the wide range from 8.6 to
10, taking into account the confidence limit of --1 to +0.26 dex. 
This might be due to the reason that the respective fractions of the 
C$^{2+}$ and O$^{2+}$ relative to C and O are very different in low-excitation 
PNe. As the models from \citet{Delgado-Inglada:2014ab} do not target 
low-excitation PNe alone, their ICF(C) does not reproduce the C/O ratio
properly using the C$^{2+}$ and O$^{2+}$ abundances.

For the above reasons, we adopt the ICF(C) and the $\epsilon$(C) derivations
based on the P-I model of M1-11, as given by the following equations:
%
\begin{eqnarray}
 \rm C         &=& \rm ICF(C)~{\cdot}~\frac{C^{2+}}{H^{+}}, \nonumber\\
 \rm ICF(C)    &=& \rm 2.46~\cdot\frac{S}{S^{2+}}\label{EQ:C3}.
\end{eqnarray}
%
We chose to write the ICF(C) as a function of the S and S$^{2+}$ abundances
(instead of writing it as a function of O abundances), as the ionisation potential
of C$^{2+}$ is similar to that of S$^{2+}$. The C$^{2+}$ fraction was
0.338 in the P-I model of M1-11. From Equation ({\ref{EQ:C3}),
we get ICF(C) = 3.96 $\pm$ 0.23 and the RL $\epsilon$(C) = 8.67 $\pm$ 0.09.

\subsubsection*{Expected C abundance from CELs \label{S:carbon}}

\begin{table*}
\centering
\caption{Effective temperature of the central star ($T_{\rm eff}$), nebular radius
 ($r$), and nebular elemental abundances in SMC C$_{60}$ PNe. 
$\epsilon$(C) are derived from C CELs. $\epsilon$(C) in SMC1 was estimated using the O
 abundances of \citet{Leisy:2006aa} and the C/O ratios of
 \citet{Vassiliadis:1998aa}. The $\epsilon$(C) in \mbox{Lin49} is an
 {\it expected} value when we adopted the average CEL C/O ratio
 amongst the other SMC C$_{60}$ PNe. We excluded the $\epsilon$(C) of
 \mbox{Lin49} to calculate the average $\epsilon$(C) amongst these PNe.}
\begin{tabu} to \textwidth {@{}l@{\hspace{24pt}}c@{\hspace{24pt}}c@{\hspace{24pt}}c@{\hspace{24pt}}c@{\hspace{24pt}}c@{\hspace{24pt}}c@{\hspace{24pt}}c@{\hspace{24pt}}c@{\hspace{24pt}}c@{\hspace{24pt}}l@{}}
\hline
C$_{60}$ PNe & $T_{\rm eff}$ (K) & $r$ (${\arcsec}$) & $\epsilon$(He) 
& $\epsilon$(C) & $\epsilon$(N) & $\epsilon$(O) & $\epsilon$(Ne) & $\epsilon$(S) & $\epsilon$(Ar) & References \\ 
\hline
SMC1 & 37\,000 & 0.15 & 10.83 & 8.00 & 7.16 & 7.86 & 6.42 & $<$~6.94 & 5.71 &(1),(2),(3),(4),(10)\\ 
SMC13 & 31\,300 & 0.19 & 11.11 & 8.73 & 7.30 & 8.06 & 7.35 & 5.96 & 5.46&(4),(5),(6),(7)  \\ 
SMC15 & 58\,000 & 0.17 & 11.03 & 8.26 & 7.71 & 8.07 & 7.32 & 7.67 & 5.72 &(1),(6),(8),(9)  \\ 
SMC16 & 37\,000 & 0.18 & 10.69 & 8.19 & 6.55 & 7.85 & 6.37 & 6.39 & 5.46 &(1),(6),(8),(9)  \\ 
SMC18 & 31\,500 & 0.15 & 11.06 & 8.31 & 7.11 & 7.90 & 7.57 & 6.18 & 5.67 &(4),(5),(6),(7)  \\ 
SMC24 & 37\,800 & 0.20 & 11.13 & 8.18 & 7.17 & 8.06 & 7.36 & 6.11 & 5.58 &(4),(5),(6),(7)  \\ 
Average & 38\,770 & 0.17 & 10.98 & 8.28 & 7.17 & 7.97 & 7.07 & 6.54 &
 5.64 &  \\ 
\hline
\mbox{Lin49} &30\,500 &0.23 &10.8-11.01 &8.46&6.93&8.11&7.18&6.02&5.48&(10)\\ 
\hline
\end{tabu}
\begin{threeparttable}
\begin{minipage}{\textwidth}
References -- 
(1) \citet{Leisy:2006aa} for abundances;
(2) \citet{Vassiliadis:1998aa} for the C/O ratios of 1.38 in SMC1;
(3) \citet{Herald:2007aa} for $T_{\rm eff}$;
(4) \citet{Stanghellini:2003aa} for $r$;
(5) \citet{Shaw:2010aa} for abundances except $\epsilon$(C);
(6) \citet{Stanghellini:2009aa} for $\epsilon$(C);
(7) \citet{Villaver:2004aa} for $T_{\rm eff}$;
(8) \citet{Shaw:2006aa} for $r$;
(9) \citet{Dopita:1991ab} for $T_{\rm eff}$;
(10) This work.
\\
\end{minipage}
\label{T:compSMCXS}
\end{threeparttable}
\end{table*}

In the field of PN research, it is well known that the C, N, O, and Ne
abundances derived from RLs are larger than those derived from
CELs. Several explanations for the abundance discrepancies
have been proposed, and consensus has yet to be reached, see 
e.g. \citet{Liu:2006aa} for the historical background and the abundance
discrepancy problem. We believe that the C abundance derived from the C\,{\sc
ii}\,4267\,{\AA} line would be reasonable and acceptable as
the C abundance for a SMC PN. \citet{Otsuka:2010aa} argued that the
emissivities of the C\,{\sc iii}$]$\,1906/09\,{\AA} lines are very
sensitive to {\te} because of the energy difference of these lines 
between upper and lower level, $\Delta\,E$ = $k\,\Delta\,T$ ($k$: the Boltzmann constant), where $\Delta\,T$ =
75\,380 K and the C$^{2+}$ abundances from RLs may be more reliable than
those from CELs if one cannot find representative {\te} values in the CEL
C$^{2+}$ emitting zone. However, it is unclear whether our measured RL C
abundance is representative for \mbox{Lin49}; the RL abundances might 
represent those in high-density zones, hydrogen-deficient cold
components, or stellar wind whereas the CEL abundances might indicate
the average in the nebula \citep[See][and references therein]{Otsuka:2010aa}.

For the above reasons, we estimate the CEL C abundance in
\mbox{Lin49} as follows. In the measurement of the CEL C abundances for extended objects, 
the flux normalisation issue would be raised due to the different sizes 
and shapes of the slits used in UV (to obtain the UV C\,{\sc 
iii}$]$\,1906/09\,{\AA} and $[$C\,{\sc
ii}$]$\,2320-30\,{\AA} lines) and optical spectroscopy (to obtain
e.g., Balmer lines) and the different slit positions putting 
on the targets. As a consequence, the measured CEL C abundances
may be largely inconsistent with the RL C values, whereas in objects
compact enough for the slit dimension, such as MC PNe, the flux
normalisation issue can be avoided. Using the Hubble Space Telescope 
(\emph{HST})/Faint Object Spectrograph and the Space Telescope
Imaging Spectrograph (STIS), the CEL C abundances have been measured in
the SMC C$_{60}$ PNe SMC1, 13, 15, 16, 18, and 24. In
Table~\ref{T:compSMCXS}, the abundances of these PNe, the nebula radii, and the effective temperatures are
compiled, with the last line the average value of each parameter. The O abundances
in this Table are measured from O CELs. The average C/O abundance
ratio is 2.28 (with a standard deviation of 1.27) amongst these six
PNe. Supposing that these six PNe and
\mbox{Lin49} evolved from stars with similar initial masses (because the
elemental abundances of all these PNe are very similar) and that their
current evolutionary stage is also similar (because both the effective
temperature of the central star and the radius of the nebula are consistent with similar ages after the AGB phase),  
we estimate the CEL C abundance for \mbox{Lin49} to be (2.91 $\pm$ 1.63)$\times$10$^{-4}$, or CEL 
$\epsilon$(C) = 8.46 $\pm$ 0.24 using this C/O ratio and the
observed O abundance. Hereafter, we regard this CEL $\epsilon$(C) as a representative C abundance in
\mbox{Lin49} and used this value in subsequent SED modelling.

\subsubsection{Metallicity \label{S:metallicity}}

In comparison to $\alpha$-elements 
S and Ar, Fe (a refractory element) is highly depleted. The extremely
low [Fe/H] abundance indicates that most iron atoms are 
trapped in dust grains. As a consequence, the Fe nebular abundance does
not reflect the metallicity of \mbox{Lin49}.

For the purpose of this study we wonder how much Fe is depleted onto dust grains,
and correspondingly what is the true metallicity of \mbox{Lin49}.
The SMC is an irregular galaxy formed through strong
interactions between the LMC and the Milky Way
Galaxy. \citet{Mucciarelli:2014aa} reported that the typical metallicity
of the old stellar populations in the SMC is $\sim$--0.9 in [Fe/H].
Although the chemical evolution of the SMC would be incompatible with
that of the Milky Way, we attempt to estimate the metallicity of
\mbox{Lin49} using the chemical evolution model of the Milky Way halo by
e.g., \citet{Kobayashi:2011aa} taking current circumstance that the
chemical evolution of the SMC based on the observed abundances remains
unclear but a typical [Fe/H] in the SMC
is close to a typical [Fe/H] in the Milky Way halo.
\citet{Kobayashi:2011aa} reported that the 
[S/Fe] and [Ar/Fe] are $\sim$+0.4 and $\sim$+0.3 in the [Fe/H] $<$~--1, 
respectively. By applying this prediction and from the [S/H] and [Ar/H] observed in \mbox{Lin49}, 
we obtain [Fe/H] = --1.55 and --1.32, respectively. By comparing the average [Fe/H] of = --1.40 
with the observed [Fe/H] = --2.91, we conclude that 96\,$\%$ of the iron
atoms in the \mbox{Lin49} nebula are trapped in dust grains. 
Although the value has large uncertainty, we estimate that
the metallicity $Z$ of \mbox{Lin49} is $\sim$0.0006 or
$\sim$0.04\,Z$_{\odot}$. Here, Z$_{\odot}$ is the solar metallicity. 
In the following discussion, we adopt $Z$ = 0.0006 (0.04\,Z$_{\odot}$).

\subsubsection{Comparison with the AGB nucleosynthesis model \label{S:compAGB}}

In the last two columns of Table~\ref{element-icf}, we list the predicted abundances 
in the AGB nucleosynthesis models for 1.0\,M$_{\odot}$,
1.25\,M$_{\odot}$, and 1.50\,M$_{\odot}$ main-sequence mass stars with
$Z$ = 0.001 by \citet{Fishlock:2014aa}. Our observed nebular
abundances are in excellent agreement with these predictions except for
O.

In the comparison between the model results and the observed abundances in LMC post-AGB
stars, \citet{Fishlock:2014aa} found that the model predicted [O/Fe] is
overabundant relative to the observed values. They discussed the
possibility that the initial O abundances in these post-AGB stars are
greater than the scaled-solar initial abundance set in the model. The enhancement of the
O abundance in \mbox{Lin49} could not be explained by the extra
$^{13}$C($\alpha$,$n$)O$^{16}$ reaction in the He-rich shell; if that
were the case, we should have observed more enhanced C, O, and
$n$-capture element abundances. The O abundance in \mbox{Lin49} could be
not polluted by local events such as Type II supernovae
($\alpha$-elements producers) and is not different from the nearby PNe. For
instance, Lin45 (the nearest PN from \mbox{Lin49}; the linear distance
projected on sky is 438$''$ (or 131\,pc at 61.9\,kpc) from the position of
\mbox{Lin49}) shows similar O and $\alpha$-elemental abundances \citep[He = 10.93, N = 6.52,
O = 8.20, Ne = 7.55, S = 6.28, Ar = 5.79;][]{Costa:2000aa} (the
line-of-sight depths toward \mbox{Lin49} and \mbox{Lin45} are unknown,
though). Therefore, we think that the initial O
abundance in \mbox{Lin49} is larger than we expected. This could be the
case for the other C$_{60}$ PNe listed in Table~\ref{T:compSMCXS}.

Although our abundance determinations depend on models of H\,{\sc ii}
regions (for He) and M1-11 (for He, C, Ne, Cl,
and Ar) and that there might be some issues with the C and O
abundances, the 1.25\,M$_{\odot}$ model fits to the \mbox{Lin49} 
abundances better. From the view of elemental abundances, the initial mass
of the progenitor in \mbox{Lin49} and the other SMC C$_{60}$ PNe would be around 1-1.25\,M$_{\odot}$.

\subsection{Characterising the central star through the analysis of absorption lines  \label{S:absorption}}

 \begin{figure*}
\centering
\includegraphics[height=1.0\textwidth,angle=-90]{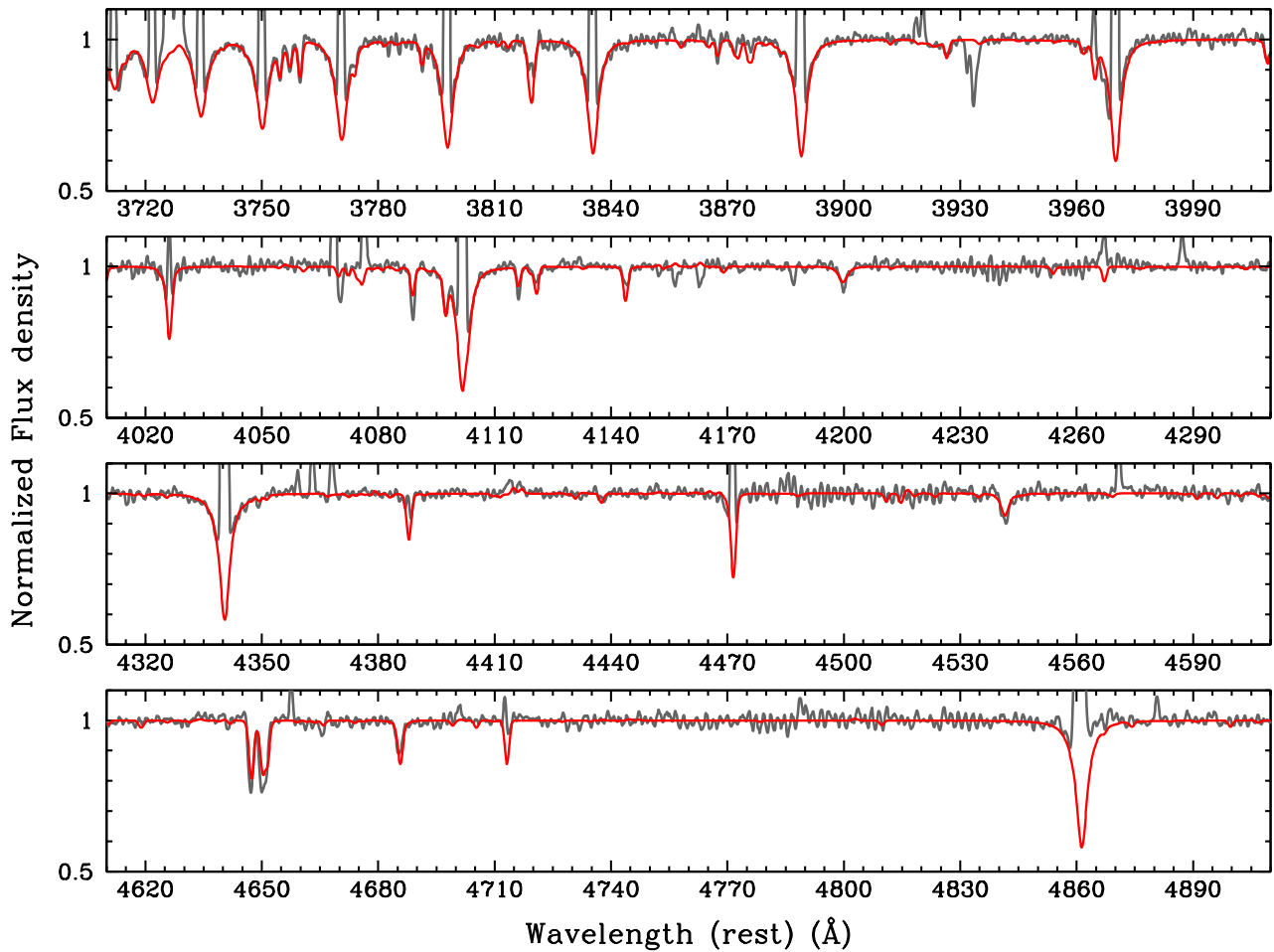}
\caption{The synthesised spectrum of \mbox{Lin49} in the range between 3720\,{\AA} and
  4910\,{\AA} as given by our {\sc TLUSTY} modelling (red line) 
and the observed XSHOOTER spectrum (grey line, after 9 pixel median smoothing). 
The FWHM of the synthesised spectrum was set to be constant and equal to
  1.2\,{\AA}.}
\label{synspec}
 \end{figure*}

 \begin{table}
\centering
\caption{The results of the {\sc TLUSTY} modelling for the stellar
  spectrum. \label{sum-tlusty}}
\begin{tabular}{@{}lr@{}}
\hline
Parameter   &Derived value \\
\hline
$T_{\rm eff}$ (K)&30\,500 $\pm$ 500~         \\
$\log$\,$g$ (cm s$^{-2}$)  &3.29 $\pm$ 0.06 \\
$\epsilon$(He)            &10.88 $\pm$ 0.30               \\
$\epsilon$(C)             &9.02  $\pm$ 0.30              \\
$\epsilon$(N)             &7.60  $\pm$ 0.30              \\
$\epsilon$(O)             &8.61  $\pm$ 0.10              \\
$\epsilon$(Si)            &6.76  $\pm$ 0.30              \\
\hline
\end{tabular}
 \end{table}

We produced a synthetic spectrum to fit the observed XSHOOTER spectrum after 9-pixel median smoothing in order to reduce the
fringe amplitude (See section \ref{S:XSHOOTER}). With this 
model spectrum, we derived the photospheric abundances, $T_{\rm eff}$, 
and surface gravity ($\log$\,$g$) of the central star. 
We used the O-type star grid model OSTAR2002 by \citet{Lanz:2003aa}
using the non-local thermodynamic equilibrium (non-LTE) stellar
atmosphere modelling code {\sc TLUSTY}\footnote{See \url{
http://nova.astro.umd.edu}}
\citep{Hubeny:1988aa}. The OSTAR2002 grid consists of 690 metal
line-blanketed, non-LTE, plane-parallel, and hydrostatic model
atmospheres. We fitted absorption lines of He, C, N, O, and Si as we
identified absorption lines of these elements in the observed spectrum.

Based on the assumption that the metallicity of the central star is the same as that
of the nebula, we adopt a metallicity $Z$ = 0.04\,$Z_{\odot}$ as 
determined in section \ref{S:metallicity}. 
We set the instrumental line broadening determined by measuring Th-Ar comparison
lines. In reference to the stellar absorption fitting report for 
the Galactic C$_{60}$ PN IC418 by \citet{Morisset:2009aa}, we set the microturbulent velocity 
to 5~{\kms} and the rotational velocity to 20~{\kms}; the synthesised
spectra using {\sc SYNSPEC}\footnote{See
\url{http://nova.astro.umd.edu/Synspec49/synspec.html}} 
with these values can fit the observed absorption line profile.

To determine $T_{\rm eff}$ and $\log$\,$g$, 
we first run photoionisation models using {\sc CLOUDY} with a stellar
atmosphere by {\sc TLUSTY} OSTAR2002 in order to find the ranges of $T_{\rm
eff}$ and $\log$\,$g$ because we do not detect any $T_{\rm eff}$
diagnostic lines with high a S/N ratio. These models keep the photospheric abundances at 
He/H = 0.1 and the metallicity at 0.04\,$Z_{\odot}$. In the {\sc CLOUDY} 
models, our initial guess for $T_{\rm eff}$ is 32\,000\,K, as given by 
equation (3.1) of \citet{Dopita:1991aa}, which was established from 
optically thick MC PNe. $T_{\rm Z}$(H\,{\sc i}) was 32\,950\,K by the
Zanstra method.  The initial guess for $\log$\,$g$ was determined 
by fitting to the profiles of the H$\gamma$ line (we blocked
the portion of the nebular line in the fitting process), and He\,{\sc
ii}\,4686\,{\AA} line with He/H = 0.1. We obtain 
a range of $\log$\,$g$ between 3.2 and 3.4\,cm s$^{-2}$. From these 
initial guesses for $T_{\rm eff}$ and $\log$\,$g$, we run {\sc CLOUDY} 
models to match the observed nebular emission line fluxes and 
abundances, and to further constrain the $T_{\rm eff}$ 
and $\log$\,$g$ ranges. Within these ranges, 
we perform profile fitting of 
the H$\delta$, H$\gamma$, and He\,{\sc ii} lines again. Finally, 
we derive $T_{\rm eff}$ = 30\,500 $\pm$ 500\,K 
and $\log$\,$g$ = 3.29 $\pm$ 0.05\,cm s$^{-2}$.

Adopting these values for $T_{\rm eff}$ and $\log$\,$g$, we fit 
the He\,{\sc ii}\,4686\,{\AA} line profile 
to determine the He abundance. Since the weak He\,{\sc ii} absorption 
lines were partially affected by fringes in the spectrum, the 
He/H abundance derived by the line fitting method presents a large 
uncertainty. Subsequently, we determined the C, N, O, and Si abundances to match the
observed line profiles. The C abundance was obtained using
the C\,{\sc iv}\,5801/5811\,{\AA} lines. The O
abundance was derived using the O\,{\sc
iii}\,3755/3774/3791\,{\AA} lines, and the Si abundance was derived
using the Si\,{\sc iv}\,4089/4116\,{\AA} lines. Finally, the N abundance
was obtained by fitting the N\,{\sc iii} + O\,{\sc
ii}\,4097\,{\AA} line complex after we determined the O
abundance. Some of the absorption 
lines, e.g., the C\,{\sc iii}\,4152/4156/4163\,{\AA} and the
O\,{\sc ii}\,4189\,{\AA} could not be fitted by the best model 
with $\log$\,$g$ = 3.29 cm s$^{-2}$. This might be because we could not
determine $T_{\rm eff}$ and $\log$\,$g$ with considerable accuracy.
However, if we set $\log$\,$g$ $\gtrsim$ 3.4 cm s$^{-2}$, 
we were able to reproduce these C\,{\sc iii} lines as absorption
lines. However, with such a high surface gravity, we cannot fit the
line-profiles of the He\,{\sc i,ii} and H\,{\sc i} lines.

We display the synthesised stellar spectrum in the range between
3720\,{\AA} and 4910\,{\AA} in Fig.~\ref{synspec}.
In Table~\ref{sum-tlusty}, we list the derived quantities with their 
1-$\sigma$ uncertainties. With the exception of He, the stellar abundances are 
systematically larger than the nebular abundances by $\sim$0.6 dex; 
the stellar abundance could reflect the latest nucleosynthesis result. 
The stellar C/O ratio (2.57 $\pm$ 1.90) supports a C-rich
classification for \mbox{Lin49} and our adopted nebular C/O ratio (2.28) for the CEL C 
derivation could be appropriate.

\subsection{Fitting the broad 30 micron feature \label{S:30um}}

\begin{figure}
\includegraphics[width=\columnwidth,bb= 44 357 523 699]{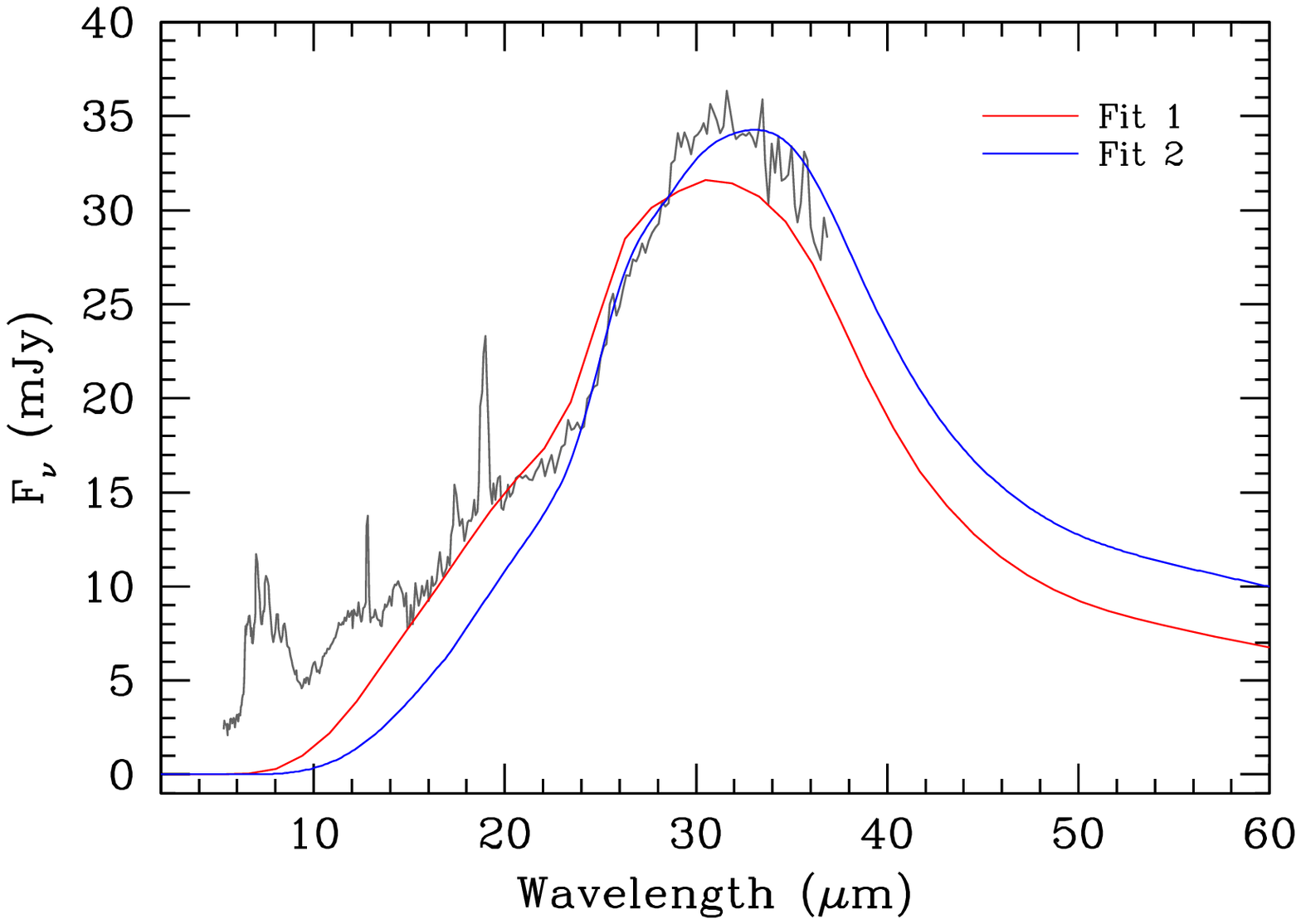}
\caption{Fits of the broad 30\,$\mu$m feature (indicated by the
 red and blue lines) overlaid on the \emph{Spitzer}/IRS
   spectrum of \mbox{Lin49} (grey line). Fits 1 and 2 are different in terms of the fitting wavelength range. 
   The fitting results are summarised in Table~\ref{T:30um}. 
\label{F:30um}}
 \end{figure}

  \begin{table}
    \centering
\caption{Fitting results for the broad 30\,$\mu$m feature and the 
predicted flux densities at 65, 90, and 120\,$\mu$m. The uncertainty
   of the predicted flux densities is $\sim$3\,$\%$. 
\label{T:30um}}
\begin{tabular}{@{}l@{\hspace{8pt}}c@{\hspace{8pt}}c@{\hspace{8pt}}c@{\hspace{6pt}}c@{\hspace{8pt}}c@{}} 
\hline
Model   &Fit range &$T_{d}$(max) &$F_{\nu}$(65\,$\mu$m) &$F_{\nu}$(90\,$\mu$m) &$F_{\nu}$(120\,$\mu$m) \\
        &($\mu$m)           & (K)         &(mJy)                 &(mJy)                 &(mJy) \\
	\hline
Fit1    &15-16,20-36      &155.5 $\pm$ 1.4&5.90&3.38&1.80\\
Fit2    &24-36            &126.0 $\pm$ 1.4&8.93&5.56&3.13\\
   \hline
\end{tabular}
\end{table}

\citet{Otsuka:2014aa} fitted the 13-160\,$\mu$m SED of 11 Galactic
C$_{60}$ PNe using  synthesised absorption efficiency
($Q_{\mathrm{abs},\lambda}$) based on the spectral data set of IC418,
and concluded that strength of the broad 30\,$\mu$m feature 
with respect to the underlying continuum in these objects is constant. 
The carrier for this feature remains unclear 
and is under debate \citep[e.g., see][for details]{Otsuka:2014aa}. 
We use the same approach to fit the broad 30\,$\mu$m feature in
\mbox{Lin49} using $Q_{\mathrm{abs},\lambda}$ from
\citet{Otsuka:2014aa}. We utilised equations (2) and (3) 
of \citet{Otsuka:2014aa} with $p$=$q$=2 and a lower 
limit on the dust temperature of 20\,K, as adopted in 
\citet{Otsuka:2014aa}. The model of \citet{Otsuka:2014aa} 
assumes that the dust density, as a function of the 
distance from the CSPN $r$, is distributed
around the CSPN with a power-law ($\propto$~$r^{-p}$) and 
that the dust temperature distribution $T_{d}$($r$) also follows a power-law 
($\propto$~$r^{-q}$). As listed in Table~\ref{T:30um}, we 
performed two fits, Fit1 and Fit2, where the difference between them is the
wavelength range over which the fit is performed. 
Fit1 (fitting region is 15--16\,$\mu$m and
20--36\,$\mu$m) is an entire fit for the
broad 16--24\,$\mu$m and 30\,$\mu$m features to verify the conclusion of
\citet{Otsuka:2014aa}. The resulting maximum dust temperatures
($T_{d}$(max)) are listed in Table~\ref{T:30um}.

As presented in Fig.~\ref{F:30um}, the SED predicted by Fit1 (indicated by the red line)
can explain the SED except for $\lambda$ $\gtrsim$28\,$\mu$m
where the model underestimates the observed flux density. At this moment,
we have two explanations for this underestimation; one might be the high noise level in the data around 
the wavelength range 28-36\,$\mu$m (a gap or a bump around 30\,$\mu$m is
seen). In fitting for the broad 30 $\mu$m feature,
while the other could be the resulting effect of the contribution from 
other dust components to the 30\,$\mu$m feature, e.g., iron-rich magnesium sulphides
such as Mg$_{0.5}$Fe$_{0.5}$S\footnote{\url{
http://www.astro.uni-jena.de/Laboratory/OCDB/sulfides.html}} \citep{Begemann:1994aa}, as 
\mbox{Lin49} is an extremely Fe deficient ([Fe/H] = --2.91) PN. Although there are
no reports of the detection of iron-rich magnesium sulphides
or iron dust in IC418 and C$_{60}$ PN M1-20, the nebular Fe abundances in these PNe
are extremely depleted, according to the results of 
\citet{Delgado-Inglada:2014aa}, who reported that the respective nebular O and
Fe abundances in IC418 are 8.52 and 4.36-4.56, corresponding to the 
[O/H] = --0.21 and [Fe/H] = --3.1 to --2.9 
\citep[See][about the nebular O and Fe abundances in M1-20]{Delgado-Inglada:2014aa}. 
In M1-11, \citet{Otsuka:2013aa} reported the nebular [O/H] = --0.07 and [Fe/H] = --2.42. 
The Fe-depletion will differ from object to object. Therefore, the strength of
the 30\,$\mu$m feature with respect to local dust continuum would be
different in each PN if any iron-rich magnesium sulphides contribute to
this feature. We will give a possible
explanation of the extreme Fe-depletion in \mbox{Lin49} later.

The Fit2 for the wavelength range 24--36\,$\mu$m (the blue line in Fig.~\ref{F:30um}) 
is a complementary test for the same hypothesis. The predicted SED 
underestimates the 16--24\,$\mu$m flux density. Taking into account the Fit1
result, at this moment is difficult to completely agree with the conclusion of
\citet{Otsuka:2014aa} based on the current data quality.

We list the predicted flux densities at 65, 90, and 120\,$\mu$m. Fit1 and Fit2
give a lower limit and an upper limit in these far-IR wavelengths. We
use these average flux densities to constrain the 
SED fitting (See section~\ref{photomodel}).

\subsection{Photoionisation modelling}
\label{photomodel}

  \begin{figure}
   \includegraphics[width=\columnwidth]{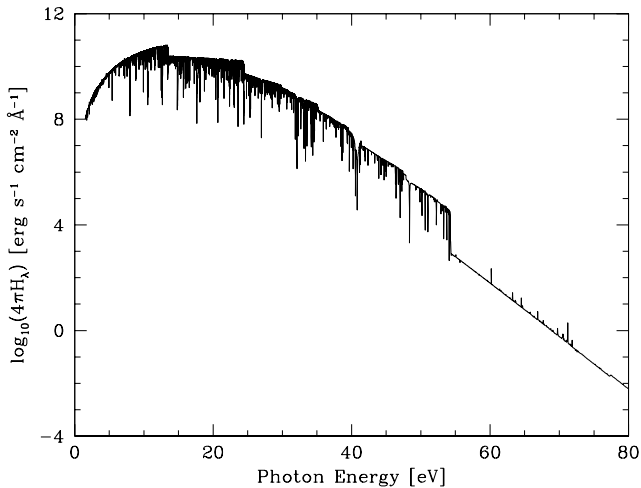}
  \caption{The SED of the CSPN synthesised by our {\sc TLUSTY} modelling.
  We used this  for the {\sc CLOUDY} modelling as the incident SED of
  the CSPN.}
  \label{F:CSPNsed}
  \end{figure}

Using a modified code based on {\sc
CLOUDY} \citep[][version C13.03]{Ferland:1998aa}, we fit the SED and 
investigate the physical conditions of the gas and dust grains 
in the nebula, and derive their masses ($m_{g}$ and $m_{d}$, respectively). In this modified code, 
we substituted the transition probabilities and effective collision
strengths of CELs by the same values used in our plasma diagnostics and
nebular abundance determinations for consistency.

The effective collision strengths of several lines such as [Ar\,{\sc ii}]\,6.99\,$\mu$m in the
original {\sc CLOUDY} were constant values not
functions of the {\te}. 
The constant collision strengths of these ions could lead to the
over-prediction of line fluxes as reported in \citet{Otsuka:2013aa}, and could affect
the P-I model results of the gas temperature and ion fraction
of each element inside the nebula.
For example, the predicted
$I$([Ar\,{\sc ii}]\,6.99\,$\mu$m)/$I$({\hb}) was $\sim$3/100
when we adopted the effective collision strength of this line used in
the original {\sc CLOUDY}. After revising its collision strength, we
obtained $\sim$1.3/100. Some of the atomic lines may contaminate 
the C$_{60}$ band fluxes. Our photoionisation model helps to estimate 
how the C$_{60}$ bands are contaminated by the atomic lines. In
particular, the C$_{60}$\,7.0\,$\mu$m flux is contaminated by the
[Ar\,{\sc ii}]\,6.99\,$\mu$m line. We should bear in mind
that the C$_{60}$\,7.0\,$\mu$m flux is important to discuss the excitation
mechanism of C$_{60}$ \citep[See e.g.,][in
details]{Bernard-Salas:2012aa}. Therefore, first we need to 
correct the effective collision strength of the [Ar\,{\sc
ii}]\,6.99\,$\mu$m. In the low-resolution
\emph{Spitzer}/IRS spectra, the C$_{60}$\,18.9\,$\mu$m flux is
contaminated by the [S\,{\sc iii}]\,18.67\,$\mu$m line. This
is in case of \mbox{Lin49}. As we discuss later, the contamination 
of the C$_{60}$ band fluxes except for the 
C$_{60}$\,18.9\,$\mu$m seems to be small.

In this modelling, we determine the intrinsic luminosity ($L_{\ast}$),
the stellar radius ($R_{\ast}$), and the core-mass ($M_{\ast}$) of the CSPN. We 
estimate the initial mass of the progenitor star by plotting the
$L_{\ast}$ and $T_{\rm eff}$ on theoretical evolutionary tracks of post AGB stars. 
We also compare the ICFs from the P-I model with those calculated in section~\ref{S:neb-abund}.

\subsubsection{Modelling approach}

\begin{table}
\centering
\caption{Input parameters for the best fitting and the derived 
properties by the {\sc CLOUDY} model. \label{model}}
 \begin{tabu} to \columnwidth {@{}l@{\hspace{15pt}}l@{}}
\hline
{Parameters of the Central star}      &\multicolumn{1}{c}{Values}\\
\hline
$L_{\ast}$       &5916~L$_{\odot}$\\
$R_{\ast}$       &2.73~R$_{\odot}$\\
$M_{\ast}$       &0.53~M$_{\odot}$\\
$T_{\rm eff}$    &30\,500\,K        \\
$M_{V}$          &--1.62\\
$m_{V}$          &17.34\\
$\log\,g$        &3.29~cm s$^{-2}$\\
Distance         &61.9~kpc\\
\hline
{Parameters of the Nebula}      &\multicolumn{1}{c}{Values}\\
\hline
Boundary condition    &Ionisation bound\\
$\epsilon$(X)         &He:10.80/C:8.46/N:7.06/O:8.03,\\
                      &Ne:7.22/S:5.88/Cl:4.09/Ar:5.21, \\
                      &Fe:4.79/the others:\citet{Fishlock:2014aa}\\ 
Geometry              &Spherical\\
Shell size            &$R_{\rm in}$ =  0.00063 pc (130 AU)\\
                      &$R_{\rm out}$ = 0.068 pc (1.42(+4) AU)\\         
$n_{\rm H}$           &5080 cm$^{-3}$\\
Filling factor        &0.50\\
$\log_{10} I$({\hb})  &--12.89~erg s$^{-1}$ cm$^{-2}$\\
$m_{\rm g}$           &0.11\,M$_{\odot}$ \\
\hline
{Parameters of the Dust}      &\multicolumn{1}{c}{Values}\\
\hline
Grain               &graphite only\\
Grain radius        &0.005-0.10\,$\mu$m\\
$T_{d}$             &See Fig.~\ref{F:cloudy_td}\\
$m_{d}$             &4.29(--5)\,M$_{\odot}$ \\
$m_{d}$/$m_{\rm g}$ &3.97(--4) \\
  \hline
\end{tabu}
\end{table}

The distance to \mbox{Lin49} is necessary for the comparison of the
model with the observed fluxes and flux densities. Recent distance
measurements to the SMC are 60.6 $\pm$ 2.9~kpc \citep{Hilditch:2005aa}, 
62.1 $\pm$1.9~kpc \citep{Graczyk:2014aa}, and 62.0 
$\pm$ 0.6~kpc \citep[][the distance was calculated from their distance
modulus ($m$-$M$) = 18.96 $\pm$ 0.02]{de-Grijs:2015aa}. Using photometric data of red clump stars, \citet{Subramanian:2009aa} 
investigated the line-of-sight (LOS) depth in the MCs. From their LOS depth map of 
the SMC and the location of \mbox{Lin49} (this PN would be a bar member), the LOS 1-sigma 
depth toward \mbox{Lin49} is in the range from 4 to 6~kpc, assuming an average value
for the distance of 60~kpc toward the SMC. Here we adopt the distance
toward \mbox{Lin49} to be 61.9~kpc, the average of the values above, 
weighted by the respective uncertainties, with a 1-sigma error in the
average of $\pm$5.0~kpc from the LOS depth.

We used the {\sc TLUSTY} synthetic spectrum of the central star to
define the ionising/heating source as displayed in Fig.~\ref{F:CSPNsed}
($H_{\lambda}$ is the flux density of the stellar photosphere), while 
$L_{\ast}$ is a free parameter.

Except for C, we adopt the results listed in Table~\ref{element-icf} 
as initial guesses for the nebular elemental 
abundances, and refine these to match the observed line intensities 
of each element. As we explained in section~\ref{S:carbon}, we adopt and keep the {\it expected} CEL
$\epsilon$(C) of 8.46 throughout the model because we do not detect any C
CELs constraining the CEL C abundance. 
For elements for which abundances could 
not be determined from nebular line analysis, we adopt the 
AGB nucleosynthesis model result 
of \citet{Fishlock:2014aa} for stars with initial mass 1.25\,M$_{\odot}$
and $Z$ = 0.001.

Following the definition of \citet{Stanghellini:1999aa} and 
\citet{Shaw:2006aa} applied to MC PNe, we measure the photometric radius of \mbox{Lin49} to be 0.23$\arcsec$, 
corresponding to the size of a circular aperture that contains 85\,\% of 
the flux in the $z'$-band. We naturally consider 
the point spread function (FWHM$\sim$0.69$''$). We adopt a spherical 
shell nebula with uniform hydrogen density ($n_{\rm H}$). 
Thus, we set the outer radius ($R_{\rm out}$) to be 0.23$\arcsec$, 
where we define the ionisation front.

A definition of the filling 
factor is the ratio of a RMS density derived from an observed hydrogen 
line flux (e.g., {\ha} and {\hb}), $T_{\rm e}$, and nebula radius to the
$n_{\rm e}$(CELs) \citep[See e.g.,][for detail]{Mallik:1988aa,Peimbert:2000aa}. 
We calculate a RMS density of 3600 cm$^{-3}$ from the observed
$I$({\hb}), $T_{\rm e}$ = 11\,000\,K, the radius = 0.23$\arcsec$, 
and a constant $n_{\rm e}$/$n({\rm H^{+}})$ = 1.15. Thus, we estimate the 
filling factor to be around 0.5 using this RMS density and the {\Ne}({\oii}).

We assume that the underlying continuum is due to graphite grains based 
on the fact that the nebula in \mbox{Lin49} shows the spectral signature
of carbon-rich species (i.e., fullerene). We use the optical data of
\citet{Martin:1991aa} for randomly oriented graphite spheres, and assume 
the "1/3-2/3" approximation \citep[for more details of this approximation see,][]{Draine:1993aa}. 
We adopt an MRN $a^{-3.5}$ size distribution
\citep{Mathis:1977aa} with the smallest grain radius ($a_{-}$) = 
0.005\,$\mu$m and the largest radius ($a_{+}$) = 
0.1\,$\mu$m. We resolved the size
distribution into 20 bins. We have not attempted to reproduce the 6-9\,$\mu$m band and the broad
11\,$\mu$m/16-24\,$\mu$m/30\,$\mu$m features
because the carriers of these features and their optical properties are not well known.

To find the best model, we use the 
{\sc vary} command of {\sc CLOUDY}. This command  allows us to vary parameter 
within a given range in order to match the observed values.
In total, we varied 12 free parameters; $L_{\ast}$, the
He/N/O/Ne/S/Cl/Ar/Fe abundances, $R_{\rm in}$, $n_{\rm H}$, and grain abundance 
until the $\chi^{2}$ value calculated from the 51 gas emission
fluxes, 4 broad band fluxes (2MASS $JHKs$ and IRAC bands were excluded), 
3 far-IR flux densities at 65, 90, 120\,$\mu$m, and the $I$({\hb}) 
was minimised. The final $\chi^{2}$ was 22.3.

In Table~\ref{model}, we list the input parameters for the best 
fit and the properties derived by applying the model. 
The 1-$\sigma$ confidence interval of each elemental abundance is as follows;
$\epsilon$(He) = 0.02, $\epsilon$(N) = 0.03, $\epsilon$(O) = 0.02,
$\epsilon$(Ne) = 0.11, $\epsilon$(S) = 0.02, $\epsilon$(Cl) = 0.09,
$\epsilon$(Ar) = 0.06, and $\epsilon$(Fe) = 0.13, respectively. 
We estimate the 1-$\sigma$ uncertainty of the $m_{g}$, $m_{d}$, and
$m_{d}$/$m_{g}$ to be 0.02\,M$_{\odot}$, 1.4(--6)\,M$_{\odot}$, and 6.8(--5) by taking
the absolute \emph{Spitzer}/IRS flux calibration uncertainty 
of $\sim$17 $\%$ \citep{Decin:2004aa} and the uncertainty of the distance.

In appendix Table~\ref{model2}, we compare the observed and the model 
predicted values and list the predicted fluxes of important 
diagnostic lines such as the [C\,{\sc ii}]\,157.6\,$\mu$m for the more ionised plasma and the 
photodissociation region (PDR) for the future studies. A discussion on the model 
results is presented in the following sections.

\subsubsection{Comments on the model results}

\begin{figure}
\includegraphics[width=\columnwidth]{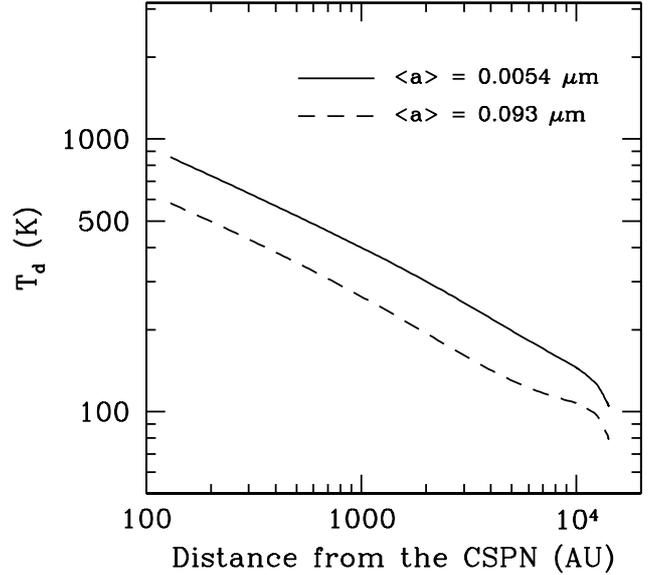}
\caption{Radial temperature profiles of graphite grains in the 
smallest and largest size bins. }
\label{F:cloudy_td}
\end{figure}

\begin{figure*}
\includegraphics[width=0.8\textwidth]{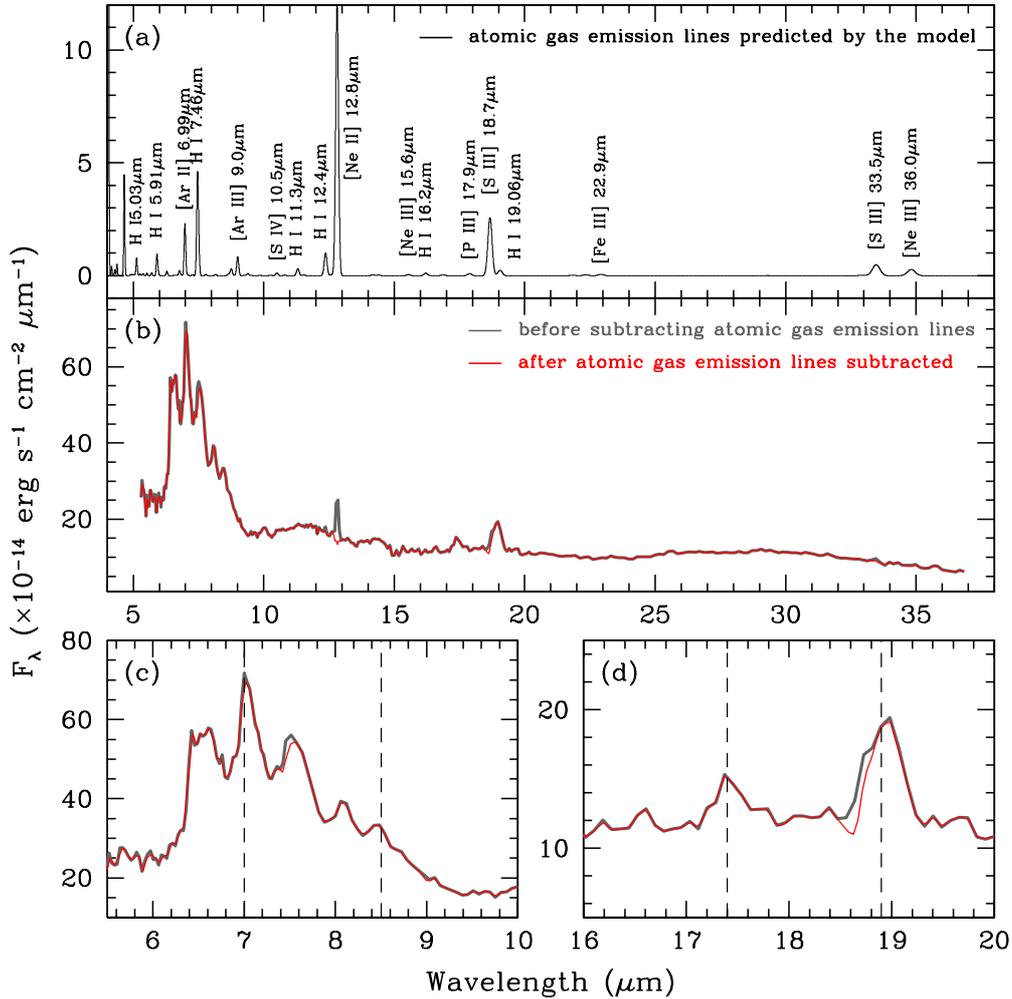}
\caption{({\it a}) The predicted atomic gas emission lines by
 the {\sc CLOUDY} model. The resolution of this synthesised
 spectrum is 100, corresponding to that of \emph{Spitzer}/IRS.
 ({\it b}) The original 
\emph{Spitzer}/IRS spectrum (grey-line) and the atomic line subtracted 
spectrum (red-line). ({\it c} and {\it d}) closed-up plot 
of the four C$_{60}$ bands, whose central wavelengths are indicted by
 the dashed vertical lines. 
}
\label{F:cloudy_emission}
\end{figure*}

\begin{figure}
\includegraphics[width=\columnwidth]{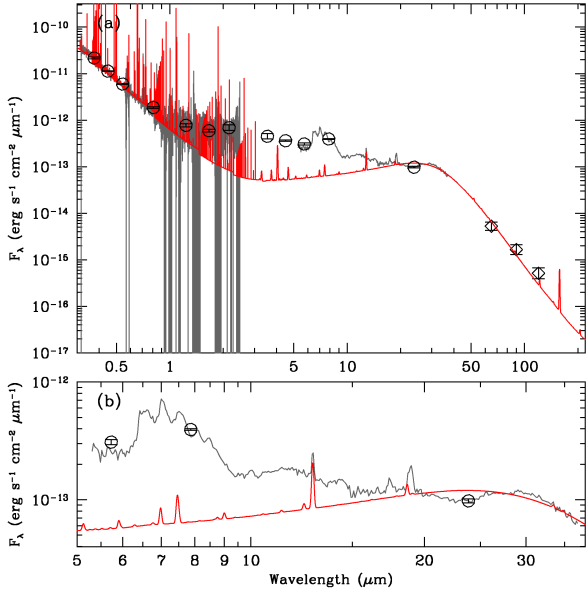}
\caption{Spectral energy distribution (SED) of \mbox{Lin49}.  ({\it panel
 a}) Comparison between the {\sc CLOUDY} model SED (red line) and the
 observational data.  The spectral resolution of the gas emission lines is constant and assumes
 the following values: 1000 in $\lambda$ $<$~0.3\,$\mu$m, 9200 in
 $\lambda$ = 0.3-1.0\,$\mu$m, 4800 in $\lambda$ = 1.0-2.5\,$\mu$m, 
and 100 in $\lambda$~$>$~2.5\,$\mu$m. The observed spectral and 
photometric data (XSHOOTER, MCPS $UBVI$ bands, 2MASS $JHKs$ bands, 
and \emph{Spitzer}/IRAC 4 bands and MIPS 24\,$\mu$m band) are 
also plotted (grey lines and black circles). The XSHOOTER, 
MCPS, and 2MASS data are de-reddened values with $c$({\hb}) 
= 0.11 and $R_{\rm V}$ = 3.1. The squares at 65, 90, 120\,$\mu$m 
are the average {\it expected} flux density obtained from 
Fits 1 and 2 in Section~\ref{S:30um}. 
({\it panel b}) Closed-up plot for mid-IR wavelengths. See the text for details.}
\label{F:sed}
\end{figure}

\begin{table}
\centering
\caption{The volume average ionisation fraction of each element and
 comparison of the ICFs from the {\sc CLOUDY} model and ones
 adopted in section~\ref{S:neb-abund}. The ICF(Obs) in He is derived
 from equation (\ref{he-icf}). The ICF(Obs) in He based on the P-I model of M1-11
 is 3.81 $\pm$ 0.22. The ICF(Obs) in C is for
 the RL $\epsilon$(C).}
\begin{tabu} to \columnwidth {@{}lccccc@{\hspace{5pt}}c@{}}
\hline
X & X$^{0}$/X & X$^{+}$/X & X$^{2+}$/X &
 X$^{3+}$/X & ICF({\sc CLOUDY}) & ICF(Obs) \\ 
\hline
He & 0.632 & 0.368 &   &   & 2.72 & 5.33$\pm$0.14 \\ 
C &        & 0.759 & 0.241 &   & 4.15 & 3.96$\pm$0.23 \\ 
N & 0.004 & 0.841 & 0.156 &   & 1.19 & 1.08$\pm$0.04 \\ 
O & 0.017 & 0.948 & 0.034 &   & 1.02 & 1.00 \\ 
Ne & 0.013 & 0.984 & 0.003 &   & 1.02 & 1.00 \\ 
S &       & 0.243 & 0.757 &   & 1.00 & 1.00 \\ 
Cl &      & 0.372 & 0.627 &   & 1.00 & 1.00 \\ 
Ar & 0.004 & 0.668 & 0.327 &   & 3.05 & 4.62$\pm$1.11 \\ 
Fe &       & 0.073 & 0.871 & 0.056 & 1.15 & 1.22$\pm$0.05 \\ 
\hline
\end{tabu}
\label{T:ICF}
\end{table}

 \begin{table*}
  \centering
\caption{Measurements of the four C$_{60}$ bands before(``B'')/after(``A'')
 subtracting the synthesised mid-IR spectrum presented in Fig.~\ref{F:cloudy_emission}(a).}
\begin{tabu} to \textwidth {@{}rrccrcc@{\hspace{-4pt}}r@{}}
 \hline
Band&\multicolumn{1}{c}{$\lambda_{\rm c}$(B)} &FWHM(B) &$F$(B)&\multicolumn{1}{c}{$\lambda_{\rm c}$(A)} &FWHM(A)&$F$(A)&\multicolumn{1}{c}{($F$(A) - $F$(B))}\\
    &\multicolumn{1}{c}{($\mu$m)}&($\mu$m)&(erg s$^{-1}$ cm$^{-2}$)
 &\multicolumn{1}{c}{($\mu$m)}&($\mu$m)&(erg s$^{-1}$
 cm$^{-2}$)&\multicolumn{1}{c}{/$F$(B) ($\%$)}\\
 \hline
7.0\,$\mu$m &7.04 $\pm$ 0.01 & 1.97(--1) $\pm$ 3.26(--3) & 5.14(--14) $\pm$ 9.50(--16) & 7.04 $\pm$ 0.01 & 1.98(--1) $\pm$ 3.22(--3) & 4.99(--14) $\pm$ 9.80(--16) & 3.0 \\ 
8.5\,$\mu$m&8.47 $\pm$ 0.01 & 1.61(--1) $\pm$ 2.29(--2) & 7.41(--15) $\pm$ 7.43(--16) & 8.47 $\pm$ 0.01 & 1.57(--1) $\pm$ 2.14(--2) & 7.18(--15) $\pm$ 8.54(--16) & 3.1 \\ 
17.4\,$\mu$m&17.42 $\pm$ 0.02 & 3.27(--1) $\pm$ 3.90(--2) & 1.05(--14) $\pm$ 1.30(--15) & 17.42 $\pm$ 0.01 & 3.23(--1) $\pm$ 5.67(--2) & 1.05(--14) $\pm$ 1.70(--15) & 0.5 \\ 
18.9\,$\mu$m&18.92 $\pm$ 0.01 & 3.76(--1) $\pm$ 2.42(--2) & 2.86(--14) $\pm$ 2.00(--15) & 18.95 $\pm$ 0.01 & 3.18(--1) $\pm$ 1.68(--2) & 2.51(--14) $\pm$ 1.40(--15) & 12.3 \\ 
\hline
\end{tabu}
\label{T:C60}
 \end{table*}

Assuming that the CSPN is in the midst of H-burning, comparing 
the estimated $L_{\ast}$ and the measured $T_{\rm eff}$ on
H-burning post-AGB evolution tracks with the initial $Z$=0.001 
of \citet{Vassiliadis:1994aa} indicates a progenitor mass of 
1.0-1.5\,M$_{\odot}$, which is consistent with our interpretation in
section~\ref{S:compAGB}, where we concluded that \mbox{Lin49} evolved
from a 1.0-1.25\,M$_{\odot}$ star based on elemental abundances. 
The conclusion on the initial mass 
does not change even in the two density composite model discussed in
section~\ref{S:highdens}.

In Table~\ref{T:ICF}, we list the volume-averaged ionisation fraction of
each element predicted by the {\sc CLOUDY} model and the ICFs derived 
from those. The ICF(He) calculated from the equation~(\ref{he-icf}), which
is tuned for the H\,{\sc ii} regions, is overestimated and 
the ICF(He) = S/S$^{2+}$ adopted for the PN M2-24 \citep{Zhang:2003aa} may not be correct for
\mbox{Lin49} (see appendix section \ref{S:He}). For low-ionised PNe, it is better to use the ICF(He) calculated by the
photoionisation model. Since we do not include O$^{0}$ and N$^{0}$ in the elemental
O and N abundances and we assume that Ne$^{0}$ is very small, 
their ICFs are slightly different from those of the model (0.02 in O and Ne and 0.11
in N).

The simulated dust temperature radial profiles are displayed in 
Fig.~\ref{F:cloudy_td}, where we plot the grain temperatures ($t_{\rm d}$)
of the smallest and largest size bins. The size range of the smallest and
largest bins is 0.0050-0.058\,$\mu$m ($\langle{a}\rangle$ = 0.0054\,$\mu$m) and
0.086-0.10\,$\mu$m ($\langle{a}\rangle$ = 0.093\,$\mu$m). 
The maximum and minimum $t_{\rm d}$ are 825\,K and 79\,K.

Fig.~\ref{F:cloudy_emission}(a) shows the synthesised mid-IR spectrum 
composed of atomic gas emission lines only based on the model result. 
By subtracting this generated spectrum from the observed \emph{Spitzer}/IRS 
spectrum, we can see how the four C$_{60}$ bands are contaminated with 
atomic lines, in particular, [Ar\,{\sc ii}]\,6.99\,$\mu$m and
[S\,{\sc iii}]\,18.67\,$\mu$m lines and we can obtain more
accurate measurements of these C$_{60}$ bands. Fig.~\ref{F:cloudy_emission}(b) shows the original 
\emph{Spitzer}/IRS spectrum (grey-line) and the atomic line subtracted 
spectrum (red-line). We focus on the C$_{60}$ bands in Figs.~\ref{F:cloudy_emission}(c) and
(d). In Table~\ref{T:C60}, we summarise the measurements of the four
C$_{60}$ bands before(``B'')/after(``A'') subtracting the synthesised
mid-IR spectrum presented in Fig.~\ref{F:cloudy_emission}(a).
Accordingly, we confirm that the line contribution is very small 
except for the 18.9\,$\mu$m C$_{60}$ band; compared to the result
before subtracting atomic gas emission contributions (mainly
{\siii}\,18.67\,$\mu$m), its central
wavelength ($\lambda_{\rm c}$) is shifted towards red wavelengths, 
FWHM is much narrower, and the flux is smaller.

In Fig.~\ref{F:sed}(a), we present the observed SED plots
(grey circles and lines) and the modelled SED (red line). The three grey
diamonds in far-IR are the fluxes predicted by the broad 30\,$\mu$m
fitting. In Fig.~\ref{F:sed}(b), we display a close-up of the SED in 
the wavelength range covered by the \emph{Spitzer}/IRS spectrum. 
The simulated SED can not fit the near-IR observed data. 
This strong near-IR excess in \mbox{Lin49} is also impossible 
to fit by models with amorphous carbon. The near-IR excess 
is better revealed in Fig.~\ref{F:nir}, where we 
show the residual spectra (grey lines) and photometry 
points (blue filled circles) obtained by making the 
difference between the observations and the {\sc CLOUDY} model values.

\section{Discussion\label{S:S4}}
\subsection{Interpretations for the near infrared excess \label{S:nir}}

\begin{figure}
\centering
\includegraphics[width=\columnwidth]{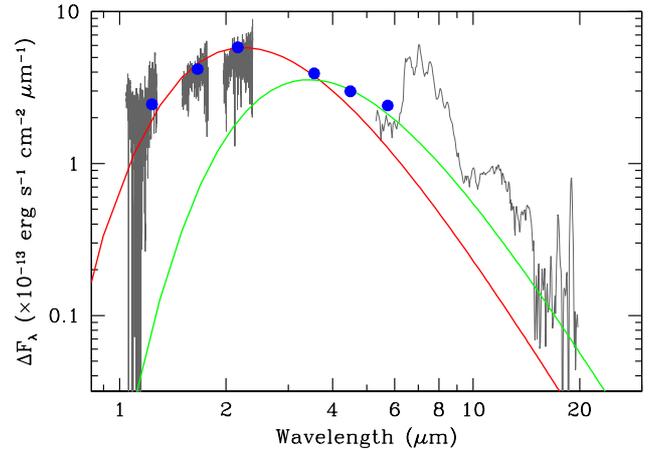}
\caption{Near-IR excess. The grey lines and the blue dots are the
 residual flux densities 
($\Delta$\,$F_{\lambda}$) between the observed XSHOOTER/IRS spectra and 
2MASS $JHKs$, IRAC 3.6/4.5/5.8\,$\mu$m photometry bands and the corresponding 
values obtained from the {\sc CLOUDY} model. In the XSHOOTER 
and IRS spectra, we block the spectral regions except for
 1.046--1.271\,$\mu$m ($J$), 1.508--1.778\,$\mu$m ($H$), 
1.974-2.377\,$\mu$m ($Ks$), and 5.31--19.74\,$\mu$m. The
 red line is the best fit of this near-IR excess with a Plank function
 with a single temperature of 1250\,K. The
 green line is another Plank function fit with the fixed 861\,K. 
The $\Delta$\,$F_{\lambda}$ in the photometry bands are listed in
 Table~\ref{T:nir-IR}. See the text in section~\ref{S:nir}.
\label{F:nir}
}
 \end{figure}

\begin{table}
\centering
\caption{The residual flux densities ($\Delta$\,$F_{\lambda}$) between
 the observed 2MASS $JHKs$ and \emph{Spitzer}/IRAC 3.6/4.5/5.8\,$\mu$m bands and 
the corresponding values obtained from the {\sc CLOUDY} model. 
\label{T:nir-IR}}
\begin{tabu} to \columnwidth {@{}LCC@{}}
\hline
Band&$\lambda_{\rm c}$ &$\Delta$\,$F_{\lambda}$\\
&($\mu$m)          &(erg s$^{-1}$ cm$^{-2}$ $\mu$m$^{-1}$)\\
\hline
2MASS $J$  &1.235  &2.83(--13)  \\
2MASS $H$  &1.662  &4.32(--13)  \\
2MASS $Ks$ &2.159  &5.87(--13)  \\
IRAC-Band1 &3.600  &3.96(--13)  \\
IRAC-Band2 &4.500  &3.06(--13)  \\
IRAC-Band3 &5.800  &2.51(--13)  \\
\hline
\end{tabu}
\end{table}

\subsubsection{Stochastic heating of extremely small particles}

For the usual dust grain sizes (e.g., $\gtrsim$ 0.01\,$\mu$m), dust 
temperatures are determined by solving an energy balance equation
between the radiative heating owing to the central star and the cooling
of grains. For such grain sizes, individual quantum events are not
important. However, for very small grains, which are composed of 100
atoms or less, single photons would cause them to heat up significantly
for very short time scales. This mechanism is known as
stochastic heating (or quantum heating), and has been proposed to explain the spectra of the
reflection nebulae NGC7023 and NGC2023 \citep{Sellgren:1984aa}, and the PNe
IC418 \citep{Phillips:1984aa} and Abell 58 \citep{Koller:2001aa}. 
Interestingly, these reflection nebulae and IC418 show mid-IR C$_{60}$
band emission. We included the stochastic heating mechanism 
in {\sc CLOUDY} model, as is default for {\sc
CLOUDY}. However, our model cannot fit the observed SED in the 
$\sim$1-5\,$\mu$m wavelength range at all.

For the residual data plots in Fig.~\ref{F:nir}, it is
possible to fit the excess with a Planck function with a single temperature 
of 1250 $\pm$ 42\, K (indicated by the red line). The 
luminosity and the minimum emitting radius of this component are
290 $\pm$ 80\,L$_{\odot}$ and $\sim$2\,AU. According to 
\citet{Sellgren:1984aa} and \citet{Whittet:2003aa}, the thermal
properties of solids are described by Debye's theory and the heat capacity
$C_{\rm V}$ for $T$ over Debye temperature $\Theta$ (in graphite, $\Theta$ is
$\sim$500\,K) is 3$N$$k$ where $N$ is the number of the atoms in a
molecule and 3$N$ is its number of degrees of freedom. For extremely small grains,
an average absorbed photon energy $E_{\rm ph}$ produces the difference between the maximum and minimum 
temperatures $\Delta~{T}$ written by the following equation: 
\begin{equation}
\Delta~{T} = \frac{E_{\rm ph}}{3Nk}.
\end{equation}
The $T_{\rm eff}$ of the central
star is 30\,500 K, so the photon energy at the radiation peak is 13.05
eV  (Fig.~\ref{F:CSPNsed}). $E_{\rm ph}$ could be
lower than 13.05\,eV. Thus, we obtained $\Delta~{T}$ $<$ 5.05(+4)/$N$.

By adopting the maximum and minimum temperatures of 1250\,K and 20\,K, 
we obtain a value for $N$ of $\lesssim$~39, although our estimation
is very optimistic and also depends on the minimum temperature. 
If such a molecule formed as a honeycomb structure sheet is distributed in the nebula, the molecule's dimension is roughly $\sim$8(--4)\,$\mu$m $\times$ 8(--4)\,$\mu$m square. If the molecule is a cage not a sheet, .e.g, fullerene C$_{36}$, the size would be small; in the case of C$_{36}$, the approximate diameter is 5(--5)\,$\mu$m \citep{Piskoti:1998aa}. 
The SED with $\Delta~T$ of 841\,K derived by keeping $N$ of 60
(indicated by the green line) gives a better fit to the differential spectrum in $\gtrsim$~3.6\,$\mu$m. 

A top-down mechanism has been suggested for the formation of C$_{60}$; 
C$_{60}$ could be formed from the shrinkage of larger
molecules, e.g., from larger clusters of PAHs \citep{Zhen_14_Laboratory,Berne_15_Top}, or
HAC \citep{Duley_12_FULLERENES}. PAH clusters could form from HACs. 
\citet{Scott:1997ab} showed that C$_{50,60,70}$ may be produced 
by the decomposition of HACs. As explained 
in Section~\ref{S:Spitzer}, the 6-9\,$\mu$m band profile in Lin49 is
very similar to the thermal emission profile of HAC 
as presented in Fig.~2 of \citet{Scott:1997aa}. Molecules composing of $\lesssim$~39
C-atoms might be a by-product in the decomposition process of HACs.

As a pragmatic problem, with grains composing of
$\lesssim$~39 C-atoms only, it could be difficult to reproduce the
observed broad continuous near-IR
excess feature seen in \mbox{Lin49}; to get a continuum like behaviour,
enough interacting vibrational modes are necessary. Therefore, we need 
to examine other possible explanations for near-IR excess in \mbox{Lin49}.

\subsubsection{High density structure nearby the CSPN}
\label{S:highdens}

\begin{figure}
\includegraphics[width=\columnwidth]{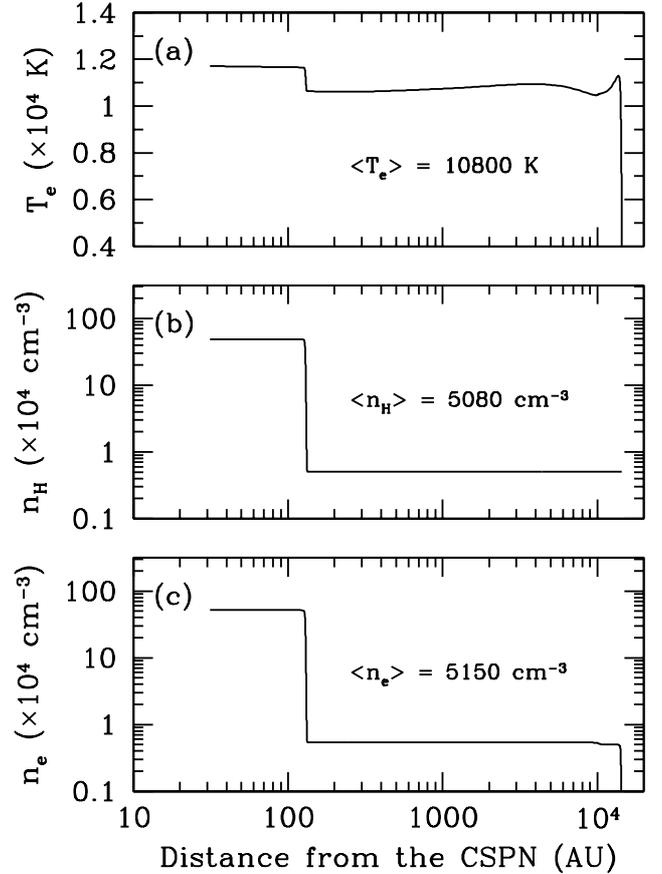}
\caption{Radial profiles of the {\te} ({\it panel a}), $n$(H)
 ({\it panel b}), and {\Ne} ({\it panel c}) predicted by the
 two density shell model. The averaged value of each physical parameter 
is indicated in each panel. }
\label{F:physical}
\end{figure}

 \begin{figure}
\includegraphics[width=\columnwidth]{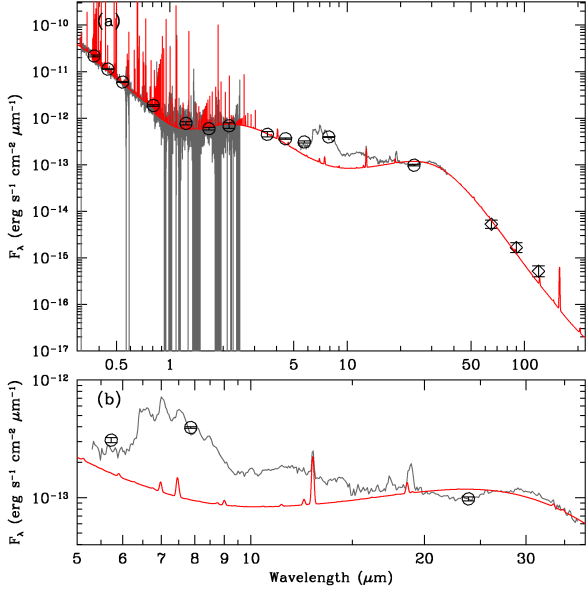}
\caption{({\it panel a}) Comparison between the observed SED plots and the predicted SED
  by the two density shell model. The
 spectral resolution of the gas emission lines is a constant 1000 in
 $<$~0.3\,$\mu$m, 9200 in 0.3-1.0\,$\mu$m, 4800 in 1.0-2.5\,$\mu$m, 
and 100 in $>$~2.5\,$\mu$m.  ({\it panel b}) Close-up plot for mid-IR
  wavelengths. The lines and symbols in both of panels are as defined in
  Fig.~\ref{F:sed}.}
\label{F:sed2}
 \end{figure}

\begin{figure}
\includegraphics[width=\columnwidth]{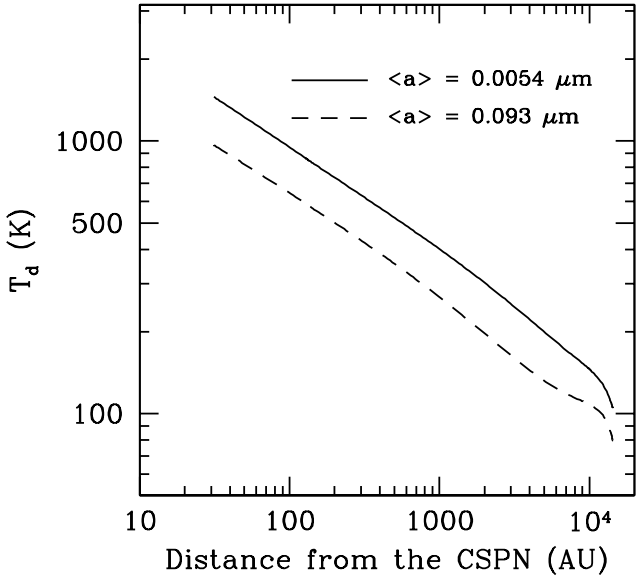}
\caption{Radial temperature profiles of graphite grains in the 
smallest and largest size bins predicted by the two density shell model. 
}
\label{F:cloudy_td2}
\end{figure}

\begin{table}
\centering
\caption{Input parameters of the best fitting in the two density shell
 model and the derived properties. \label{modelT}}
 \begin{tabu} to \columnwidth {@{}l@{\hspace{15pt}}l@{}}
\hline
{Parameters of the Central star}      &\multicolumn{1}{c}{Values}\\
\hline
$L_{\ast}$       &6333~L$_{\odot}$\\
$R_{\ast}$       &2.84~R$_{\odot}$\\
$M_{\ast}$       &0.57~M$_{\odot}$\\
$T_{\rm eff}$    &30\,500\,K        \\
$M_{V}$          &--1.70\\
$m_{V}$          &17.26\\
$\log\,g$        &3.29~cm s$^{-2}$\\
Distance         &61.9~kpc\\
\hline
{Parameters of the Nebula}      &\multicolumn{1}{c}{Values}\\
\hline
Boundary condition    &Ionisation bound\\
$\epsilon$(X)         &He:10.80/C:8.46/N:7.06/O:8.03,\\
                      &Ne:7.27/S:5.88/Cl:4.08/Ar:5.22, \\
                      &Fe:4.71/the others:\citet{Fishlock:2014aa}\\ 
Geometry              &Spherical\\
Shell size            &$R_{\rm in}$ =  0.00015 pc (31 AU)\\
                      &$R_{\rm out}$ = 0.068 pc (1.42(+4) AU)\\         
$n_{\rm H}$           &See Fig.~\ref{F:physical}\\
Filling factor        &0.50\\
$\log_{10} I$({\hb})  &--12.89~erg s$^{-1}$ cm$^{-2}$\\
$m_{\rm g}$           &0.11\,M$_{\odot}$ \\
\hline
{Parameters of the Dust}      &\multicolumn{1}{c}{Values}\\
\hline
Grain               &graphite only\\
Grain radius        &0.005-0.10\,$\mu$m\\
$T_{d}$             &See Fig.~\ref{F:cloudy_td2}\\
$m_{d}$             &4.15(--5)\,M$_{\odot}$ \\
$m_{d}$/$m_{\rm g}$ &3.86(--4) \\
  \hline
\end{tabu}
\end{table}

In section~\ref{photomodel}, we assumed 
that \mbox{Lin49} does not have any sub-structures 
surrounding the CSPN but that this PN has a normal density nebula. However, 
the minimum emitting radius of the 1250\,K blackbody component
suggests that the near-IR component 
could be emitted by a sub-structure near the central star. A similar idea was
proposed for the near-IR excess at the central position of 
IC418; \citet{Hora:1993aa} took the near-IR $JHK$ images of this PN and
found excess at the central position after subtracting the
contribution from the central star. The authors argued that the excess
indicates a possible compact shell interior to the main shell. 

\mbox{Lin49} may also have a central dense structure, which will
be responsible for its near-IR excess. To test this hypothesis, we
construct a two-shell model. The model is composed of an outer low density shell 
and an inner high density shell. 
For the dust distribution on the high density shell we assume the dust
is composed of graphite grains with an $a^{-3.5}$ size distribution,
where $a$ = 0.005-0.1\,$\mu$m. 
$R_{\rm out}$ for this shell corresponds to the $R_{\rm in}$ for the outer, low density shell.
The photoionisation model for the low density shell has already been constructed
in section~\ref{photomodel}, with the residual SED indicating the near-IR
excess as presented in Fig.~\ref{F:nir}.

We fit the residual
SED in the wavelength range from 1 to 5\,$\mu$m using {\sc CLOUDY}. In
this process, we keep the following parameters derived in the low
density shell model: the filling factor, the elemental 
abundances and the dust composition/size distribution/abundance, and
$L_{\ast}$. As the luminosity of the near-IR excess component is
small compared to that of the central star ($<$~5\,$\%$ of $L_{\ast}$),
we will minimise the number of free parameters by initially 
fixing $L_{\ast}$ (later this value will be fine-tuned). The free
parameters are $R_{\rm in}$ and $n$(H), which are determined through 
fitting the residual 1-5\,$\mu$m SED.

We combine the radial $n$(H) profiles of the low and high density
shells into one (See Fig. \ref{F:physical}), and run the model with this $n$(H) profile to match the
observed SED plots in UV to far-IR wavelength. For fine-tuning, we allow an
increase of $L_{\ast}$ by 10\,$\%$ and a slight increase of the elemental abundances, except for
C.

Finally, we obtained the predicted SED as presented in
Fig.~\ref{F:sed2}, which better fits the 1-5\,$\mu$m wavelength rage, 
compared to Fig.~\ref{F:sed}. The $\chi^{2}$ was 37 calculated from 
the 51 gas emission fluxes, 9 broad band fluxes, 3 far-IR flux densities at 65, 
90, 120\,$\mu$m, and the $I$({\hb}). In the fourth and ninth
columns of appendix Table~\ref{model2}, we compare the observed and model predicted values and 
list the predicted fluxes. The input parameters for the best-fitting
model and the derived parameters are summarised in Table~\ref{modelT}.

Since we increased $L_{\ast}$ to obtain the best fit, the core-mass of the
CSPN is 0.57\,M$_{\odot}$, accordingly. As we argued before, even in
this core-mass, our conclusion on the initial mass of
the progenitor star does not change. The predicted elemental
abundances in the two density shell model are almost consistent with the single shell model in
section~\ref{photomodel}. The 1-$\sigma$ of the elemental abundances and
the $m_{g}$, $m_{d}$, and $m_{d}$/$m_{g}$ is almost same value
discussed in section~\ref{photomodel}. The radial profiles of {\te}, $n$(H), and {\Ne}
are presented in Fig.~\ref{F:physical}. The value for {\Ne} of the high density shell is very high. 
However, the emitting volume of the high density shell is very
small. Therefore, the volume-averaged {\Ne} (5150\,cm$^{-3}$) is not over the
observed one\footnote{XSHOOTER could not resolve Lin49 and this
instrument looks at the average light of this PN in each
wavelength. The Paschen line analysis could suggest the presence of such
high density shell indirectly.}. The radial $t_{d}$ profile is presented
in Fig.~\ref{F:cloudy_td2}. The maximum temperature of
the $\langle{a}\rangle$ = 0.0054\,$\mu$m grains (1450\,K) meets the
requirement to fit the residual 1-5\,$\mu$m SED plots but is not over
the evaporation temperature of graphite.

Thus, we succeeded in explaining the near-IR excess by postulating
the existence of a high density structure nearby the CSPN. If we can
believe that the near-IR excess emits from this high-density structure,
how did the progenitor form this during its evolution?

\subsubsection{Does Lin49 have a disc?}

We suggest that the near-IR emission does not originate from the nebular
shell, but from a disc around the central star. According to the review of
\citet{van-Winckel:2003aa}, the hot dust component ($\sim$1000\,K) in
some post-AGB stars \citep[first noticed by][]{Trams:1991aa} 
was interpreted as evidence for significant post-AGB mass loss. However,
this interpretation became untenable because dusty post-AGB mass loss
would speed up the evolution such that very few objects would be observable
\citep{Trams:1989aa}. At present, the most accepted formation mechanism
to produce a disc around an evolved star is the binary model
\citep[e.g.,][and references therein]{Kwok:2000aa}. This model states
that PNe with a disc around the central star evolve from a
binary system that went through a common envelope phase. During this
phase, the secondary star induces the mass loss of the AGB star to occur
preferentially in the orbital plane, which gives birth to a disc. Thus, if the
presence of a disc is confirmed around the central star of \mbox{Lin49}, this is
a strong indication that this PN evolved from a binary system. The
binary disc could stably harbour the near-IR emitters 
near the central star for a long time.

\citet{Kamath:2014aa} classified 63 SMC objects into post-AGB/red giant
branch (RGB) candidates based on their SEDs in the optical to mid-IR, 
and they reported that 21 objects out of these are post-AGB stars and 27
show a strong near-IR excess interpreted as the presence of a
circumbinary disc. \citet{Maas:2005aa} investigated 
elemental abundances of the Galactic 12 post-AGB stars showing near-IR
excess (they interpreted as the presence of a disc); 
nine of them are affected by the depletion process, that is, 
elements with a high condensation temperature (e.g., Fe) is largely depleted and get 
locked in dust grains whereas elements with a low temperature remain 
in gas phase. Since the temperature in the disc decreases with
increasing radius, at the inner radius of the disc only the elements with a high condensation 
temperature are caught in grains, given that the inner temperature 
of the disc is high. \citet{Maas:2005aa} indicated that the presence 
of of the depletion process and the presence of a disc are linked.

From the observational results of \citet{Kamath:2014aa}
\citet{Maas:2005aa}, the strong near-IR excess and the strongly 
depleted {\it nebular} Fe
abundance in \mbox{Lin49} might be explained by the presence of a
disc. Most of the Fe-atoms might be tied up in dust grains 
(e.g., FeO) within a disc. If \mbox{Lin49} has a disc, the large 
carbon molecules such as fullerene were relatively easily formed.

\subsubsection{Near-IR excess in SMC C$_{60}$ PNe and counterparts}

\begin{figure}
\includegraphics[width=1.0\columnwidth]{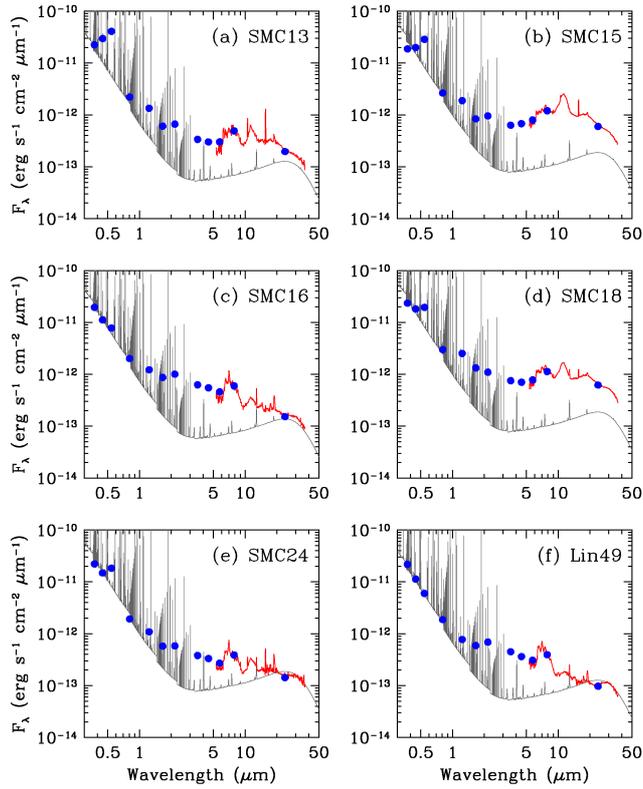}
\caption{SED plots of SMC C$_{60}$ PNe SMC13, 15, 16, 18, 24
 and Lin49. The blue filled circles are the de-reddened photometric 
data.  In each panel, we compare with the resultant
SED of Lin49 synthesised in section~\ref{photomodel}. Note that in each
 panel Lin49's SED is scaled to the observed de-reddened
flux density of each PN at the $I_{c}$ band.}
\label{F:sedSMC}
\end{figure}

\begin{figure}
\includegraphics[width=1.0\columnwidth]{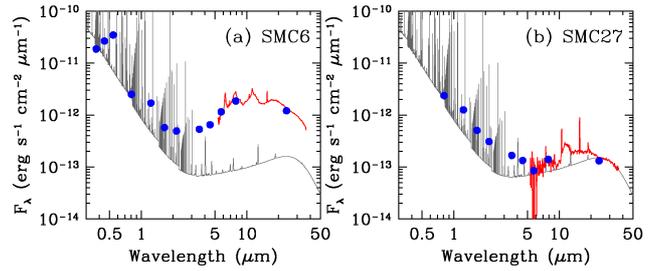}
\caption{SED plots of non-C$_{60}$ C-rich SMC PNe SMC6 and 27. 
The lines and symbols in both of panels are as defined in
 Fig.~\ref{F:sedSMC}.}
\label{F:sedSMC2}
\end{figure}

Is the near-IR excess seen in other SMC C$_{60}$ PNe? 
Amongst SMC C$_{60}$ PNe, SMP24 also shows a near-IR excess. 
By applying an SED fit over the range from $B$-band to $\sim$1.337 
GHz, \citet{Bojicic:2010aa} find that a hotter dust component 
($\sim$1000 K) is necessary to fit the observed SED down to 
1\,$\mu$m, apart from the hot dust component ($\sim$270\,K).

The photometric data from MCPS, 2MASS, and \emph{Spitzer}/IRAC and 
MIPS, and the \emph{Spitzer}/IRS spectra of 
the SMC C$_{60}$ PNe are plotted in Fig.~\ref{F:sedSMC}. As no optical 
data are available for SMC1, this source was not included in the 
figure. According to Table~\ref{T:compSMCXS}, the effective temperatures of the SMC 
C$_{60}$ PNe are in the range between 30\,500\,K and 58\,000\,K and 
the average $T_{\rm eff}$ amongst these PNe except for SMC15 
(58\,000\,K) is 34\,100\,K. The observed UV-optical wavelength 
SEDs in these PNe are not largely different from each other, except for SMC15. 
Thus, we plot the resultant synthesised SED of Lin49 (section
\ref{photomodel}) as the comparison; in each
panel this Lin49's SED is scaled to the observed de-redden
flux density of each PN at the $I_{c}$ band.

Amongst  C$_{60}$ PNe, the SEDs of SMC16 and SMC24 are very similar 
to that of Lin49; their SEDs have a flux density peak around 2MASS
$Ks$-band and the near-IR excess features are 
apparently as broad as that seen in \mbox{Lin49}. SMC13, 15, and 18 
do not show such a broad near-IR excess as seen in
\mbox{Lin49}. However, their SED slope in the range from $U$ to 
$Ic$ band wavelength is different from those in the range 
from 2MASS to IRAC band wavelength.

In Fig.~\ref{F:sedSMC2}, we show the SEDs of non-C$_{60}$ 
C-rich PNe, SMC6 and SMC27. From appendix Table~\ref{T:compSMCXS2}, they are 
selected as comparison objects to the C$_{60}$ SMC PNe because the $T_{\rm eff}$s 
of SMC6 and SMC27 are close to those of SMC C$_{60}$ PNe. We excluded
SMC11 as a comparison because \citet{Villaver:2004aa} reported $T_{\rm
eff}$ of 40\,900\,K whereas we confirmed that its SED in the range from
MCPS $B$ (no $m_{U}$) to IRAC bands has a peak around $Ic$-band 
and it can be well expressed by a single Planck function with the
temperature of 3722\,K, indicating that this object could not be a PN.

Although it is hard to draw a strong conclusion based on the photometric
data points and mid-IR spectra only, SMC C$_{60}$ PNe 
seem to show a near-IR excess component to lesser or greater degree. 
This suggests that these C$_{60}$ PNe might maintain a structure 
near their central star. Meanwhile, we do not find a flux density peak
around 2MASS $Ks$ in SMC non-C$_{60}$ PNe SMC6 and 27. To reach a firm
conclusion, we need to obtain near-IR spectra in order to check 
whether or not each PN displays a near-IR excess.

\subsection{Comparison of physical properties between 
C$_{60}$ PNe and non-C$_{60}$ C-rich PNe}

 \begin{table}
 \centering
\caption{The average elemental abundances of C$_{60}$-containing PNe and  
  non-C$_{60}$ C-rich PN in the SMC and the Milky Way (MW). The data of the SMC PNe
  are taken from Table~\ref{T:compSMCXS} and appendix Table~\ref{T:compSMCXS2}. The
  data of the MW PNe are from Tables~3 and 4 of \citet{Otsuka:2014aa}. 
}
 \begin{tabu} to \columnwidth {@{}lcccccc @{}}
\hline
PNe                 &$\epsilon$(C)&$\epsilon$(N)&$\epsilon$(O)&$\epsilon$(Ne)&$\epsilon$(S)&$\epsilon$(Ar)\\
\hline
SMC C$_{60}$     PNe&8.28 &7.17&7.97&7.07&6.54&5.64\\
Lin49               &8.46 &6.93&8.11&7.18&6.02&5.48\\   
SMC C-rich PNe&8.67 &7.37&8.08&7.28&6.69&5.64\\
\hline
MW C$_{60}$ PNe     &8.73 &7.81 &8.48 &7.85 &6.45&5.98\\
MW C-rich PNe       &8.85 &8.23 &8.56 &8.06 &8.05&6.85 \\
\hline
\end{tabu}
\label{T:abundS}
\end{table}

In Table~\ref{T:abundS}, we summarise the average elemental abundances
of SMC and Milky Way (MW) C-rich PNe. In section~\ref{S:compAGB}, from
the view of elemental abundances, we concluded that \mbox{Lin49} and the 
other SMC C$_{60}$ PNe evolved from the initially 1.0-1.25\,M$_{\odot}$ stars by comparing with the AGB nucleosynthesis
results of \citet{Fishlock:2014aa}. The lowest S and Ar abundances of \mbox{Lin49} amongst SMC C$_{60}$ PNe indicate that this PN is an older population in the SMC. While there are no large differences
in S and Ar (because these are Type II SN products), we found that the 
C, N, and Ne in SMC non-C$_{60}$ PNe are greater than those in SMC 
C$_{60}$ PNe, indicating that non-C$_{60}$ PNe evolved from
more massive stars. The C and Ne are synthesised in the He-rich intershell
during the thermal pulse AGB phase and these elements together with N and $n$-capture elements
are brought up to the stellar surface by the third dredge-up (TDU). The 
efficiency of TDU depends on the initial mass and composition and 
increases as larger initial mass \citep[e.g.,][]{Karakas:2010aa}. The rich Ne in 
non-C$_{60}$ PNe would be due to the double $\alpha$ capturing by rich 
$^{14}$N. Certainly, the average abundances in SMC non-C$_{60}$ PNe are
close to the AGB model result for the 1.25\,M$_{\odot}$ and
1.50\,M$_{\odot}$ stars rather for the 1.0\,M$_{\odot}$ and 1.25\,M$_{\odot}$ stars.

\citet{Otsuka:2014aa} argued that elemental abundances of MW C$_{60}$
PNe can be explained by AGB models for 1.5-2.5\,M$_{\odot}$ stars 
with the SMC metallicity (i.e., $Z$ = 0.004). Although there 
is a sample selection bias, the C, N, and Ne abundances in 
the non-C$_{60}$ MW C-rich PNe indicate that these PNe evolved
from more massive stars. Thus, at this moment, we might conclude that
the progenitors of C$_{60}$ PNe in both the SMC and the MW are not
greater than those of non-C$_{60}$ PNe.

\citet{Otsuka:2013aa,Otsuka:2014aa} reported that the MW 
C$_{60}$ PNe have cool central stars. The average $T_{\rm eff}$ amongst
MW C$_{60}$ PNe calculated from Table~3 of \citet{Otsuka:2014aa} 
is 37\,780\,K, which is in excellent agreement with the average $T_{\rm
eff}$ = 38\,770\,K amongst the SMC C$_{60}$ PNe (See
Table~\ref{T:compSMCXS}). The $T_{\rm eff}$ of \mbox{Lin49} is the
coolest amongst SMC C$_{60}$ PNe. Taking their average 
ionised nebula radius $r$ of 0.17$''$ (0.05\,pc in linear scale), 
SMC C$_{60}$ PNe are slower evolving objects than SMC 
non-C$_{60}$ PNe, where the average $T_{\rm eff}$ is 72\,100\,K and 
the average $r$ is 0.34$''$ (0.10\,pc in linear scale).

\section{Summary and future works\label{S:S5}}

We performed a detailed spectroscopic analysis of the fullerene 
C$_{60}$-containing PN \mbox{Lin49} in the SMC. 
We derived the nebular abundances of the nine elements. Compared to the
[S,Ar/H] abundances, the  [Fe/H] value is extremely low. Applying the predictions 
of the [S,Ar/Fe] abundances at [Fe/H]~$<$~--1 by the chemical 
evolution model for the Milky Way halo, the [Fe/H] in \mbox{Lin49} 
was originally $\sim$--1.4, indicating that the metallicity ($Z$) was 
$\sim$0.0006 ($\sim$0.04\,Z$_{\odot}$) and that $>$~96\,$\%$ of the
Fe-atoms are trapped in Fe-rich dust grains. The nebular abundances are
in good agreement with the AGB nucleosynthesis model for stars 
with an initial mass 1.25\,M$_{\odot}$ and $Z$=0.001 
of \citet{Fishlock:2014aa}, even taking into 
account that the {\it expected} CEL $\epsilon$(C) is 8.46.

We derived stellar abundances, effective temperature, and surface
gravity. From the nebular and stellar C/O ratio abundances and the observed 
dust features, \mbox{Lin49} is certainly a C-rich PN. We construct a
photoionisation model in order to investigate the 
physical conditions of the central star, nebula, and dust grains and
derive the gas and dust masses. The current core-mass of the CSPN is
0.53-0.57\,M$_{\odot}$, and theoretical evolution tracks of post-AGB stars
indicate that it was initially 1.0-1.5\,M$_{\odot}$. Taking the elemental
abundances into account, we conclude that \mbox{Lin49} evolved from
1.0-1.25\,M$_{\odot}$. Our 
model with the 0.005-0.1\,$\mu$m radius graphite grains and 
a constant hydrogen density shell cannot fit the $\sim$1-5\,$\mu$m part 
of the SED due to a prominent near-IR excess, whereas in the other 
wavelengths the model gave a reasonable fit to the observed 
fluxes of nebular lines and broad band fluxes/flux densities.

The near-IR excess might possibly be due to either (1) the presence 
of small carbon clusters, small graphite sheet, and also 
fullerene precursors, or (2) the presence of high-density structure 
surrounding the central star. 
Taking the observational results of C$_{60}$ PN IC418 in the MW 
and post-AGB/RGB stars in the SMC, and the extremely Fe-depletion in
\mbox{Lin49} into account, the latter option seems to be a
better interpretation for \mbox{Lin49}'s near-IR excess.

In addition to \mbox{Lin49}, we find that other SMC C$_{60}$ PNe also 
show a near-IR excess component to lesser or greater degree based on 
their UV to mid-IR photometry data and \emph{Spitzer}/IRS spectra. 
We suggest that these C$_{60}$ PNe might maintain a structure near their central 
star as a stable producing source of the near-IR excess in these PNe. 
Such a structure might be a circumbinary
disc and it might play a role in C$_{60}$ formation in evolved stars. Near-IR spectroscopy and monitoring 
observations of SMC C$_{60}$ PNe might confirm the near-IR excess and whether these PNe have binary central stars. 
Such observations would be important to 
understand the nature of the C$_{60}$ PNe, the evolution of their
central stars, and the formation of the C$_{60}$ in the circumstellar environment.

   \section*{Acknowledgements}
We are grateful to the anonymous referee for carefully reading the
manuscript and making useful suggestions which greatly improved this article. 
MO was supported by the research fund 104-2811-M-001-138 and 
104-2112-M-001-041-MY3 from the Ministry of Science and Technology
(MOST), R.O.C. FK also acknowledges financial support from MOST,
under grant numbers MOST103-2112-M-001-033- and 
MOST104-2628-M-001-004-MY3. MO and FK thank the IAA ICSM group members for fruitful discussions. MLL-F was supported by CNPq, 
Conselho Nacional de Desenvolvimento Cient\'ifico e Tecnol\'ogico - 
Brazil, process number 248503/2013-8. Studies of interstellar 
chemistry at Leiden Observatory are supported through the 
advanced-ERC grant 246976 from the European Research Council, 
through a grant by the Dutch Science Agency, NWO, as part 
of the Dutch Astrochemistry Network, and through the Spinoza 
prize from the Dutch Science Agency, NWO. JBS wishes to acknowledge the
support of a Career Integration Grant within the 7th European
Community Framework Program, FP7-PEOPLE-2013-CIG-630861-SYNISM. This work was partly based on archival data obtained with 
the \emph{Spitzer} Space Telescope, which is operated by the Jet 
Propulsion Laboratory, California Institute of Technology under a 
contract with NASA. Support for this work was provided by an award 
issued by JPL/Caltech. A portion of this work was based on 
the use of the ASIAA clustering computing system.

\bibliographystyle{mnras}

\appendix

\section{Comments on elemental abundance derivations using ICFs }
\label{AS:abundances}

\subsection*{He abundance \label{S:He}}

We removed the enhancement of the {\hei} triplet by collisional excitation 
from the 2$^{3}$$S$ level to derive He$^{+}$/H$^{+}$. Within their
uncertainties, we do not find significant differences between 
He$^{+}$/H$^{+}$ derived 
from the triplet 10830\,{\AA} and from other triplet lines except for 
He\,{\sc i}\,12528\,{\AA} (See below), which indicates 
a small radiative transfer effect \citep[e.g.,][]{Robbins:1968aa,Benjamin:2002aa,Olive:2004aa} 
in the lower transition lines (i.e.,
$\tau$({\hei}\,3889\,{\AA}) would be small). 
Therefore, we do not correct the radiative transfer effect for 
triplets. The abundance result obtained from the He\,{\sc
i}\,12528\,{\AA} is below the 3-$\sigma$ limit from the average
He$^{+}$/H$^{+}$ abundance (possibly due to the earth atmospheric
absorption correction). Therefore, we exclude this result 
when calculating the adopted He$^{+}$ abundance.

Estimating the He elemental abundance is not straightforward. 
Its derivation with the usage of the ICF(He) = 
(S$^{+}$+S$^{2+}$)/S$^{2+}$, proposed by \citet{Zhang:2003aa} 
for a low-excitation PN M2-24 , provides $\epsilon$(He) = 10.46, which 
is 1.6 lower than the predictions for the SMC primordial 
helium abundance \citep[e.g., $Y_{p}$ = 0.2477 $\pm$ 0.0029,
corresponding to $\epsilon$(He)$\sim$10.9,][]{Peimbert:2007aa}. 
This inconsistent result indicates that the He$^{0}$ abundance 
has not been satisfactorily taken into account by this approach.

To overcome this problem, we tested the ICF(He) = 2.37~$\cdot$~(S/S$^{2+}$). This 
ICF(He) was determined by the P-I model of M1-11, and is a 
consequence of the prediction that the respective fractions of the 
He$^{+}$ and S$^{2+}$ to He and S in M1-11 are 0.352 and 0.832. 
The selection of the S and S$^{2+}$ abundances to correlate with the He 
and He$^{+}$ abundances was based on the fact that the ionisation
potential of S$^{2+}$ is close to that of He$^{+}$. The usage of this
ICF(He) (3.81 $\pm$ 0.22) improved our results, leading to 10.83 $\pm$ 0.03. Here, 
we derived the S abundance using the equation (12) of \citet{Wesson:2005aa}
as follows. We obtain ICF(S) = 1.0. 
\begin{eqnarray}
\rm \frac{S}{H} &=&
\rm ICF(S)~\cdot~\left(\frac{S^{+}}{H^{+}}+\frac{S^{2+}}{H^{+}}\right), \nonumber\\ 
\rm ICF(S) &=& \rm \left[1-\left(1-\frac{O^{+}}{O}\right)^{3}\right]^{-1/3}.
\end{eqnarray}
From all elements in the neutral stage, Helium is the one with the
highest ionisation potential, $\gtrsim$~24~eV. 
Even a small difference in the $T_{\rm eff}$ of the central star of
\mbox{Lin49} in
comparison to the central star of M1-11, can lead to very
different fractions of He$^{0}$ and He$^{+}$ in each PN. Therefore, we
should treat $\epsilon$(He) = 10.83 $\pm$ 0.03 as a lower limit for the He abundance.

Finally, we tested the ICF(He) used in H\,{\sc ii} regions 
ionised by soft radiation sources, as written by \citet{Peimbert:1969aa} as: 
\begin{equation}
\rm \frac{He}{H} = \left(0.13\frac{O}{O-O^{+}}~+~0.87\frac{S}{S-S^{+}}\right)~\cdot~\frac{He^{+}}{H^{+}}.
\label{he-icf}
\end{equation}
\noindent Here, the O elemental abundance is the sum of the O$^{+}$ and
O$^{2+}$ abundances, and we excluded the neutral O abundance from the elemental O abundance.
Equation (\ref{he-icf}) gives $\epsilon$(He) = 10.99 $\pm$ 0.02, where
the term in parenthesis corresponds to ICF(He) = 5.53 $\pm$ 0.14. This
seems to be a much more reasonable He abundance result for an SMC PN, as
it is comparable to the median value amongst the 14 SMC PNe analysed by
\citet[][$\epsilon$(He) = 11.11]{Shaw:2010aa} and it is higher than
$\epsilon$(He) = 10.90, which is the mean He abundance of the SMC
H\,{\sc ii} regions compiled by \citet{Shaw:2010aa} based on the results
of \citet{Dennefeld:1989aa}. Thus, we adopt a range from 10.80 to 11.01 for the 
value of $\epsilon$(He).

\subsection*{N, Ne, Cl, Ar, and Fe abundances}

We exclude the neutral N abundance from the elemental N abundance.
The ICF(N) and N abundance were derived by \citet{Delgado-Inglada:2014ab} from:
\begin{eqnarray}
 \rm N         &=& \rm ICF(N)~{\cdot}~\frac{N^{+}}{H^{+}}, \nonumber\\
 \rm ICF(N)    &=& \rm 10^{0.64{\omega}}~\cdot~\frac{O}{O^{+}}.\label{EQ:N1}
\end{eqnarray}
while the P-I model of M1-11 indicates 
\begin{equation}
 \rm ICF(N)    = 1.14~\cdot~\frac{O}{O^{+}}.\label{EQ:N2}
\end{equation}
The ICF(N)s given by Equations (\ref{EQ:N1}) and (\ref{EQ:N2}) are very similar, 
1.08 $\pm$ 0.04 and 1.18 $\pm$ 0.05, respectively. 
For \mbox{Lin49}, we adopt the ICF(N) calculated by Equation (\ref{EQ:N1}).

The ICFs of Ne, Cl, and Ar are based on the P-I model of M1-11 \citep{Otsuka:2013aa}. 
Following this model, the Ne and Cl elemental abundances are derived as:
\begin{eqnarray}
 \rm Ne        &=& \rm ICF(Ne)~{\cdot}~\frac{Ne^{+}}{H^{+}}, \nonumber\\
 \rm Cl        &=& \rm ICF(Cl)~{\cdot}~\left(\frac{Cl^{+}}{H^{+}} + \frac{Cl^{2+}}{H^{+}}\right).
\end{eqnarray}
The model predicts an Ne$^{+}$ fraction of 0.995, which supports 
the adopted ICF(Ne) = 1.0. The predicted fractions of Cl$^{+}$ and Cl$^{2+}$ 
(0.169 and 0.832, respectively) were consistent with those of S$^{+}$ and S$^{2+}$. 
Since we confirm that the ICF(S) = 1.0 in \mbox{Lin49}, we also adopt a value 
equal to unity for ICF(Cl). 

With the similarity of the Cl$^{2+}$ and Ar$^{2+}$ ionisation potentials and 
the predicted fraction of Ar$^{2+}$ (0.308), we adopt 
\begin{eqnarray}
 \rm ICF(Ar)   &=& \rm 2.7~\cdot~\frac{Cl}{Cl^{2+}}, \nonumber \\ 
 \rm Ar        &=& \rm ICF(Ar)~{\cdot}~\frac{Ar^{2+}}{H^{+}}.
\end{eqnarray}

Finally, from the P-I model of M1-11, we obtain the following expression for 
the Fe elemental abundance: 
\begin{eqnarray}
 \rm ICF(Fe)   &=& \rm 1.02~\cdot~\frac{O}{O^{+}}, \nonumber \\ 
  \rm Fe        &=& \rm ICF(Fe)~{\cdot}~\frac{Fe^{2+}}{H^{+}},
\end{eqnarray}
which gives ICF(Fe) = 1.05 $\pm$ 0.04. This result is similar to the 
one given by the ICF(Fe) from \citet{Delgado-Inglada:2014aa}:
\begin{eqnarray}
\rm ICF(Fe)   &=& \rm 0.9~\cdot~\left(\frac{O^{+}}{O^{2+}}\right)^{0.08}~\cdot~\frac{O}{O^{+}}.
\end{eqnarray}
which gives ICF(Fe) = 1.22 $\pm$ 0.05. We adopt the latter ICF(Fe) in the present work.

\section{Tables}

\begin{table*}
\centering
\footnotesize
\caption{The detected and identified emission lines in the VLT/XSHOOTER
 spectrum of Lin49.}
\begin{tabular}{@{}
c@{\hspace{4pt}}c@{\hspace{4pt}}r@{\hspace{4pt}}c@{\hspace{4pt}}r@{\hspace{4pt}}r@{\hspace{15pt}}
c@{\hspace{4pt}}c@{\hspace{4pt}}r@{\hspace{4pt}}c@{\hspace{4pt}}r@{\hspace{4pt}}r@{\hspace{15pt}}
r@{\hspace{4pt}}r@{\hspace{4pt}}r@{\hspace{4pt}}c@{\hspace{4pt}}r@{\hspace{4pt}}r@{}}
\hline
$\lambda_{\rm obs.}$ &$\lambda_{\rm lab.}$&$f$($\lambda$)&Ion&$I$($\lambda$)&$\delta~I$($\lambda$)&
$\lambda_{\rm obs.}$ &$\lambda_{\rm lab.}$&$f$($\lambda$)&Ion&$I$($\lambda$)&$\delta~I$($\lambda$)&
\multicolumn{1}{c}{$\lambda_{\rm obs.}$} &\multicolumn{1}{c}{$\lambda_{\rm lab.}$}&$f$($\lambda$)&Ion&$I$($\lambda$)&$\delta~I$($\lambda$)\\
({\AA})&({\AA}) & & & \multicolumn{2}{c}{($I$({\hb})=100)}&
({\AA})&({\AA}) & & & \multicolumn{2}{c}{($I$({\hb})=100)}&
\multicolumn{1}{c}{({\AA})}&\multicolumn{1}{c}{({\AA})} & & & \multicolumn{2}{c}{($I$({\hb})=100)}\\
\hline
3662.97 & 3661.25 & 0.335 & H31 & 0.199 & 0.021 & 6315.25 & 6312.10 & --0.264 & [S\,{\sc iii}] & 0.306 & 0.010 & 8708.09 & 8703.87 & --0.564 & [Ni\,{\sc ii}] & 0.022 & 0.003 \\ 
3664.07 & 3662.26 & 0.335 & H30 & 0.252 & 0.023 & 6350.26 & 6347.10 & --0.269 & Si\,{\sc ii} & 0.111 & 0.017 & 8716.13 & 8711.70 & --0.565 & N\,{\sc i} & 0.020 & 0.006 \\ 
3665.27 & 3663.40 & 0.335 & H29 & 0.312 & 0.022 & 6366.98 & 6363.78 & --0.271 & [O\,{\sc i}] & 0.537 & 0.014 & 8731.47 & 8727.12 & --0.566 & [C\,{\sc i}]? & 0.036 & 0.005 \\ 
3666.53 & 3664.68 & 0.334 & H28 & 0.346 & 0.031 & 6551.39 & 6548.04 & --0.296 & [N\,{\sc ii}] & 16.754 & 0.300 & 8754.93 & 8750.47 & --0.568 & P12 & 1.014 & 0.032 \\ 
3667.95 & 3666.09 & 0.334 & H27 & 0.492 & 0.030 & 6566.12 & 6562.80 & --0.298 & H3 & 284.700 & 4.423 & 8867.31 & 8862.78 & --0.578 & P11 & 1.287 & 0.038 \\ 
3669.53 & 3667.68 & 0.334 & H26 & 0.705 & 0.035 & 6581.36 & 6578.05 & --0.300 & C\,{\sc ii} & 0.261 & 0.008 & 9019.51 & 9014.91 & --0.590 & P10 & 1.742 & 0.053 \\ 
3671.26 & 3669.46 & 0.334 & H25 & 0.751 & 0.041 & 6586.78 & 6583.46 & --0.300 & [N\,{\sc ii}] & 50.106 & 0.844 & 9073.62 & 9068.60 & --0.594 & [S\,{\sc iii}] & 2.680 & 0.090 \\ 
3673.31 & 3671.48 & 0.333 & H24 & 0.814 & 0.038 & 6681.56 & 6678.15 & --0.313 & He\,{\sc i} & 0.920 & 0.019 & 9222.94 & 9218.25 & --0.604 & Mg\,{\sc ii} & 0.036 & 0.006 \\ 
3675.55 & 3673.76 & 0.333 & H23 & 0.895 & 0.038 & 6719.86 & 6716.44 & --0.318 & [S\,{\sc ii}] & 1.749 & 0.030 & 9233.76 & 9229.01 & --0.605 & P9 & 2.208 & 0.070 \\ 
3678.09 & 3676.36 & 0.332 & H22 & 1.006 & 0.145 & 6734.24 & 6730.81 & --0.320 & [S\,{\sc ii}] & 3.450 & 0.058 & 9248.79 & 9244.26 & --0.606 & Mg\,{\sc ii} & 0.058 & 0.008 \\ 
3681.07 & 3679.35 & 0.332 & H21 & 1.203 & 0.169 & 7005.70 & 7002.12 & --0.356 & O\,{\sc i} & 0.192 & 0.006 & 9535.88 & 9530.60 & --0.625 & [S\,{\sc iii}] & 5.950 & 0.200 \\ 
3684.66 & 3682.81 & 0.331 & H20 & 1.277 & 0.150 & 7068.81 & 7065.18 & --0.364 & He\,{\sc i} & 1.640 & 0.033 & 9550.86 & 9545.97 & --0.626 & P8 & 2.949 & 0.096 \\ 
3688.69 & 3686.83 & 0.330 & H19 & 1.463 & 0.138 & 7103.36 & 7099.80 & --0.369 & [Pb\,{\sc ii}]? & 0.074 & 0.004 & 9829.20 & 9824.13 & --0.643 & [C\,{\sc i}] & 0.051 & 0.009 \\ 
3693.33 & 3691.55 & 0.329 & H18 & 1.608 & 0.152 & 7139.42 & 7135.80 & --0.374 & [Ar\,{\sc iii}] & 0.788 & 0.023 & 9855.19 & 9850.26 & --0.644 & [C\,{\sc i}] & 0.110 & 0.009 \\ 
3698.94 & 3697.15 & 0.328 & H17 & 1.759 & 0.138 & 7234.77 & 7231.33 & --0.387 & C\,{\sc ii} & 0.074 & 0.013 & 10032.74 & 10027.72 & --0.655 & He\,{\sc i} & 0.047 & 0.008 \\ 
3705.63 & 3703.85 & 0.327 & H16 & 1.888 & 0.174 & 7240.11 & 7236.42 & --0.387 & C\,{\sc ii} & 0.170 & 0.010 & 10054.52 & 10049.37 & --0.656 & P7 & 4.325 & 0.147 \\ 
3713.76 & 3711.97 & 0.325 & H15 & 1.970 & 0.134 & 7258.02 & 7254.38 & --0.390 & O\,{\sc i} & 0.103 & 0.011 & 10293.30 & 10286.73 & --0.668 & [S\,{\sc ii}] & 0.300 & 0.034 \\ 
3723.76 & 3721.94 & 0.323 & H14 & 2.849 & 0.148 & 7285.09 & 7281.35 & --0.393 & He\,{\sc i} & 0.231 & 0.010 & 10343.05 & 10336.41 & --0.671 & [S\,{\sc ii}] & 0.329 & 0.025 \\ 
3727.88 & 3726.03 & 0.322 & [O\,{\sc ii}] & 112.976 & 3.685 & 7322.91 & 7318.92 & --0.398 & [O\,{\sc ii}] & 3.248 & 0.120 & 10377.16 & 10370.49 & --0.673 & [S\,{\sc ii}] & 0.156 & 0.028 \\ 
3730.63 & 3728.81 & 0.322 & [O\,{\sc ii}] & 51.359 & 1.890 & 7323.90 & 7319.99 & --0.398 & [O\,{\sc ii}] & 6.491 & 0.158 & 10404.42 & 10397.74 & --0.674 & [N\,{\sc i}] & 0.051 & 0.015 \\ 
3736.20 & 3734.37 & 0.321 & H13 & 3.078 & 0.155 & 7333.45 & 7329.66 & --0.400 & [O\,{\sc ii}] & 4.046 & 0.107 & 10611.83 & 10605.00 & --0.684 & Ca\,{\sc i}? & 0.116 & 0.009 \\ 
3751.99 & 3750.15 & 0.317 & H12 & 3.695 & 0.173 & 7334.55 & 7330.73 & --0.400 & [O\,{\sc ii}] & 3.471 & 0.085 & 10660.29 & 10653.04 & --0.687 & N\,{\sc i}? & 0.036 & 0.015 \\ 
3772.45 & 3770.63 & 0.313 & H11 & 4.473 & 0.184 & 7446.02 & 7442.30 & --0.415 & N\,{\sc i} & 0.049 & 0.005 & 10837.35 & 10829.89 & --0.695 & He\,{\sc i} & 29.610 & 0.733 \\ 
3799.74 & 3797.90 & 0.307 & H10 & 5.862 & 0.262 & 7472.13 & 7468.31 & --0.418 & N\,{\sc i} & 0.073 & 0.004 & 10945.37 & 10938.10 & --0.700 & P6 & 8.927 & 0.273 \\ 
3837.23 & 3835.38 & 0.299 & H9 & 7.508 & 0.261 & 7755.07 & 7751.10 & --0.455 & [Ar\,{\sc iii}] & 0.176 & 0.007 & 12296.82 & 12288.69 & --0.751 & C\,{\sc ii}? & 0.029 & 0.010 \\ 
3890.88 & 3889.05 & 0.286 & H8 & 13.409 & 0.400 & 7820.14 & 7816.13 & --0.464 & He\,{\sc i} & 0.019 & 0.002 & 12336.58 & 12328.40 & --0.752 & Fe\,{\sc i} & 0.038 & 0.006 \\ 
3920.73 & 3918.97 & 0.279 & C\,{\sc ii} & 0.105 & 0.015 & 8220.45 & 8215.90 & --0.513 & N\,{\sc ii} & 0.061 & 0.004 & 12389.56 & 12381.63 & --0.754 & N\,{\sc i}? & 0.035 & 0.013 \\ 
3922.48 & 3920.68 & 0.279 & C\,{\sc ii} & 0.189 & 0.018 & 8246.37 & 8242.39 & --0.516 & N\,{\sc i} & 0.071 & 0.005 & 12535.55 & 12527.49 & --0.759 & He\,{\sc i} & 0.074 & 0.007 \\ 
3966.61 & 3964.73 & 0.267 & He\,{\sc i} & 0.212 & 0.022 & 8247.89 & 8243.69 & --0.516 & P43 & 0.027 & 0.003 & 12574.74 & 12566.50 & --0.760 & Si\,{\sc iii}? & 0.057 & 0.012 \\ 
3971.99 & 3970.07 & 0.266 & H7 & 16.743 & 0.463 & 8249.89 & 8245.64 & --0.516 & P42 & 0.042 & 0.003 & 12742.46 & 12733.89 & --0.765 & Fe\,{\sc ii}]? & 0.149 & 0.022 \\ 
4028.11 & 4026.18 & 0.251 & He\,{\sc i} & 0.288 & 0.039 & 8251.95 & 8247.73 & --0.516 & P41 & 0.057 & 0.003 & 12793.08 & 12784.91 & --0.766 & He\,{\sc i} & 0.167 & 0.033 \\ 
4070.56 & 4068.60 & 0.239 & [S\,{\sc ii}] & 1.689 & 0.055 & 8254.21 & 8249.97 & --0.517 & P40 & 0.054 & 0.008 & 12798.77 & 12790.35 & --0.767 & He\,{\sc i} & 0.065 & 0.034 \\ 
4078.32 & 4076.35 & 0.237 & [S\,{\sc ii}] & 0.581 & 0.022 & 8256.41 & 8252.40 & --0.517 & P39 & 0.034 & 0.006 & 12826.31 & 12818.08 & --0.767 & P5 & 16.370 & 0.579 \\ 
4103.70 & 4101.73 & 0.230 & H6 & 26.380 & 0.614 & 8259.19 & 8255.02 & --0.517 & P38 & 0.058 & 0.006 & 13173.28 & 13164.55 & --0.777 & Ca\,{\sc i}? & 0.570 & 0.021 \\ 
4269.24 & 4267.26 & 0.180 & C\,{\sc ii} & 0.121 & 0.018 & 8262.09 & 8257.85 & --0.517 & P37 & 0.057 & 0.005 & 14857.58 & 14848.14 & --0.816 & Br30 & 0.043 & 0.010 \\ 
4289.33 & 4287.39 & 0.173 & [Fe\,{\sc ii}] & 0.128 & 0.012 & 8265.09 & 8260.93 & --0.518 & P36 & 0.063 & 0.007 & 14921.34 & 14911.45 & --0.818 & Br27 & 0.045 & 0.011 \\ 
4342.53 & 4340.46 & 0.157 & H5 & 46.910 & 0.773 & 8268.59 & 8264.28 & --0.518 & P35 & 0.088 & 0.008 & 14947.28 & 14937.73 & --0.818 & Br26 & 0.064 & 0.011 \\ 
4361.46 & 4359.33 & 0.151 & [Fe\,{\sc ii}] & 0.064 & 0.012 & 8272.03 & 8267.94 & --0.519 & P34 & 0.058 & 0.005 & 14977.28 & 14967.33 & --0.819 & Br26 & 0.070 & 0.012 \\ 
4365.29 & 4363.21 & 0.149 & [O\,{\sc iii}] & 0.145 & 0.012 & 8276.12 & 8271.93 & --0.519 & P33 & 0.069 & 0.009 & 15010.76 & 15000.86 & --0.819 & Br24 & 0.091 & 0.014 \\ 
4389.95 & 4387.93 & 0.142 & He\,{\sc i} & 0.066 & 0.007 & 8280.64 & 8276.31 & --0.520 & P32 & 0.071 & 0.009 & 15143.43 & 15133.22 & --0.822 & Br21 & 0.113 & 0.008 \\ 
4473.62 & 4471.47 & 0.115 & He\,{\sc i} & 0.964 & 0.028 & 8285.29 & 8281.12 & --0.520 & P31 & 0.091 & 0.015 & 15201.58 & 15191.84 & --0.823 & Br20 & 0.062 & 0.013 \\ 
4573.19 & 4571.10 & 0.084 & Mg\,{\sc i}]? & 0.175 & 0.010 & 8290.73 & 8286.43 & --0.521 & P30 & 0.101 & 0.013 & 15270.63 & 15260.54 & --0.824 & Br19 & 0.124 & 0.010 \\ 
4660.31 & 4658.05 & 0.058 & [Fe\,{\sc iii}] & 0.091 & 0.008 & 8296.46 & 8292.31 & --0.521 & P29 & 0.107 & 0.015 & 15351.91 & 15341.79 & --0.826 & Br18 & 0.119 & 0.008 \\ 
4715.38 & 4713.20 & 0.042 & He\,{\sc i} & 0.070 & 0.007 & 8303.07 & 8298.83 & --0.522 & P28 & 0.108 & 0.011 & 15448.82 & 15438.92 & --0.828 & Br17 & 0.172 & 0.010 \\ 
4863.62 & 4861.33 & 0.000 & H4 & 100.000 & 1.029 & 8310.36 & 8306.11 & --0.523 & P27 & 0.126 & 0.010 & 15566.22 & 15556.45 & --0.830 & Br16 & 0.115 & 0.011 \\ 
4883.55 & 4881.00 & --0.005 & [Fe\,{\sc iii}] & 0.053 & 0.007 & 8318.35 & 8314.26 & --0.524 & P26 & 0.122 & 0.010 & 15710.86 & 15700.66 & --0.832 & Br15 & 0.256 & 0.014 \\ 
4924.21 & 4921.93 & --0.016 & He\,{\sc i} & 0.212 & 0.012 & 8327.67 & 8323.42 & --0.525 & P25 & 0.131 & 0.014 & 15891.00 & 15880.54 & --0.835 & Br14 & 0.205 & 0.020 \\ 
4961.25 & 4958.91 & --0.026 & [O\,{\sc iii}] & 5.367 & 0.075 & 8338.03 & 8333.78 & --0.526 & P24 & 0.185 & 0.015 & 16119.79 & 16109.31 & --0.839 & Br13 & 0.402 & 0.020 \\ 
5009.20 & 5006.84 & --0.038 & [O\,{\sc iii}] & 15.996 & 0.311 & 8349.84 & 8345.55 & --0.527 & P23 & 0.173 & 0.012 & 16417.76 & 16407.19 & --0.844 & Br12 & 0.502 & 0.026 \\ 
5018.03 & 5015.68 & --0.040 & He\,{\sc i} & 0.603 & 0.014 & 8363.28 & 8359.00 & --0.529 & P22 & 0.195 & 0.013 & 16668.39 & 16657.53 & --0.847 & Ca\,{\sc i}? & 0.058 & 0.018 \\ 
5043.35 & 5041.02 & --0.046 & Si\,{\sc ii} & 0.087 & 0.011 & 8378.74 & 8374.48 & --0.531 & P21 & 0.205 & 0.016 & 16817.58 & 16806.52 & --0.850 & Br11 & 0.658 & 0.029 \\ 
5058.43 & 5055.98 & --0.050 & Si\,{\sc ii} & 0.115 & 0.010 & 8396.70 & 8392.40 & --0.533 & P20 & 0.246 & 0.016 & 17373.85 & 17362.11 & --0.857 & Br10 & 0.910 & 0.050 \\ 
5200.38 & 5197.90 & --0.082 & [N\,{\sc i}] & 0.130 & 0.023 & 8417.63 & 8413.32 & --0.535 & P19 & 0.274 & 0.018 & 17432.15 & 17429.49 & --0.858 & Mg\,{\sc i}]? & 0.268 & 0.031 \\ 
5202.72 & 5200.26 & --0.083 & [N\,{\sc i}] & 0.060 & 0.011 & 8442.22 & 8437.95 & --0.537 & P18 & 0.316 & 0.015 & 19457.63 & 19445.56 & --0.881 & Br8 & 1.122 & 0.050 \\ 
5272.88 & 5270.40 & --0.098 & [Fe\,{\sc iii}] & 0.059 & 0.007 & 8450.77 & 8446.48 & --0.538 & O\,{\sc i} & 1.750 & 0.048 & 20595.26 & 20581.28 & --0.891 & He\,{\sc i} & 1.003 & 0.048 \\ 
5301.44 & 5298.83 & --0.104 & [Fe\,{\sc ii}] & 0.050 & 0.007 & 8471.60 & 8467.25 & --0.541 & P17 & 0.359 & 0.015 & 20640.28 & 20626.70 & --0.892 & Si\,{\sc i}? & 0.164 & 0.036 \\ 
5540.35 & 5537.89 & --0.149 & [Cl\,{\sc iii}] & 0.056 & 0.011 & 8506.86 & 8502.48 & --0.544 & P16 & 0.442 & 0.015 & 21669.28 & 21655.29 & --0.900 & Br7 & 2.752 & 0.123 \\ 
5698.76 & 5695.92 & --0.175 & C\,{\sc iii} & 0.206 & 0.021 & 8549.75 & 8545.38 & --0.549 & P15 & 0.523 & 0.017 &  &  &  &  &  &  \\ 
5757.49 & 5754.64 & --0.185 & [N\,{\sc ii}] & 1.241 & 0.047 & 8583.08 & 8578.69 & --0.552 & [Cl\,{\sc ii}] & 0.115 & 0.006 &  &  &  &  &  &  \\ 
5878.61 & 5875.60 & --0.203 & He\,{\sc i} & 3.772 & 0.044 & 8602.77 & 8598.39 & --0.554 & P14 & 0.661 & 0.023 &  &  &  &  &  &  \\ 
5961.53 & 5958.54 & --0.215 & O\,{\sc i} & 0.133 & 0.016 & 8669.44 & 8665.02 & --0.560 & P13 & 0.847 & 0.026 &  &  &  &  &  &  \\ 
5981.85 & 5978.93 & --0.218 & Si\,{\sc ii} & 0.051 & 0.015 & 8684.69 & 8680.53 & --0.562 & Ca\,{\sc iii} & 0.043 & 0.005 &  &  &  &  &  &  \\ 
6049.37 & 6046.44 & --0.228 & O\,{\sc i} & 0.139 & 0.011 & 8687.91 & 8683.40 & --0.562 & N\,{\sc i} & 0.034 & 0.005 &  &  &  &  &  &  \\ 
6303.48 & 6300.30 & --0.263 & [O\,{\sc i}] & 1.608 & 0.026 & 8690.65 & 8686.15 & --0.562 & N\,{\sc i} & 0.014 & 0.004 &  &  &  &  &  &  \\ 
\hline
\end{tabular}
\label{xstab}
\end{table*}

\begin{table}
\centering
\caption{The adopted {\te} and {\Ne} pairs for each ionic abundance
  derivation. \label{T:tene}}
\begin{tabu} to \columnwidth {@{}LLL@{}}
\hline
{\te} & {\Ne}       &Ions\\
\hline
{\te}({\sii})  &{\Ne}({\sii})&N$^{0}$, O$^{0}$, S$^{+}$\\
{\te}(PJ)      &10\,000 cm$^{-3}$ &C$^{2+}$\\
{\te}({\oii})  &{\Ne}({\oii})&O$^{+}$\\
{\te}({\siii})&{\Ne}({\oii})&Ne$^{+}$, S$^{2+}$, Cl$^{2+}$, Ar$^{2+}$\\
10\,860 $\pm$ 210$^{\rm a}$ K\tnote{a}&{\Ne}({\oii})&Fe$^{2+}$, Cl$^{+}$\\ 
{\te}({\oiii})&{\Ne}({\oii})&O$^{2+}$\\
11\,360 $\pm$ 840$^{\rm b}$ K\tnote{b}&10\,000 cm$^{-3}$&He$^{+}$\\
{\te}({\nii})&{\Ne}({\oii})&N$^{+}$\\
\hline     
\end{tabu}
\begin{threeparttable}
\begin{tablenotes}
\footnotesize
\item[a] the average value between {\te}({\oii}) and {\te}({\nii}).
\item[b] the average value between two {\te}({\hei}) derived from the
 {\hei} $I$(7281\,{\AA})/$I$(6678\,{\AA}) and
 $I$(6678\,{\AA}/$I$(5876\,{\AA}). 
\end{tablenotes}
\end{threeparttable}
\end{table}

\begin{table*}
\caption{Ionic and elemental abundances derived from recombination
 lines and collisional excitation lines. The ICF means the ionisation
 correction factor for the unseen ionisation stage ions in the XSHOOTER
 and \emph{Spitzer}/IRS spectra. The ICF(He) determined based on the P-I
 model of M1-11 is 3.81	$\pm$ 0.22, leading He/H = 6.73(--2) $\pm$
 4.29(--3). \label{T:ionica}
}
\begin{tabular}{@{}lcccclcccc@{}}
\hline
Elem.    &Ion&$\lambda_{\rm lab.}$&$I$($\lambda_{\rm
 lab.}$)&X$^{\rm m+}$/H$^{+}$&Elem.    &Ion&$\lambda_{\rm lab.}$&$I$($\lambda_{\rm
 lab.}$)&X$^{\rm m+}$/H$^{+}$\\
(X)      &(X$^{\rm m+}$)&({\AA} or {$\mu$}m)&[$I$({\hb})=100]&&(X) &(X$^{\rm m+}$)&({\AA} or {$\mu$}m)&[$I$({\hb})=100]\\
\hline

He & He$^{+}$ & 3964.73 & 2.12(--1) $\pm$ 2.20(--2) & 1.79(--2) $\pm$ 2.14(--3) & O & O$^{2+}$ & 4363.21 & 1.45(--1) $\pm$ 1.23(--2) & 4.03(--6) $\pm$ 8.04(--7) \\
 &   & 4471.47 & 9.64(--1) $\pm$ 2.77(--2) & 1.83(--2) $\pm$ 4.10(--3) &  &   & 4958.91 & 5.37(+0) $\pm$ 7.45(--2) & 3.90(--6) $\pm$ 3.54(--7) \\
 &   & 4921.93 & 2.12(--1) $\pm$ 1.24(--2) & 1.55(--2) $\pm$ 1.32(--3) &  &   & 5006.84 & 1.60(+1) $\pm$ 3.11(--1) & 4.03(--6) $\pm$ 3.69(--7) \\
 &   & 5015.68 & 6.03(--1) $\pm$ 1.39(--2) & 1.94(--2) $\pm$ 1.22(--3) &
		     &  &  &    & {\bf 3.97(--6) $\pm$ 2.43(--7)} \\
 &    & 5875.60 & 3.77(+0) $\pm$ 4.40(--2) & 2.39(--2) $\pm$ 5.60(--3) &  &  &  & ICF(O)   & 1   \\
 &   & 6678.15 & 9.20(--1) $\pm$ 1.88(--2) & 2.29(--2) $\pm$ 1.58(--3) &
		     &   &  &    & {\bf 1.28(--4) $\pm$ 3.66(--6)} \\
 &   & 7281.35 & 2.31(--1) $\pm$ 1.03(--2) & 2.02(--2) $\pm$ 1.38(--3) & Ne & Ne$^{+}$ & 12.81 & 1.12(+1) $\pm$ 1.16(+0) & 1.49(--5) $\pm$ 1.56(--6) \\
 &   & 10829.9 & 2.96(+1) $\pm$ 7.33(--1) & 1.57(--2) $\pm$ 1.05(--3) &  &  &  & ICF(Ne)    & 1 \\
 &   & 12527.5 & 7.44(--2) $\pm$ 7.13(--3) & 4.47(--2) $\pm$ 4.87(--3) &
		     &   &  &    & {\bf 1.49(--5) $\pm$ 1.56(--6)} \\
 &   & 12784.9 & 1.67(--1) $\pm$ 3.30(--2) & 2.33(--2) $\pm$ 5.11(--3) & S & S$^{+}$ & 4068.60 & 1.69(+0) $\pm$ 5.47(--2) & 4.22(--7) $\pm$ 7.03(--8) \\
 &   & 12790.4 & 6.46(--2) $\pm$ 3.38(--2) & 2.74(--2) $\pm$ 1.45(--2) &  &   & 4076.35 & 5.81(--1) $\pm$ 2.24(--2) & 4.47(--7) $\pm$ 7.51(--8) \\
 &   & 20581.3 & 9.98(--1) $\pm$ 4.83(--2) & 1.45(--2) $\pm$ 1.21(--3) &  &   & 6716.44 & 1.75(+0) $\pm$ 3.00(--2) & 3.94(--7) $\pm$ 9.92(--9) \\
 &  &  &     & {\bf 1.77(--2) $\pm$ 4.93(--4)} &  &   & 6730.81 & 3.45(+0) $\pm$ 5.83(--2) & 3.92(--7) $\pm$ 6.74(--9) \\
 &  &  & ICF(He)   & 5.53 $\pm$ 0.14 &  &   & 10286.7 & 3.00(--1) $\pm$ 3.39(--2) & 3.20(--7) $\pm$ 6.36(--8) \\
 &   &  &    & {\bf 9.78(--2) $\pm$ 3.63(--3)} &  &   & 10336.4 & 3.29(--1) $\pm$ 2.52(--2) & 3.56(--7) $\pm$ 6.43(--8) \\
C & C$^{2+}$ & 3918.97 & 1.05(--1) $\pm$ 1.53(--2) & 5.31(--3) $\pm$ 3.51(--3) &  &   & 10370.5 & 1.56(--1) $\pm$ 2.77(--2) & 3.54(--7) $\pm$ 8.57(--8) \\
 &   & 3920.68 & 1.89(--1) $\pm$ 1.85(--2) & 5.31(--3) $\pm$ 9.59(--4) &
		     &  &  &    & {\bf 3.92(--7) $\pm$ 5.49(--9)} \\
 &   & 4267.26 & 1.21(--1) $\pm$ 1.82(--2) & 1.17(--4) $\pm$ 2.29(--5) &  & S$^{2+}$ & 6312.10 & 3.06(--1) $\pm$ 1.05(--2) & 6.44(--7) $\pm$ 6.41(--8) \\
 &   & 6578.05 & 2.61(--1) $\pm$ 7.61(--3) & 3.36(--4) $\pm$ 5.32(--5) &  &   & 9068.60 & 2.68(+0) $\pm$ 8.97(--2) & 6.43(--7) $\pm$ 3.57(--8) \\
 &   & 7236.42 & 1.70(--1) $\pm$ 9.54(--3) & 2.32(--4) $\pm$ 3.64(--5) &
		     &  &  &     & {\bf 6.43(--7) $\pm$ 3.12(--8)} \\
 &   & 7231.33 & 7.41(--2) $\pm$ 1.25(--2) & 1.82(--4) $\pm$ 4.06(--5) &  &  &  & ICF(S)   & 1   \\
 &  &   &     & {\bf 1.17(--4) $\pm$ 2.29(--5)} &  &   &  &    & {\bf 1.04(--6) $\pm$ 3.17(--8)} \\
 &  &  & ICF(C)   & 3.96 $\pm$ 0.23 & Cl & Cl$^{+}$ & 8578.69 & 1.15(--1) $\pm$ 5.52(--3) & 4.50(--9) $\pm$ 2.72(--10) \\
 &   &  &    & {\bf 4.62(--4) $\pm$ 9.43(--5)} &  & Cl$^{2+}$ & 5537.89 & 5.65(--2) $\pm$ 1.12(--2) & 6.32(--9) $\pm$ 1.31(--9) \\
N & N$^{0}$ & 5197.90 & 1.30(--1) $\pm$ 2.33(--2) & 6.15(--7) $\pm$ 1.12(--7) &  &  &  & ICF(Cl)   & 1 \\
 &   & 5200.26 & 5.96(--2) $\pm$ 1.15(--2) & 6.63(--7) $\pm$ 1.28(--7) &
		     &   &  &    & {\bf 1.08(--8) $\pm$ 1.33(--9)} \\
 &   & 10397.7 & 5.11(--2) $\pm$ 1.48(--2) & 5.75(--7) $\pm$ 2.10(--7) & Ar & Ar$^{2+}$ & 7135.80 & 7.88(--1) $\pm$ 2.33(--2) & 6.77(--8) $\pm$ 3.92(--9) \\
 &  &  &    & {\bf 6.28(--7) $\pm$ 7.81(--8)} &  &   & 7751.10 & 1.76(--1) $\pm$ 7.03(--3) & 6.32(--8) $\pm$ 4.03(--9) \\
 & N$^{+}$ & 5754.64 & 1.24(+0) $\pm$ 4.66(--2) & 7.29(--6) $\pm$
		 9.10(--7) &  &  &  &    & {\bf 6.69(--8) $\pm$ 3.94(--9)} \\
 &   & 6548.04 & 1.68(+1) $\pm$ 3.00(--1) & 7.18(--6) $\pm$ 2.45(--7) &  &  &  & ICF(Ar)   & 4.62 $\pm$ 1.11 \\
 &   & 6583.46 & 5.01(+1) $\pm$ 8.44(--1) & 7.25(--6) $\pm$ 2.43(--7) &
		     &   &  &    & {\bf 3.03(--7) $\pm$ 7.48(--8)} \\
 &   &  &    & {\bf 7.22(--6) $\pm$ 1.70(--7)} & Fe & Fe$^{2+}$ & 4658.05 & 9.13(--2) $\pm$ 7.78(--3) & 2.95(--8) $\pm$ 2.76(--9) \\
 &  &  & ICF(N)   & 1.08 $\pm$ 0.04 &  &   & 4881.00 & 5.35(--2) $\pm$ 6.97(--3) & 2.38(--8) $\pm$ 3.58(--9) \\
 &   &  &    & {\bf 7.80(--6) $\pm$ 3.69(--7)} &  &   & 5270.40 & 5.94(--2) $\pm$ 7.23(--3) & 3.92(--8) $\pm$ 4.98(--9) \\
O & O$^{0}$ & 6300.30 & 1.61(+0) $\pm$ 2.55(--2) & 5.30(--6) $\pm$
		 1.25(--6) &  &  &  &    & {\bf 2.92(--8) $\pm$ 2.00(--9)} \\
 &   & 6363.78 & 5.37(--1) $\pm$ 1.42(--2) & 5.53(--6) $\pm$ 1.31(--6) &  & &  &ICF(Fe)    & 1.22 $\pm$ 0.05 \\
 &  &  &    & {\bf 5.41(--6) $\pm$ 9.03(--7)} &  &   &  &    & {\bf 3.57(--8) $\pm$ 2.85(--9)} \\
 & O$^{+}$ & 3726.03 & 1.13(+2) $\pm$ 3.68(+0) & 1.23(--4) $\pm$ 4.30(--6) &  &  &  &    &    \\
 &   & 3728.81 & 5.14(+1) $\pm$ 1.89(+0) & 1.23(--4) $\pm$ 8.70(--6) &  &  &  &    &    \\
 &   & 7318.92 & 3.25(+0) $\pm$ 1.20(--1) & 1.76(--4) $\pm$ 3.16(--5) &  &  &  &    &    \\
 &   & 7319.99 & 6.49(+0) $\pm$ 1.58(--1) & 1.14(--4) $\pm$ 1.99(--5) &  &  &  &    &    \\
 &   & 7329.66 & 4.05(+0) $\pm$ 1.07(--1) & 1.33(--4) $\pm$ 2.35(--5) &  &  &  &    &    \\
 &   & 7330.73 & 3.47(+0) $\pm$ 8.53(--2) & 1.16(--4) $\pm$ 2.03(--5) &  &  &  &    &    \\
 &  &  &    & {\bf 1.24(--4) $\pm$ 3.65(--6)} &  &  &  &    &    \\
\hline
\end{tabular}
\label{}
\end{table*}

\begin{table*}
\caption{Comparison between SED modellings and observations. In the third
 and eighth columns, the predicted values by the single shell model as
 we discussed in section~\ref{photomodel} are given. The values in the fourth and
 ninth columns are the predictions by the two density shell model in
 section~\ref{S:highdens}. \label{model2}}
\begin{tabu} to \textwidth {@{}LLRRRLLRRR@{}}
\hline
Ion & $\lambda_{\rm lab.}$ & $I$(ModelS) & $I$(ModelT) & $I$(Obs) & Ion &
 $\lambda_{\rm lab.}$ & $I$(ModelS) & $I$(ModelT) & $I$(Obs) \\ 
 &  & [$I$({\hb})=100] & [$I$({\hb})=100] & [$I$({\hb})=100] &  &  & [$I$({\hb})=100] & [$I$({\hb})=100] & [$I$({\hb})=100] \\ 
\hline
{\oiii} & 1661/65\,{\AA} & 0.077 & 0.091 & $\cdots$ & {\ariii} & 7135\,{\AA} & 0.766 & 0.773 & 0.788 \\ 
N\,{\sc iii}$]$ & 1747/49\,{\AA} & 0.251 & 0.242 & $\cdots$ & {\hei} & 7281\,{\AA} & 0.245 & 0.242 & 0.231 \\ 
C\,{\sc iii}$]$ & 1906/09\,{\AA} & 49.742 & 49.380 & $\cdots$ & {\oii} & 7323\,{\AA} & 10.961 & 10.811 & 9.740 \\ 
$[$C\,{\sc ii}$]$ & 2326\,{\AA} & 182.426 & 184.421 & $\cdots$ & {\oii} & 7332\,{\AA} & 8.754 & 8.635 & 7.517 \\ 
{\oii} & 2471\,{\AA} & 14.829 & 14.627 & $\cdots$ & {\ariii} & 7751\,{\AA} & 0.185 & 0.186 & 0.176 \\ 
{\oii} & 3726\,{\AA} & 140.734 & 138.440 & 112.976 & {\clii} & 8579\,{\AA} & 0.114 & 0.113 & 0.115 \\ 
{\oii} & 3729\,{\AA} & 68.410 & 67.314 & 51.359 & {\sii} & 9069\,{\AA} & 4.269 & 4.249 & 2.680 \\ 
{\hei} & 3965\,{\AA} & 0.331 & 0.330 & 0.212 & {\sii} & 1.029\,$\mu$m & 0.280 & 0.279 & 0.300 \\ 
{\sii} & 4070\,{\AA} & 1.198 & 1.197 & 1.689 & {\sii} & 1.034\,$\mu$m & 0.274 & 0.274 & 0.329 \\ 
{\sii} & 4078\,{\AA} & 0.388 & 0.387 & 0.581 & {\sii} & 1.037\,$\mu$m & 0.131 & 0.131 & 0.156 \\ 
{\hi} & 4102\,{\AA} & 26.729 & 26.739 & 26.380 & {\hi} & 1.094\,$\mu$m & 9.069 & 9.009 & 8.927 \\ 
C\,{\sc ii} & 4267\,{\AA} & 0.073 & 0.072 & 0.121 & {\hei} & 1.253\,$\mu$m & 0.068 & 0.071 & 0.074 \\ 
{\hi} & 4340\,{\AA} & 47.483 & 47.500 & 46.910 & {\hei} & 1.278\,$\mu$m & 0.191 & 0.190 & 0.167 \\ 
{\oiii} & 4363\,{\AA} & 0.122 & 0.185 & 0.145 & {\hei} & 1.279\,$\mu$m & 0.064 & 0.063 & 0.065 \\ 
{\hei} & 4388\,{\AA} & 0.156 & 0.155 & 0.066 & {\hi} & 1.282\,$\mu$m & 16.126 & 16.125 & 16.370 \\ 
{\hei} & 4471\,{\AA} & 1.249 & 1.243 & 0.964 & {\hi} & 1.736\,$\mu$m & 0.929 & 0.930 & 0.910 \\ 
{\feiii} & 4659\,{\AA} & 0.110 & 0.106 & 0.091 & {\hei} & 2.058\,$\mu$m & 1.137 & 1.121 & 0.998 \\ 
{\feiii} & 4881\,{\AA} & 0.040 & 0.039 & 0.053 & {\hi} & 2.166\,$\mu$m & 2.742 & 2.743 & 2.752 \\ 
{\hei} & 4922\,{\AA} & 0.337 & 0.335 & 0.212 & {\arii} & 6.99\,$\mu$m & 1.320 & 1.352 & $\cdots$ \\ 
{\oiii} & 4959\,{\AA} & 4.703 & 4.779 & 5.367 & {\hi} & 7.46\,$\mu$m & 2.562 & 2.561 & $\cdots$ \\ 
{\oiii} & 5007\,{\AA} & 14.155 & 14.384 & 15.996 & {\ariii} & 9.00\,$\mu$m & 0.626 & 0.623 & $\cdots$ \\ 
{\hei} & 5016\,{\AA} & 0.817 & 0.812 & 0.603 & {\siv} & 10.52\,$\mu$m & 0.019 & 0.018 & $\cdots$ \\ 
{\NI} & 5198\,{\AA} & 0.028 & 0.028 & 0.130 & {\hi} & 11.31\,$\mu$m & 0.311 & 0.310 & $\cdots$ \\ 
{\NI} & 5200\,{\AA} & 0.017 & 0.017 & 0.060 & {\hi} & 12.37\,$\mu$m & 0.992 & 0.991 & $\cdots$ \\ 
{\feiii} & 5271\,{\AA} & 0.062 & 0.060 & 0.059 & {\neii} & 12.81\,$\mu$m & 12.645 & 14.168 & 11.157 \\ 
{\cliii} & 5538\,{\AA} & 0.080 & 0.078 & 0.056 & {\neiii} & 15.55\,$\mu$m & 0.068 & 0.079 & $\cdots$ \\ 
{\nii} & 5755\,{\AA} & 1.127 & 1.127 & 1.241 & {\siii} & 18.67\,$\mu$m & 3.958 & 3.913 & $\cdots$ \\ 
{\hei} & 5876\,{\AA} & 3.305 & 3.799 & 3.772 & {\hi} & 19.06\,$\mu$m & 0.438 & 0.437 & $\cdots$ \\ 
{\oi} & 6300\,{\AA} & 1.367 & 1.374 & 1.608 & {\siii} & 33.47\,$\mu$m & 1.341 & 1.326 & $\cdots$ \\ 
{\siii} & 6312\,{\AA} & 0.352 & 0.352 & 0.306 & {\neiii} & 36.01\,$\mu$m & 0.005 & 0.006 & $\cdots$ \\ 
{\oi} & 6363\,{\AA} & 0.436 & 0.438 & 0.537 & {\oiii} & 51.80\,$\mu$m & 0.902 & 0.821 & $\cdots$ \\ 
{\nii} & 6548\,{\AA} & 19.214 & 19.164 & 16.754 & $[$N\,{\sc iii}$]$ & 57.21\,$\mu$m & 0.273 & 0.259 & $\cdots$ \\ 
{\hi} & 6563\,{\AA} & 286.554 & 286.602 & 284.700 & {\oi} & 63.17\,$\mu$m & 0.211 & 0.212 & $\cdots$ \\ 
{\nii} & 6584\,{\AA} & 56.701 & 56.552 & 50.106 & {\oiii} & 88.33\,$\mu$m & 0.146 & 0.133 & $\cdots$ \\ 
{\hei} & 6678\,{\AA} & 1.025 & 1.018 & 0.920 & {\nii} & 121.7\,$\mu$m & 0.108 & 0.108 & $\cdots$ \\ 
{\sii} & 6716\,{\AA} & 1.881 & 1.879 & 1.749 & {\oi} & 145.5\,$\mu$m & 0.012 & 0.012 & $\cdots$ \\ 
{\sii} & 6731\,{\AA} & 3.325 & 3.319 & 3.450 & $[$C\,{\sc ii}$]$ & 157.6\,$\mu$m & 0.722 & 0.724 & $\cdots$ \\ 
{\hei} & 7065\,{\AA} & 1.550 & 1.576 & 1.431 & {\nii} & 205.4\,$\mu$m & 0.012 & 0.012 & $\cdots$ \\ 
\hline
Band & $\lambda_{\rm c}$ & $I$(ModelS) &$I$(ModelT)  & $I$(Obs) & X &  & $\epsilon$(X)$_{\rm ModelS}$ &$\epsilon$(X)$_{\rm ModelT}$  &  $\epsilon$(X)$_{\rm Obs}$ \\ 
  &   & [$I$({\hb})=100] & [$I$({\hb})=100] & [$I$({\hb})=100] &  &  &  &  &  \\ 
\hline
MCPSV & 5450\,{\AA} & 42.396 & 46.315 & 38.095 & He &  & 10.80 & 10.80 & 10.80-11.01 \\ 
2MAJ & 1.235\,$\mu$m & 61.385 & 91.351 & 96.770 & C &  & 8.46 & 8.46 & 8.46 \\ 
2MAH & 1.662\,$\mu$m & 31.233 & 131.036 & 114.920 & N &  & 7.06 & 7.06 & 6.93 \\ 
2MAK & 2.159\,$\mu$m & 20.788 & 156.257 & 139.370 & O &  & 8.03 & 8.03 & 8.11 \\ 
IRA3 & 3.600\,$\mu$m & 28.570 & 251.534 & 236.040 & Ne &  & 7.22 & 7.27 & 7.18 \\ 
IRA4 & 4.500\,$\mu$m & 37.205 & 191.422 & 240.080 & S &  & 5.88 & 5.88 & 6.02 \\ 
LIS1 & 15.00\,$\mu$m & 36.248 & 36.749 & 44.550 & Cl &  & 4.09 & 4.08 & 4.03 \\ 
LIS2 & 23.35\,$\mu$m & 64.739 & 63.899 & 53.290 & Ar &  & 5.21 & 5.22 & 5.48 \\ 
LIS3 & 36.50\,$\mu$m & 26.733 & 26.192 & 24.710 & Fe &  & 4.79 & 4.71 & 4.55 \\ 
\hline
Band & $\lambda_{\rm c}$ & $F_{\nu}$(ModelS) &$F_{\nu}$(ModelT)  & $F_{\nu}$(Obs) &  &  &  &  &  \\ 
 &  & (mJy) &(mJy)  & (mJy) &  &  &  &  &  \\ 
\hline
FIT1 & 65.00\,$\mu$m & 9.255 & 9.023 & 7.415 &  &  &  &  &  \\ 
FIT2 & 90.00\,$\mu$m & 3.419 & 3.331 & 4.470 &  &  &  &  &  \\ 
FIT3 & 120.0\,$\mu$m & 1.400 & 1.366 & 2.465 &  &  &  &  &  \\ 
\hline
\end{tabu}
\end{table*}

\begin{table*}
\centering
\caption{Effective temperature of the central star ($T_{\rm eff}$), nebular radius
 ($r$), and nebular elemental abundances in non C$_{60}$-containing C-rich
 PNe in the SMC. We recalculated the C in SMC17 using the
 $F$($[$C\,{\sc iii}\,1906/09\,{\AA}) and $F$({\hb}) of
 \citet{Aller:1987aa}, $c$({\hb}), {\te}({\oiii})=12200~K, and
 {\Ne}=2900 cm$^{-3}$ of \citet{Shaw:2010aa}, and the ICF(C)=O/O$^{2+}$
 derived from the elemental O and ionic O$^{2+}$ abundances of
 \citet{Shaw:2010aa}. 
}
\begin{tabu} to \textwidth {@{}l@{\hspace{24pt}}c@{\hspace{24pt}}c@{\hspace{24pt}}c@{\hspace{24pt}}c@{\hspace{24pt}}c@{\hspace{24pt}}c@{\hspace{24pt}}c@{\hspace{24pt}}c@{\hspace{24pt}}c@{\hspace{24pt}}l@{}}
\hline
Non-C$_{60}$ PNe & $T_{\rm eff}$ (K)& $r$ (${\arcsec}$)&$\epsilon$(He) & $\epsilon$(C) & $\epsilon$(N) & $\epsilon$(O) & $\epsilon$(Ne) & $\epsilon$(S) & $\epsilon$(Ar) & References \\
\hline
SMC2 & 111\,500 & 0.25 & 11.10 &8.74  & 7.47 & 8.01 & 7.21 & 7.32 & 5.78 &(1),(2),(3),(4)  \\ 
SMC5 & 137\,500 & 0.31 & 11.11 & 8.68 & 7.76 & 8.24 & 7.43 & 7.68 & 6.01&(1),(3),(5),(6)  \\ 
SMC6 & 28\,200 & 0.19 & 10.99  & 8.35 & 8.06 & 7.99 & 7.14 & 7.37 & 5.70 &(1),(7),(8),(9)  \\ 
SMC11 & 40\,900 & 0.99 & 11.01 &$\cdots$  & 6.52 & 8.02 & 6.90 & 6.28 & 5.97 &(7),(10),(9)  \\ 
SMC14 & 83\,500 & 0.42 & 11.13 & 9.16 & 7.36 & 8.29 & 7.65 & $>$~6.27 & 5.82 &(1),(7),(9),(10)  \\ 
SMC17 & 58\,400 & 0.25 & 11.14 & 8.63  & 7.38 & 8.21 & 7.67 & $>$~6.15 &5.56 &(4),(7),(9),(10),(11)  \\ 
SMC19 & 59\,400 & 0.30 & 11.09 & 8.97 & 7.28 & 8.19 & 7.62 & $>$~6.05 & 5.54 &(1),(7),(9),(10)  \\ 
SMC20 & 86\,500 & 0.15 & 11.14 & 8.25 & 6.95 & 7.74 & 6.91 & 5.67 & 5.03 &(7),(8),(9),(10)  \\ 
SMC27 & 43\,300 & 0.23 & 10.99 & 8.58 & 7.55 & 8.03 & 7.03 & 5.84 & 5.32 &(7),(9),(10),(12) \\ 
Average & 72\,100 & 0.34 & 11.08 & 8.67 & 7.37 & 8.08& 7.28 & 6.69 & 5.64\\
\hline
\end{tabu}
\begin{threeparttable}
\begin{minipage}{\textwidth}
References -- 
(1) \citet{Leisy:2006aa} for abundances;
(2) \citet{Shaw:2006aa} for $r$;
(3) \citet{Liu:1995aa} for $T_{\rm eff}$;
(4) \citet{Aller:1987aa};
(5) \citet{Vassiliadis:1998aa} for the C/O ratio of 2.779 in SMC5;
(6) \citet{Stanghellini:1999aa} for $r$;
(7) \citet{Stanghellini:2003aa} for $r$;
(8) \citet{Stanghellini:2009aa} for C;
(9) \citet{Villaver:2004aa} for $T_{\rm eff}$;
(10) \citet{Shaw:2010aa} for abundances except C;
(11) this work;
(12) \citet{Tsamis:2003aa} for abundances except S. 
\end{minipage}
\end{threeparttable}
\label{T:compSMCXS2}
\end{table*}

\label{lastpage}

\end{document}